\documentclass[review,3p,times,11pt]{elsarticle}

\usepackage{txfonts}
\AtBeginDocument{\mathcode`v=\varv}
\usepackage{lineno,hyperref}
\usepackage[]{hyperref}

\usepackage{mathrsfs} 

\usepackage[dvipsnames]{xcolor}
\usepackage[fleqn]{amsmath}
\usepackage{amsfonts}
\usepackage[T1]{fontenc}
\usepackage{lmodern} 
\usepackage{calligra} 
\usepackage{amsmath}

\usepackage{graphicx,epsfig}
\RequirePackage{color}
\usepackage{graphicx,subfigure}
\usepackage{indentfirst}
\hypersetup{
	colorlinks = true,
	linkcolor = blue,
	urlcolor= black,
	citecolor = blue,
	anchorcolor = black}
\usepackage{epstopdf}
\usepackage{physics}
 \makeatletter
\input epsf

\newcounter{saveeqn}

\begin{document}
	\begin{frontmatter}
		
		\title{Stability of soluble surfactant-laden falling film over a hydrophobic incline in the presence of external shear}
		\author[add1]{Dipankar Paul}
		\ead{dipankarpaul3103@gmail.com}
		\author[add1,add2]{Harekrushna Behera }
		\ead{hkb.math@gmail.com}
		
		\author[add3]{Sukhendu Ghosh}
		\ead{sukhendu.math@gmail.com}
		
		\address[add1]{Department of Mathematics, SRM Institute of Science and Technology, Kattankulathur, Tamil Nadu - 603203, India.}
		\address[add2]{Center of Excellence for Ocean Engineering, National Taiwan Ocean University, Keelung 202301, Taiwan}
		\address[add3]{Department of Mathematics, Indian Institute of Technology Jodhpur, Rajasthan-342037, India}
		
		\begin{abstract}
			The hydrodynamic stability analysis of gravity-driven, soluble surfactant-laden fluid streaming down a slippery, slanted plane in the presence of external shear force is being explored in this article. The Navier-Stokes equations are considered for the fluid flow along with the appropriate advection-diffusion equations for the concentration of different surfactant species. The monomers considered here are anticipated to dissolve in the bulk flow and can be adsorbed at the interface of air-liquid as well. Also, the adsorption-desorption kinetics of the surfactants at the free space is taken into consideration. The motivation behind this work is to extend the work of \citet{karapetsas2014role} for flow over a slippery bottom and in the presence of externally imposed shear forces and observe their impact on the flow dynamics. The Orr-Sommerfeld eigensystem is obtained, then it is solved analytically using the longwave approximation method in the longwave regime ($k \ll 1$) and subsequently, the Chebyshev spectral collocation method is employed for numerical evaluation in the arbitrary wave regime. Using the analytical method, two longwave modes, viz, surface mode and surfactant mode, are detected. Alternatively, the numerical analysis substantiated the existence of temporal surface and surfactant modes. Moreover, another temporal mode named shear mode arises in the high modified Reynolds number region at a low inclination angle. Thereafter, the modified Reynolds-Orr energy equation is deduced under the normal mode conditions, and the behaviour of different energy components is investigated for various slip parameters and imposed shear force.
		\end{abstract}
		
	\end{frontmatter}
	\newpage
	\section{Introduction}

	SURFace ACTive AgeNTS or SURFACTANTS are unique molecules with hydrophilic and hydrophobic moieties that are mostly positioned at the interfaces of two fluid phases with distinct hydrogen bonding and polarity, e.g., air-water and oil-water interfaces and so on. These properties help surfactants to reduce surface tensions and, in some cases, interfacial tensions as well and create microemulsions. As a consequence, surfactants have widespread applications such as foaming agents, emulsifiers, dispersing traits, detergents, wetting, antimicrobial activity \citep{greek1991sales, desai1997microbial, de2015review} etc tailored as per industrial requirements. Moreover, the appearance of surfactants in the biomedical field, such as naturally forming in the lungs\citep{halpern1993surfactant, guerra1984pseudomonas, craster2009dynamics}, and assisting in the treatment of respiratory distress syndrome \citep{grotberg1994pulmonary, halliday2008surfactants}, and among other miscellaneous applications\citep{banat1995biosurfactants, banat1995characterization, fiechter1992biosurfactants, santos2016biosurfactants, diez2001contact}. The earliest available theoretical exploration of the influence of surfactant on gravity-driven thin film flow over an inclined plane was done by \citet{benjamin1964effects, whitaker1964effect} using a linear stability analysis for insoluble surfactants. It is reported by both the studies that the surfactants proliferate stability of the flow by impacting the elasticity of the flow surface, which authenticates the experimental results of \citet{emmert1954study, stirba1955turbulence} for fluid surfaces contaminated by surfactant. There are numerous research articles both experimental \citep{ fulford1964flow, skelland1972effects, roche2014marangoni, kim2017solutal} as well as theoretical \citep{ji1994instabilities, blyth2004effect,pereira2008dynamics,hu2020falling} examining the effect of surfactants, mainly for thin film flow or droplet-related studies, indicating the importance of surfactant related studies.\\
	\indent In general, there are largely two types of surfactants viz., $(i)$ insoluble and $(ii)$ soluble. Insoluble surfactants are a kind of surfactants that resides on the surface of the flow and does not absorb in the bulk flow. It is well known fact that the interfacial tension between fluid layers is decreased by surfactants, and its gradient imposes Marangoni stress on the interfaces of the fluid. There are extensive investigations available in the literature on the significance of insoluble surfactants in hydrodynamics stability analysis for thin film flow over inclined planes \citep{matar2004rupture, pal2021linear, hossain2022linear}. The primary stability analysis of \citet{blyth2004effect} observes that a stable Marangoni mode (surfactant mode or concentration mode) and an unstable Yih mode (surface mode) exist for Stokes flow, and a shear mode also exists, and surfactant helps to increase the critical Reynolds number to make the flow more stable. They also mentioned that the growth rate of the Yih mode is faster than that of the Marangoni mode when the Reynolds number starts to increase, and it becomes dominant after a certain threshold value of the Reynolds number. Also, \citet{wei2005effect} choose the same problem with consideration for externally imposed shear and do hydrodynamics analysis in long wave regime and explicitly calculate the expression for the interface mode and surfactant mode, and unlike \citet{blyth2004effect}, observe that Marangoni instability can be unstable at small Reynolds numbers. Furthermore, the nonlinear analysis conducted by \citet{pereira2008dynamics} of the same problem chosen by \citet{blyth2004effect}, utilizing the integral boundary layer approximation method and travelling wave solutions reveals the existence of three modes, viz. Kapitza interfacial mode (surface mode), shear mode, and concentration mode (surfactant mode). Also, their findings suggests that in the nonlinear regime, the surfactant alleviates the formation of free surface solitary waves. Moreover, \citet{batchvarov2021three} considered a three-dimensional approach for  vertically falling film in the presence of insoluble surfactants and noticed that for a low Marangoni number, oscillating wavefronts emerge due to the spanwise surfactant concentration gradients and ultimately turn into quasi two-dimensional structures. Recently, the linear and nonlinear instability analysis for insoluble surfactant-laden thin film flows down a wavy bottom was explored by \citet{zivkov2025thin} and mentioned the stabilization effect of surfactants, although they observed a chin shape in the neutral stability curve where the flow is vulnerable to instability in the long wave regime. They also mentioned that perturbed waves can either combine into solitary waves or form wave trains, but the chin shape isn't present in the neutral stability curve at small wavenumber during nonlinear analysis.
	
	
	\indent Conversely, soluble surfactants are a class of surfactants that dissolve in the bulk flow, and their monolayers are named as Gibbs monolayers. This Gibbs monolayer defines an equilibrium point between surfactants adsorbing at the interface and dissolving in the bulk flow. A detailed analysis of different monolayers and corresponding isotherms, such as the Gibbs isotherm and Langmuir isotherm, can be found in \citet{manikantan2020surfactant}. The soluble surfactants have been a topic of interest for the past several decades, with one of the earliest available investigations being \citet{ji1994instabilities}. However, they presumed to overlook the kinetics of adsorption and the interface is at equilibrium as they focused their study more towards the volatile surfactant instigated surface instability. In contrast, \citet{shkadov2004falling} consider the adsorption-desorption kinetics for a falling aqueous solution along a vertical slope with Fick's equation for the surfactant concentration profile. They used the Galerkin method to check the nonlinear behaviour of the flow, and the linearised model is further solved for eigenvalues and four Marangoni-driven diffusion unstable modes are identified. Further, it has been shown by them that one mode approaches hydrodynamic mode with an absolute value of Marangoni number increases, another one mode is a slow mode, and the other two modes being categorized as diffusion modes.\\
	\indent On a rather disjoint development, the experimental study of \citet{georgantaki2012effect} by mixing soluble surfactants such as isopropanol (IP) and Sodium Dodecyl Sulfate (SDS) in water divulges that IP helps inlet disturbances to form solitary humps and eventually capillary ripples, and on the contrary, small-amplitude sinusoidal travelling waves are the dominant structure when SDS is used. Furthermore, the linear hydrodynamic analysis of non-volatile, soluble surfactant-filled film via numerical tools such as finite-element/Galerkin method and the analytical method of long wave approximation was done by \citet{karapetsas2013primary}. They consider the convection-diffusion equation of mass exchange of surfactants between the interface and the bulk with adsorption-desorption kinetics at the free surface as well as the bottom surface. and perceived that the instability of such flow is within the longwave regime, and the critical conditions are influenced by the solubility of the surfactant along with the concentrations of the interfacial concentration of the surfactants. Also, it is concluded that adsorption-desorption kinetics plays a major role in finite-wave regimes, and smaller values of kinetics rate made the surfactant pretend to be insoluble. The aforementioned investigation for arbitrary solubility of the surfactant is extended by them \cite{karapetsas2014role} using the long wave approximation method, and they recognized that the longitudinal perturbation creates a convective flux, which promotes an interfacial concentration gradient that is in the same phase with interfacial disturbances and generates Marangoni stress, which stabilizes the flow. Recently, the effect of soluble surfactant on liquid flowing over a compliant substrate is explored by \citet{samanta2025role}. They observed that the surface mode instability is stabilized by the soluble surfactant for a flat bottom but destabilizes when the bottom becomes a compliant substrate. Also, in the arbitrary wave regime, they identify two different modes viz, wall mode and shear mode and the solutal Marangoni number triggers wall mode to become destabilizing in the finite wave regime. They concluded that the soluble surfactant is more unstable than the insoluble one in the surface mode zone but more stable than the insoluble surfactant in the wall mode regime on the perspective of linear stability analysis. \\
	\indent A significant study on the thermocapillary effect for soluble surfactant contaminated flow was investigated by \citet{d2020marangoni} in which they carried out an investigation examining the Marangoni instabilities for gravity-driven falling film for a soluble surfactant-filled fluid over a uniformly heated sloped plane. They accounted for both the solutal as well as the thermal Marangoni effect and performed linear and nonlinear methods to deal with the problem and witnessed that if thermocapillary has a strong presence, then the surfactants intensify the inertial instabilities, but for small Reynolds numbers, the Marangoni mode displays a stabilizing effect. Also, their simulation results reveal that permanent waves form and propagate along the free surface. Lately, \citet{samanta2025effect} revisited the same problem chosen by \citet{d2020marangoni} to observe the effect of thermocapillary instability for temporal modes, viz., H-mode (surface mode), shear mode, and thermocapillary P- and S-modes in the presence of soluble surfactant. Their/his findings indicate that the solutal Marangoni number has an opposite influence on H-mode compared to the thermal Marangoni number with the former contributing to its stabilization. Also, it is disclosed that base surface surfactant concentration has a dual effect on all the modes, as a modal coalescence can be witnessed between S-mode and H-mode for some specific set of parametric values. \\
	\indent One of the initial investigation to include constant imposed shear at the free surface was \citet{smith1990mechanism} due to its association with tangential stress. He observed that the stress created due to this imposed shear is the primary source of energy, and it helps to enhances a longitudinal flow perturbation. Later, this concept was used in lots of articles\citep{samanta2014shear, choudhury2024thermocapillary}due to its prevalence in bolus-dispersal across the lungs in surfactant replacement theory\citep{espinosa1999bolus} or airway closure process \citep{halpern1993surfactant}. \citet{samanta2014shear} remarked that the imposed shear force as tangential stress behaves as a destabilizing agent for small values of the shear near the onset of instabilities but displays exactly opposite trend far from the onset of instabilities if its value of the imposed shear force crosses a certain threshold. He also found out that downstream applied shear stress helps the small-amplitude periodic waves to turn into finite-amplitude localized pulses with capillary ripples succeeded by a large hump. The effect of externally imposed shear on gravity-driven film in the existence of insoluble surfactant in the arbitrary wavenumber regime is investigated by \citet{bhat2019linear}. They identified three temporal modes, viz, surface mode, surfactant mode, and shear mode, with surface mode being quicker than the surfactant mode, and also it has been shown that the imposed shear destabilises both the surface mode and surfactant mode. \citet{choudhury2024thermocapillary} explored the thermocapillary effect on the instability of insoluble surfactant-laden thin film flow over a uniformly heated inclined plane using Pad\'e approximation along with the conventional longwave approximation and Chebyshev spectral collocation method. They detected five modes, which include two thermocapillary modes ($S$- mode and $P$- mode), with the other three modes already observed by \citet{bhat2019linear}. It is further reported by them that there exists a dynamic competition between the thermal and solutal Marangoni numbers, which systematically categorized in a table, and they derived a threshold relation using the longwave approximation method to address the competition between the effects of solutal and thermal Marangoni numbers.\\
	\indent Another important aspect of laminar flow across a surface is that the bottom of the falling film in general is not solid, which always gives rise to the concept of having different bottom topographies \cite{hossain2024instability, hossain2024hydrodynamic, miladinova2002long, d2010film, usha2005dynamics, samanta2011falling, samanta2021effect, choudhury2022linear}. One such type of bottom is a slippery bottom, which gains prominancy particularly due to the interesting development in the microfluids sector and emergence of new measurement technologies \cite{neto2005boundary, lefevre2004intrusion, denn2001extrusion}. The air confined between the bottom substrate and the liquid can have substantial slip lengths \cite{cottin2003low, choi2006large} with ranges in between $30~nm$ to $1~\mu$m, and it can have various reasons such as nanobubbles formed at the interface \cite{lauga2003effective}, formation of dipole moment of polar liquids \cite{cho2004dipole}, or molecular slip \cite{blake1990slip}. On that context, one of the earliest investigation of the interface of fluid and porous regions was done by \citet{beavers1967boundary}. They observed through their experiment that a boundary layer forms at the interface, and in the vicinity of this boundary layer, the tangential velocity component changes swiftly to account as a velocity at the interface from the Darcian mean filter velocity. However, the normal velocity doesn't change in the region. This concept was eventually taken by \citet{pascal1999linear} to define
	\begin{eqnarray}
		\nonumber \upsilonup (x, y, t) |_{y = 0} = \frac{\alpha}{\sqrt{\kappa}} \displaystyle \frac{\partial \upsilonup}{\partial y} (x, y, t) \bigg{|}_{y = 0} 
	\end{eqnarray}
	as Navier-slip boundary condition for slippery inclined plane, where $\alphaup$ is a dimensionless parameter that arises based on the structure of the porous medium and $\kappa$ is the characteristic length scale of the pore space in the porous medium. The introduction of this equation transpired a plethora of research articles \citep{ma2020effect, chakraborty2019dynamics, chattopadhyay2021weakly,ghosh2016stability, mukhopadhyay2024modeling} to follow based on this concept, with \citet{samanta2011falling} being one of the notable works. They considered thin fluid streaming atop an inclined slippery substance and computed a coupled evolution equation and came to the conclusion that the slippery bottom essentially displays a destabilizing behaviour near the threshold of the instability zone, but away from this zone, it stabilizes the flow. This work is further extended by \citet{bhat2018linear} for insoluble surfactant-contaminated fluids, and it is observed that the surfactant mode is damped for the configuration. However, they pointed out that the P\'eclet number associated with the flow is displaying a dual nature on the shear mode. Recently, \citet{agrawal2024impact} examined the effect of wall slip on soluble surfactant-doped two-layer Poiseuille flow in the creeping flow limit. They observed two short wave modes viz, $M1,~M2$, with their energy budget analysis exhibiting that the $M1$ mode is triggered by the transverse bulk convection of species and the introduction of the slip effect to the upper wall, propelling the system towards instability. However, the $M2$ mode shows a stabilizing effect if slip is being introduced in one of the walls, and if slip is being considered for both walls, then the $M2$ mode leads the system to destabilization.\\
	\indent The above discussion clearly suggests that despite of having several important contribution involving soluble surfactants, the effect of slip length with imposed shear on soluble surfactant laden flow is not addressed properly in the framework of primary instability. This fact motivates the current study to decipher a linear stability analysis of soluble surfactant-laden fluid flow over a slippery inclined plane in the presence of externally imposed shear. The current investigation extends the work of \citet{karapetsas2014role} for a slippery bottom in the existence of externally imposed shear force as tangential stress. The aim of this study is to do a detailed linear hydrodynamic analysis of the flow to comprehend the impact of slip parameters and imposed shear force in the arbitrary wavenumber regime. The well-known Orr-Sommerfeld eigenvalue problem is formulated, and then it is dealt with analytically via the longwave approximation method and numerically using the Chebyshev spectral collocation method. Also, the energy flow through different components associated with the flow is explored through energy budget analysis. In that context, the section \ref{mf} is devoted to describing the problem and equations corresponding to the problem. The linear stability analysis methods and subsequent discussions are recorded in section \ref{lsa}. The section \ref{eba} contains the energy budget analysis and pertinent discussions. Finally, the conclusion of the entire study is recorded in the section \ref{con}.

	\section{Mathematical formulation} 
	\begin{figure}[ht!]
		\begin{center} 
			\includegraphics[height=6cm,width=15cm]{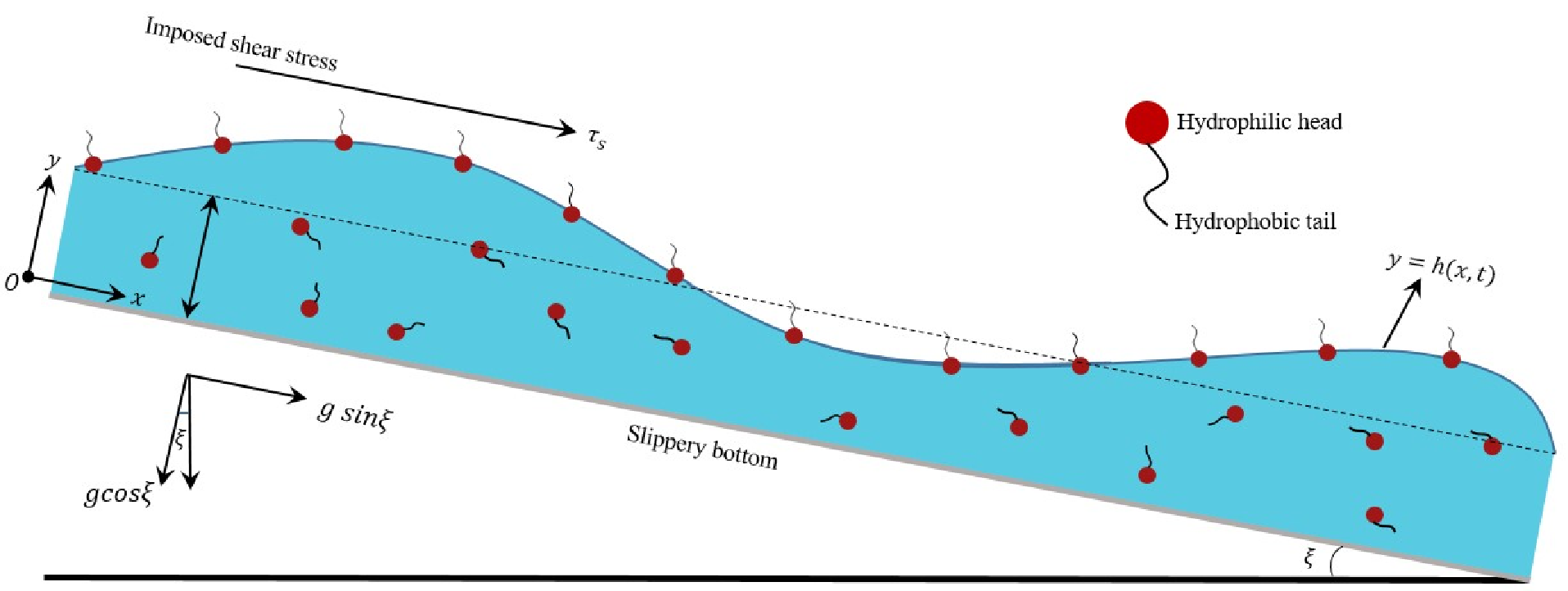}
		\end{center}\label{fig_1}
		\caption{ The depiction of the problem with the coordinate system in the two-dimensional frame. The soluble surfactant is present in the bulk as monomers, and the size shown is not to scale. Also, the moieties of one surfactant molecule are shown in close view.} \label{mf}
	\end{figure}
	This manuscript considers a gravity-driven, Newtonian fluid contaminated with non-volatile soluble surfactant flowing over an inclined slippery inclined plane making an angle $\xi$ with the horizon in the presence of externally imposed shear at the free surface see \hyperref[fig_1]{Schematic diagram}. The fluid assumed to be viscous and incompressible. The surfactant chosen here may display two types of characteristics, such as either adsorb near the air-liquid interface and disturb the surface tension or present in the fluid bulk as monomers. The fluid characteristics, like kinematic viscosity $\nu$, and density $\rhoup$ are assumed to be constant, although the surface tension $\sigma$ as stated above, behave as a function of surfactant interfacial concentration i.e., $\sigma = \sigma(\Gamma_a)$, where $\Gamma_a$ is the interfacial concentration of the surfactant. At the same time, it is important to mention the dependence of the surface tension on the interfacial surfactant concentration through the following relation known as the Sheludko equation \cite{karapetsas2013primary, sheludko1967thin, gaver1990dynamics} of state as follows
	\begin{eqnarray}
		\nonumber \sigma = \sigma_f \bigg[ 1 + \frac{\Gamma_a}{\Gamma_{a\infty}} \biggl\{ \bigg(\frac{\sigma_f}{\sigma_m}\bigg)^{1/3} - 1 \biggr\} \bigg]^{-3}
	\end{eqnarray}
	where, $\sigma_f$ and $\sigma_m$ are the respective surface tensions of a clean flow and of a flow with maximum possible surfactant concentration, and $\Gamma_{a\infty}$ is the maximum concentration of the surfactant present at the free surface. The usefulness of this model lies in the its nonlinear and asymptotic nature which makes it suitable to use even at high surfactant concentrations tending to critical micelle concentration.\\ 
	\indent It is supposed that the flow maintains its uniformity along the breadth of the flow plane, which permits the use of a two-dimensional model during the flow analysis. So, a two-dimensional cartesian coordinate system $(x,y)$ is considered where the $x$- axis is taken along the flow and $y$- axis is perpendicular to the flow plane. The velocity components along the $x$ and $y$ directions are denoted by $u$ and $v$, respectively. The free surface is $y = h(x,t)$ at time $t$ and the slippery bottom is at $y = 0$. Then, the mass and momentum conservation equations can be written as
	
	\begin{eqnarray}
		& u_x+v_y = 0 \label{eqn1},~ 0 \leqslant y \leqslant h(x,t) \\
		& \nonumber \hspace{1cm} \rhoup \begin{pmatrix} u \\ v \end{pmatrix}_t + \rhoup u \begin{pmatrix} u \\ v \end{pmatrix}_x + \rhoup~v\begin{pmatrix} u \\ v \end{pmatrix}_y = - \begin{pmatrix} p_x \\ p_y \end{pmatrix} + \muup_{visc} \begin{pmatrix} u_{xx} + u_{yy} \\ v_{xx} + v_{yy} \end{pmatrix} - \rhoup g \begin{pmatrix} - \sin{\xi} \\ \cos{\xi} \end{pmatrix},\label{eqn2}\\
		& \hspace{9cm} ~0 \leqslant y \leqslant h(x,t) 
	\end{eqnarray}
	respectively, where $p$ denotes the pressure term, $\muup_{visc}$ is the viscosity of the flow. At the air-liquid surface, the total hydrodynamic stress is equilibrated by the Marangoni stress induced due to the surfactant present at the free surface coupled with the surface stress induced by surface tension and externally imposed shear force. Thus, provide the tangential and normal stress balance equations along with the kinematic condition, which are respectively as follows
	\begin{eqnarray}
		& \muup_{visc} (u_y + v_x) (1 - h_x^2) - 2 \muup_{visc} h_x ( u_x - v_y ) = (\sigma_x + h_x \sigma_y + \tau_s) \sqrt{1 + h_x^2} \label{eqn3} \\
		& p_{atm} - p + \displaystyle \frac{2 \muup_{visc} u_x ( h_x^2 - 1) - 2 \muup_{visc} h_x (u_y + v_x)}{1 + h_x^2} = \sigma \frac{h_{xx}}{(1 + h_x^2)^{3/2}} \label{eqn4} \\
		& v = h_t + u h_x \label{eqn5}
	\end{eqnarray}

	In this current study, the concentration of the soluble surfactants is assumed to be below the concentration of critical micelle, and hence, the existence of micelles is kept out of consideration. Only monomers are considered in the bulk with concentration $\Gamma$. The underlying kinetics law between the concentrations of different layers can be found in detail in the literature \cite{karapetsas2013primary, edmonstone2006surfactant}.
	
	The flux transmission of the surfactant at the air-liquid interface can be described by 
	\begin{eqnarray}
		& \mathcal{J}_{ba} = k_1 \Gamma\bigg|_{y = h} \biggl( 1 - \displaystyle \frac{\Gamma_a}{\Gamma_{a\infty}} \biggr) - k_2 \Gamma_a, \label{eqn6} 
	\end{eqnarray}
	where the flux of the monomers between bulk and the air-liquid interface is $\mathcal{J}_{ba}$, and, $k_i~,~(i = 1,2)$ are the rates of reaction. Now, the dynamics of the free surface surfactant concentration can be described by the following convection-diffusion equation at the air-liquid interface is demonstrated by
	\begin{eqnarray}
		& \nonumber \displaystyle \Gamma_{a,t} + u \Gamma_{a,x} + \frac{\Gamma_a}{1 + h^2_x} \biggl[ u_x + h_x (u_y + v_x + h_x v_y) \biggr] = \frac{D_a}{\sqrt{1 + h^2_x}} \biggl[ \frac{\Gamma_{a,x}}{\sqrt{1 + h^2_x}} \biggr]_x + \mathcal{J}_{ba},\\
		&\hspace{9cm}~\text{at}~ y = h(x,t), \label{eqn7}
	\end{eqnarray}
	The equation for the surfactant concentration in the bulk is
	\begin{eqnarray}
		\Gamma_t + u~\Gamma_{x} + v~\Gamma_y = \displaystyle D_b (\Gamma_{xx} + \Gamma_{yy}),~ 0 \leqslant y \leqslant h(x,t) \label{eqn8}
	\end{eqnarray}
	Also, the boundary conditions at the air-liquid interface for the monomers present in the bulk follow the equation :
	\begin{eqnarray}
		& \displaystyle \frac{D_b (h_x \Gamma_x - \Gamma_y)}{\sqrt{1 + h^2_x}} = \mathcal{J}_{ba}~\text{at}~y=h(x,t),\label{eqn9}
	\end{eqnarray}
	Moreover, the accumulated mass of different species of surfactants present in the flow is $M_{tot}$, is a conserved quantity for each unit of length and obeys \cite{karapetsas2013primary, kalogirou2019role} 
	\begin{eqnarray}
		\hspace{2cm}\iint_{\Omega} \Gamma dxdy + \int_{ x = 0}^{L} \Gamma_adx = M_{tot},~\Omega := [0,L]\times[0,h],	\label{eqn10}
	\end{eqnarray}
	where $L$ is the length of the flow considered in the investigation.
	
	Finally, the bottom boundary condition follows the Navier slip condition \citep{oron1997long}, which provides the bottom to be slippery and the no penetration condition as follows
	\begin{eqnarray}
		u = \delta_s u_y~ \text{and}~ v = 0 ~\text{at}~ y = 0, \label{eqn11}
	\end{eqnarray}
	and the behaviour of the monomers at the bottom is described by 
	\begin{eqnarray}
		\Gamma_y = 0~~ \text{at}~~ y = 0. \label{eqn12}
	\end{eqnarray}

	\subsection{Scaling and non-dimensionalisation}
	The governing equations and the associated boundary conditions \eqref{eqn1}-\eqref{eqn12} are non-dimensionalised using the following transformations
	\begin{eqnarray}
		& \nonumber \langle x , y \rangle = H \langle x^{\ast}, y^{\ast} \rangle;~ h = H h^{\ast}; ~ t = (H/U_N) t^{\ast};~ \langle u, v \rangle = U_N \langle u^{\ast}, v^{\ast} \rangle;\\
		& \nonumber ~ p = p_{atm} + ( \rhoup g H \sin(\xi) ) p^{\ast}; \langle \Gamma, \Gamma_a\rangle = \langle \Gamma_{CMC} \Gamma^{\ast}, \Gamma_{a \infty} \Gamma^{\ast}_a \rangle;\\
		& \nonumber\mathcal{J}_{ba}= (U_N/H) \Gamma_{a \infty} \mathcal{J}^{\ast}_{ba},~ \sigma = \sigma_f \sigma^{\ast}; \tau = \displaystyle \frac{\tau^{\ast} \mu U_N}{H};~\delta_s = H \delta^{\ast};
	\end{eqnarray}
	where $U_N = \displaystyle \frac{g H^2 \sin(\xi)}{2 \nu}$ is the streamwise velocity component for steady flow of uniform thickness, $H$ is the thickness of the Nusselt flat film, $p_{atm}$ is the atmospheric pressure, and $\Gamma_{CMC}$ is the critical micelle concentration. Also, the following nondimensional numbers appears during the precedent transformation,
	\begin{eqnarray}
		& \nonumber \displaystyle \chi = \frac{g H^3}{\nu^2},~~ Ka = \frac{\sigma_f}{\rho g^{1/3} \nu^{4/3}},~~ Sc_j = \frac{\nu}{D_j},~ k_a = \frac{k_2 H}{U},~ \beta_a = \frac{\Gamma_{a\infty}}{H \Gamma_{CMC}},~ R_a = \frac{k_1 \Gamma_{CMC}}{k_2 \Gamma_{a\infty}} ~ \text{and}\\
		& \nonumber \displaystyle We = \frac{Ka}{\chi^{2/3} \sin(\xi)},~~ Pe_j = Re Sc_j,~~ Re = \frac{\chi \sin(\xi)}{2}
	\end{eqnarray}
	where $j $ assumes the value $a$ and $b$ to denote the respective parameters related to the free surface and bulk substrate of the soluble surfactant, respectively. The parameters are $\chi$ is modified Reynolds number, $Ka$ is the Kapitza number, $Sc_j$ is the Schmidt number, $We$ is the Weber number, $Pe_j$ is the P\'eclet number, $Re$ is the Reynolds number, $\beta_a$ is the ratio of the maximum quantity of surfactant present at the surface and that in the bulk flow, $k_a$ is the ratio of the convection rate to the reverse reaction rate, and $R_a$ is the ratio of the ultimate rate of the forward and backward reactions possible. Moreover, $\xi_a = \beta_a R_a$ is defined to provide the solubility of the surfactant present in the bulk \cite{jensen1993spreading}.
	
	\begin{table}[htbp]
		\centering
		\caption{Typical ranges of the physical parameters of the flow, surfactants. \cite{pereira2008dynamics, samanta2025effect, edmonstone2004simultaneous}}
		\label{tab1}
		\begin{tabular}{|c|c|c|}
			\hline
			\textbf{Parameters} & \textbf{Definition} & \textbf{Range} \\
			\hline
			\hline
			viscosity $(kg~m^{-1}~s^{-1})$ & $\muup_{visc}$ & $\approx 10^{-3}$ \\
			Thickness (Nusselt flat film) (mm) & $H$ & $10^{-4}-10^0$\\
			Typical inclination angle & $\xi$ & $0^{\circ}-90^{\circ}$\\
			Density $(kg~m^{-3})$ & $\rhoup$ & $\approx 10^3$ \\
			Surface tension $(mN~m^{-1})$ & $\sigma_0$ & $20-74 $ \\
			Surface surfactant concentration at maximum capacity $(mol~m^{-1})$ & $\Gamma_{a\infty}$ & $10^{-6}-10^{-5}$ \\
			Surface surfactant diffusion $(m^2~s^{-1})$ & $D_a$ & $10^{-12}-10^{-11}$ \\
			Bulk surfactant diffusion $(m^2~s^{-1})$ & $D_b$ & $10^{-11}-10^{-10}$ \\
			\hline
		\end{tabular}
	\end{table}
	
	\begin{table}[htbp]
		\centering
		\caption{The definitions and estimated magnitude of the dimensionless parameters are listed \cite{samanta2014shear, samanta2011falling, edmonstone2006surfactant, mavromoustaki2012dynamics}}
		\label{tab2}
		\begin{tabular}{|c|c|c|}
			\hline
			\textbf{Parameters} & \textbf{Definition} & \textbf{Range} \\
			\hline
			\hline
			Kapitza number  & $Ka$ & $10^2~-~10^5$ \\
			sorption kinetics rate & $k_a$ & $0.1 - 100$ \\
			Surfactant parameter & $\beta_a$ & $0.01 - \infty$ \\
			Solubility parameter & $R_a$ & $0.01 - 100$ \\
			P\'eclet number(surface) & $Pe_{ca}$ & $10^2 - 10^5$ \\
			P\'eclet number(monomers in the bulk) & $Pe_b$ & $10 - 10^4$ \\
			Mass of the surfactant & $M_{tot}$ & $0.1 - 20$ \\
			External imposed shear & $\tau$ & $(-1.0)~-~40$ \\
			Slip parameter & $\delta$ & $0.0~-~0.4$ \\
			\hline
		\end{tabular}
	\end{table}

	These aforementioned nondimensionalisation turn the governing equations of the flow and the surfactant concentration equation in the bulk into
	\begin{eqnarray}
		&u_x+v_y = 0,~~ 0 \leqslant y \leqslant h(x,t) \label{eqn13} \\
		&Re \begin{pmatrix} u \\ v \end{pmatrix}_t + Re~u \begin{pmatrix} u \\ v \end{pmatrix}_x + Re~v\begin{pmatrix} u \\ v \end{pmatrix}_y = - 2 \begin{pmatrix} p_x \\ p_y \end{pmatrix} + \begin{pmatrix} u_{xx} + u_{yy} \\ v_{xx} + v_{yy} \end{pmatrix} - 2 \begin{pmatrix} - 1 \\ \cot{\xi} \end{pmatrix},\label{eqn14}\\
		& \nonumber \hspace{9cm} 0 \leqslant y \leqslant h(x,t)\\
		\vspace{1cm}
		&Pe_b (\Gamma_t + u \Gamma_x + v \Gamma_y) = \Gamma_{xx} + \Gamma_{yy},~~ 0 \leqslant y \leqslant h(x,t), \label{eqn15}
	\end{eqnarray}
	The asterisk signs are removed from the variables after nondimensionalization. The  Sheludko equation of state for the surface tension reformulates to
	\begin{eqnarray}
		\sigma = \biggl\{ 1 + \Gamma_{a} (\Sigma^{1/3} - 1) \biggr\}^{-3}~, \label{eqn16}
	\end{eqnarray}
	where $\Sigma = \sigma_f / \sigma_m$. The dimensionless form of tangential and normal stress boundary conditions, along with the kinematic condition at the free surface $ y = h(x,t)$ are
	\begin{eqnarray}
		&(u_y + v_x) (1 - h_x^2) - 4 h_x u_x = \{ 2 We (\sigma_x + h_x \sigma_y) + \tau \} \sqrt{1 + h_x^2}, \label{eqn17} \\
		&p + \displaystyle \frac{u_x (1 - h_x^2) + h_x (u_y + v_x)}{1 + h_x^2} = - We \sigma \frac{h_{xx}}{(1 + h_x^2)^{3/2}}, \label{eqn18} \\
		&v = h_t + u h_x. \label{eqn19}
	\end{eqnarray}
	respectively.\\
	The surfactant concentration transportation equation in dimensionless form is
	\begin{eqnarray}
		& \nonumber \Gamma_{a,t} + u \Gamma_{a,x} + \displaystyle \frac{\Gamma_a}{1 + h^2_x} \biggl\{u_x + h_x (u_y + v_x + h_x v_y) \biggr\} = \frac{1}{Pe_{ca}\sqrt{1 + h^2_x}} \biggl[ \frac{\Gamma_{a,x}}{\sqrt{1 + h^2_x}} \biggr]_x + \mathcal{J}_{ba},\\
		& \hspace{10cm} \text{at}~ y = h(x,t) \label{eqn20}
	\end{eqnarray}
	and the boundary condition associated with the flux at the free surface is 
	\begin{eqnarray}
		& \displaystyle \frac{h_x \Gamma_x - \Gamma_y}{Pe_b \sqrt{1 + h^2_x}} = \beta_a \mathcal{J}_{ba}~\text{at}~ y = h(x,t), \label{eqn21} \\
		& \mathcal{J}_{ba} = k_a \biggl\{ R_a \Gamma(y=1) (1 - \Gamma_a) - \Gamma_a \biggr\}, \label{eqn22}
	\end{eqnarray}
	Finally, the bottom boundary condition consists of the following
	\begin{eqnarray}
		& u = \delta u_y, ~\text{at}~ y = 0 \label{eqn23} \\
		& v = 0, ~\text{at}~ y = 0, \label{eqn24}\\
		& \Gamma_y = 0, ~\text{at}~ y = 0. \label{eqn25}
	\end{eqnarray}

	\subsection{Base profile}
\begin{figure}
		\begin{center}
			\subfigure[]{\label{fig_2_a}\includegraphics*[width=8cm]{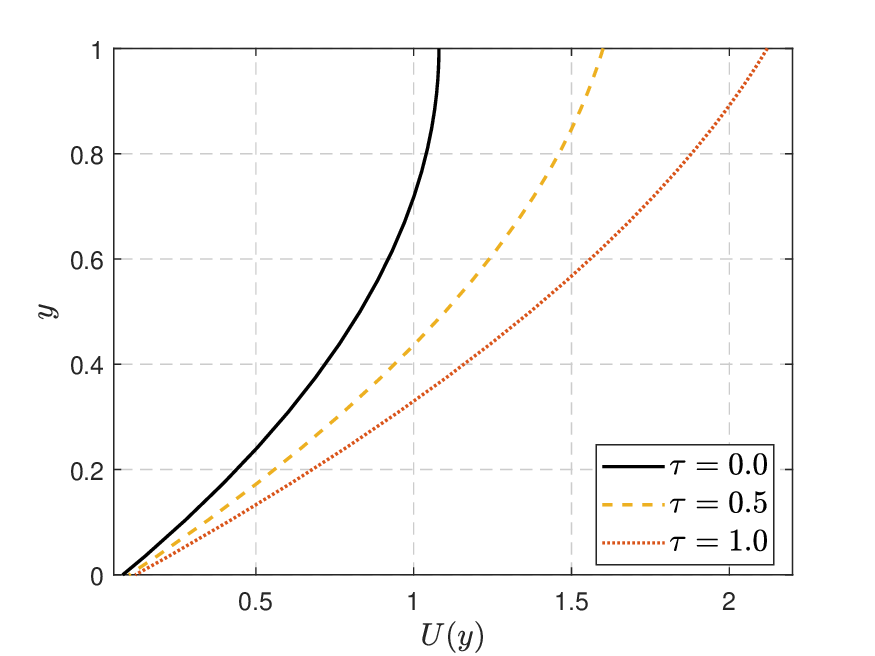}}
			\subfigure[]{\label{fig_2_b}\includegraphics*[width=8cm]{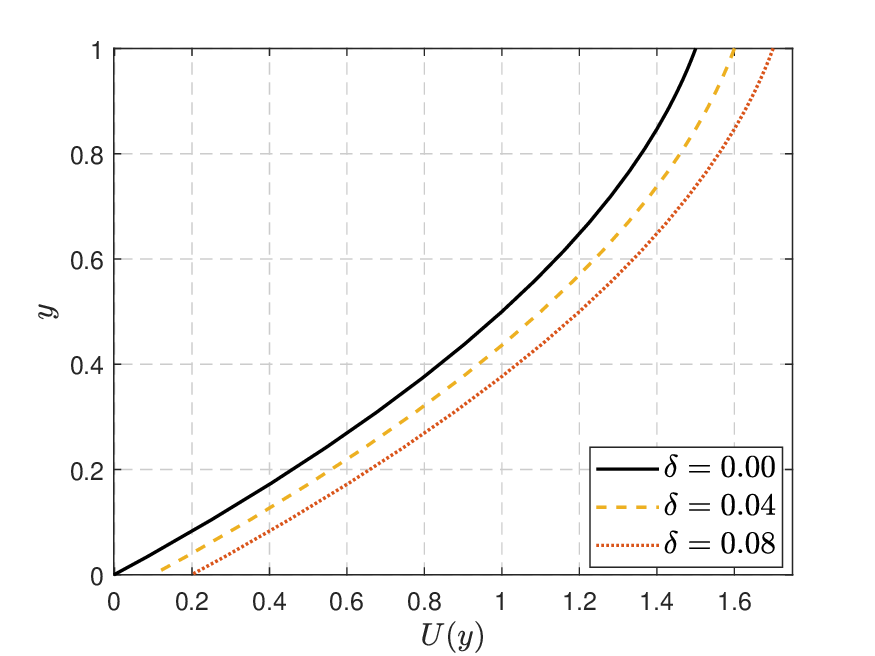}}
		\end{center} 
		\caption{The streamwise base velocity profile $U(y)$ as a function of $y$ for varying (a) imposed shear $\tau$, (b) slippery bottom $\delta$. For (a) $\delta = 0.04$ and (b) $\tau = 0.5$ are chosen as fixed values during computing the above results.} \label{fig_2}
	\end{figure}

	This subsection is dedicated to explore the base profile of the flow, as this current study deals with primary instabilities, and base flow has influence on the primary instability. On that front, a unidirectional parallel flow with uniform fluid width is considered, and then the governing equations and boundary conditions \eqref{eqn13}-\eqref{eqn25} based on that assumption are reduced, and the solution of the equations is as following 
	\begin{eqnarray}
		& U(y) = - y^2 + (2 + \tau) y + \delta (2 + \tau) ~~ \text{and} ~~ P(y) = (1 - y) \cot(\xi) \label{eqn26}
	\end{eqnarray}
	The base velocity profile matches with \citet{hossain2024odd}, where it can be seen that both imposed shear force and a slippery bottom influence the base velocity. The effects are shown in the figures (\ref{fig_2}), where it is evident that the base velocity increases with both imposed shear and a slippery bottom. The magnitude of the increment is greater in the case of a slippery bottom. So, these physical quantities play a major role in the base velocity and, subsequently, in the primary instability. On the other hand, it can be seen that base pressure is not exactly similar to \citet{hossain2024odd} because the scale chosen for pressure in the current article is slightly different from that of the \citet{hossain2024odd} article.
	
	In equilibrium condition, the concentration for the surfactants present in the bulk and at the free surface is assumed to be uniform, and they are chosen as 
	\begin{eqnarray}
		\Gamma\bigg|_{equilibrium} = \Gamma_0~; \Gamma_a \bigg|_{equilibrium} = \Gamma_{a0} \label{eqn27}
	\end{eqnarray}
	At the same time, the surface tension of the free surface at equilibrium is
	\begin{eqnarray}
		\sigma_0 = \biggl\{ 1 + \Gamma_{a0} \biggl( \Sigma^{1/3} - 1 \biggr) \biggr\}^{-3}. \label{eqn28}
	\end{eqnarray}
	Also, at the equilibrium, the flux of the free surface will be zero, i.e.$\mathcal{J}_{ba} = 0$, so the non-dimensionalized flux equation \eqref{eqn22} implies
	\begin{eqnarray}
		\Gamma_{a0} = \displaystyle \frac{R_a \Gamma_0}{1 + R_a \Gamma_0}.\label{eqn29}
	\end{eqnarray}
	The mass conservation equation \cite{karapetsas2013primary} for surfactant of all spices is
	\begin{eqnarray}
		\Gamma_0 + \beta_a ~ \Gamma_{a0} = M_{tot}. \label{eqn30}
	\end{eqnarray} 
	The exact solution for $\Gamma_0$ and subsequently $\Gamma_{a0}$ can be evaluated using equations \eqref{eqn29} and \eqref{eqn30}, which can be found in \citet{karapetsas2013primary}.
	
	\section{\bf{Linear stability analysis}} \label{lsa}
	
	This section constitutes the linear stability analysis by means of the analytical longwave approximation method and the numerical Chebyshev spectral collocation method. In order to apply those methods, the disturbance equations of the flow due to the infinitesimal perturbation are required. Hence, an infinitesimal disturbance is imposed on the flow, and the variable can be written as
	\begin{eqnarray}
		\nonumber \theta = \Theta + \widetilde{\thetaup}
	\end{eqnarray}
	Here, the perturbed variables are indicated using the overhead tilde notations. The disturbance equations are written below
	\begin{eqnarray}
		& \widetilde{u}_x + \widetilde{v}_y = 0, \label{eqn31} 
		\\
		& Re \{ \widetilde{u}_t + U(y) \widetilde{u}_x + U^{'}(y) \widetilde{v} \} = -2 \widetilde{p}_x + \widetilde{u}_{xx} + \widetilde{u}_{yy}, \label{eqn32} 
		\\
		& Re \{ \widetilde{v}_t + U(y) \widetilde{v}_x \} = -2 \widetilde{p}_y + \widetilde{v}_{xx} + \widetilde{v}_{yy}, \label{eqn33} 
		\\
		& \widetilde{u}_y + \widetilde{v}_x - 2 \widetilde{h} = 2 We \widetilde{\sigma}_x ~~ \text{at}~~y = 1, \label{eqn34} 
		\\
		& \widetilde{p} + \widetilde{u}_x - \cot(\xi) \widetilde{h} + U^{'}(1) \widetilde{h}_x = - We \sigma_0 \widetilde{h}_{xx} ~~ \text{at}~~y = 1, \label{eqn35} 
		\\
		& \widetilde{v} = \widetilde{h}_t + U(1) \widetilde{h}_x ~~\text{at}~~y = 1, \label{eqn36} 
		\\
		& \widetilde{\Gamma}_{a,t} + U(1) \widetilde{\Gamma}_{a,x} + \Gamma_{a0} \{ \widetilde{u}_x + U^{'}(1) \widetilde{h}_x \} = \displaystyle \frac{1}{Pe_{ca}} \widetilde{\Gamma}_{a,xx} + \widetilde{\mathcal{J}}_{ba} ~~ \text{at}~~y = 1, \label{eqn37} 
		\\
		& \displaystyle \frac{1}{Pe_b} \widetilde{\Gamma}_y + \beta_a \widetilde{\mathcal{J}}_{ba} = 0, ~~ \text{at}~~y = 1, \label{eqn38}
		\\
		& \widetilde{\Gamma}_t + U(y) \widetilde{\Gamma}_x = \displaystyle \frac{1}{Pe_b} (\widetilde{\Gamma}_{xx} + \widetilde{\Gamma}_{yy} ), \label{eqn39} 
		\\
		& \displaystyle \widetilde{\Gamma}_y = 0 ~~ \text{at}~~y = 0, \label{eqn40} 
		\\
		& \widetilde{u} = \delta \widetilde{u}_y~~ \text{at}~~y = 0, \label{eqn41} 
		\\
		& \widetilde{v} = 0 ~~ \text{at}~~y = 0, \label{eqn42} 
		\\
		& \widetilde{\mathcal{J}}_{ba} = k_a \bigg[ R_a \bigg\{ \widetilde{\Gamma}\big|_{y=1} (1 - \Gamma_{a0}) - \Gamma_0\widetilde{\Gamma}_a \bigg\} -\widetilde{\Gamma}_a \bigg] ~~\text{at}~~ y=1. \label{eqn43}
	\end{eqnarray}
	The prime signs represent differentiation with respect to $y$. On that note, it is important to define the surface elasticity 
	\begin{eqnarray}
		E_0 = -\Gamma_{a0} \displaystyle ~\frac{d \sigma}{d \Gamma_{a}} \label{eqn44}
	\end{eqnarray}
	as defined by \citet{karapetsas2013primary}, which helps to define disturbed surface tension as
	\begin{eqnarray}
		\widetilde{\sigma}_x = \displaystyle - \frac{E_0}{\Gamma_{a0}}~ \widetilde{\Gamma}_{a,x}. \label{eqn45}
	\end{eqnarray}
	Using the nondimensionalised Sheludko equation \eqref{eqn16} and the above equation \eqref{eqn45}, the expression for the surface elasticity can be deduced as
	\begin{eqnarray}
		E_0 = \displaystyle \frac{3 \Gamma_{a0} (\Sigma^{1/3} - 1)}{\{ 1 + \Gamma_{a0} (\Sigma^{1/3} - 1) \}^4 }. \label{eqn46}
	\end{eqnarray}
	In order to solve these disturbance equations in the normal mode analysis, the following transformations 
	\begin{eqnarray}
		& \nonumber \biggl( \widetilde{u}(x,y,t), \widetilde{v}(x,y,t), \widetilde{p}(x,y,t), \widetilde{\Gamma}(x,y,t), \widetilde{\Gamma}_a(x,t), \widetilde{h}(x,t), \widetilde{\mathcal{J}}_{ba}(x,t) \biggr) \\
		& \nonumber = \biggl( \widehat{u}(y), \widehat{v}(y), \widehat{p}(y), \widehat{\Gamma}(y), \widehat{\Gamma}_a, \widehat{h}, \widehat{\mathcal{J}}_{ba} \biggr)\exp\{ \mathrm{i} k (x - \lambdaup t) \}+~c.c.
	\end{eqnarray}
	are applied to equations \eqref{eqn31}-\eqref{eqn43}, where c.c. denotes the complex conjugates and the variables with hat symbols indicate the amplitudes of the corresponding variables. The $k$ is a real-valued wavenumber, and  $\lambdaup$ is the complex-valued eigenvalue or simply the complex wave speed of the disturbance. The following equations are obtained using the above transformations
	\begin{eqnarray}
		& \widehat{v} + \mathrm{i} k \widehat{u} = 0, \label{eqn47} \\
		& Re \bigg[ \mathrm{i} k \bigg\{ U(y) - \lambdaup \bigg\} \widehat{u} + U^{'}(y) \widehat{v} \bigg] = - 2 \mathrm{i} k \widehat{p} + \widehat{u}^{''} - k^2 \widehat{u} , \label{eqn48} \\
		& Re \bigg[ \mathrm{i} k \bigg\{ U(y) - \lambdaup \bigg\} \widehat{v} \bigg] = - 2 \widehat{p}^{'} + \widehat{v}^{''} - k^2 \widehat{v} , \label{eqn49} 
		\\
		& \widehat{u}^{'} + \mathrm{i} k \widehat{v} - 2 \widehat{h} = \displaystyle - 2 \mathrm{i} k We \frac{E_0}{\Gamma_{a0}} \widehat{\Gamma}_{a} ~~\text{at}~~ y =1, \label{eqn50} \\
		& \widehat{p} + \mathrm{i} k \widehat{u} + \bigg\{ \mathrm{i} k U^{'}(1) - \cot(\xi) \bigg\} \widehat{h} = k^2 We \sigma_0 \widehat{h} ~~\text{at}~~ y =1, \label{eqn51} \\
		& \widehat{v} = \mathrm{i} k \bigg\{ U(1) - \lambdaup \bigg\} \widehat{h} ~~\text{at}~~ y =1, \label{eqn52}\\
		& \widehat{v} = 0 ~\text{and}~ \widehat{u} = \delta \widehat{u}^{'} ~~\text{at}~~y = 0, \label{eqn53}
		\\
		& \mathrm{i} k \bigg\{ U(1) - \lambdaup \bigg\} \widehat{\Gamma}_a + \mathrm{i} k \Gamma_{a0} \bigg\{ U^{'}(1) \widehat{h} + \widehat{u} \bigg\} = \displaystyle - \frac{k^2}{Pe_{ca}} \widehat{\Gamma}_a + \widehat{\mathcal{J}}_{ba} ~~\text{at}~~ y =1, \label{eqn54} \\
		& \widehat{\Gamma}^{''} - k^2 \widehat{\Gamma} = \mathrm{i} k Pe_b \bigg\{ U(y) - \lambdaup \bigg\} \widehat{\Gamma}, \label{eqn55} \\
		& \widehat{\Gamma}^{'} + \beta_a Pe_b \widehat{\mathcal{J}}_{ba} = 0 ~~\text{at}~~ y =1, \label{eqn56} \\
		& \widehat{\mathcal{J}}_{ba} = k_a \bigg[ R_a \bigg\{ \widehat{\Gamma}\big|_{y=1} (1 - \Gamma_{a0}) - \Gamma_0 \widehat{\Gamma}_a \bigg\} -\widehat{\Gamma}_a \bigg] ~~\text{at}~~ y=1 \label{eqn57}\\
		& \widehat{\Gamma}^{'}(y) = 0,~~\text{at}~~y = 0. \label{eqn58}
	\end{eqnarray}
	
	\subsection{Linear stability analysis for longwaves ($k \ll 1$)}
	This subsection comprises asymptotic analysis of the flow instability for wavenumbers $k \rightarrow 0 $ i.e., long wave regime. Following \citet{yih1991stability, samanta2021effect}, the perturbed physical quantities are expanded with respect to wavenumber $k$ as follows. 
	
	\begin{eqnarray}
		& \widehat{u} = u_0 + k u_1 + k^2 u_2 + k^3 u_3 + O(k^4), \label{eqn59} \\
		& \widehat{v} = k v_1 + k^2 v_2 + k^3 v_3 + O(k^4), \\
		& \widehat{p} = p_0 + k p_1 + k^2 p_2 + k^3 p_3 + O(k^4), \\
		& \widehat{\Gamma} = \gamma_0 + k \gamma_1 + k^2 \gamma_2 + k^3 \gamma_3 + O(k^4), \\
		& \lambdaup = \lambdaup_0 + k \lambdaup_1 + k^2 \lambdaup_2 + k^3 \lambdaup_3 +O(k^4), \\
		& \widehat{\Gamma}_a = A_0 + k A_1 + k^2 A_2 + k^3 A_3 + O(k^4), \\
		& \widehat{h} = h_0 + k h_1 + k^2 h_2 + k^3 h_3 + O(k^4), \\
		& \widehat{\mathcal{J}}_{ba} = \widehat{\mathcal{J}}_{ba,0} + k \widehat{\mathcal{J}}_{ba,1} + k^2 \widehat{\mathcal{J}}_{ba,2} + k^3 \widehat{\mathcal{J}}_{ba,3} + O(k^4). \label{eqn66}
	\end{eqnarray}
	
	Now, substituting the precedent equations \eqref{eqn59}-\eqref{eqn66} in the equations \eqref{eqn47}-\eqref{eqn58} and collecting the like powers of $k$, the equations for each order can be found. It is noteworthy to mention that the Weber number `$We$' and the P\'eclet numbers are presumed to be of unit order, i.e., $We \sim O(1),~Pe_b \sim O(1),~ Pe_{ca} \sim O(1)$. Hence, the Weber number `$We$' is absent from the zero$^{th}$ order normal stress boundary condition. The same reason is attributed to the fact that P\'eclet numbers $Pe_{ca}$ do not appear in the zero$^{th}$ surfactant transport equations.
	
	\subsubsection{Zero$^{th}$ order analysis}
	The zero$^{th}$ order equations are 
	\begin{eqnarray}
		& u_0^{''} = 0; \label{eqn67} 
		\\
		& p_0^{'} = 0; \label{eqn68} 
		\\
		& u_0{'} = 2 h_0 ~~ \text{at}~~ y=1; \label{eqn69} 
		\\
		& p_0 = h_0 \cot(\xi) ~~ \text{at}~~ y=1; \label{eqn70} 
		\\
		& u_0 = \delta u_0{'} ~~ \text{at}~~ y=0; \label{eqn71} 
		\\
		& J_{ba,0} = 0 ~~ \text{at}~~ y=1; \label{eqn72} 
		\\
		& \gamma_0^{''} = 0; \label{eqn73} 
		\\
		& \gamma^{'}_0 = - \beta_a Pe_b \mathcal{J}_{ba,0} ~~ \text{at}~~ y=1; \label{eqn74} 
		\\
		& \gamma^{'}_0 = 0~~\text{at}~~ y = 0;\label{eqn75}
		\\
		& \mathcal{J}_{ba,0} = k_a \bigg[ R_a \bigg\{ (1 - \Gamma_{a0}) \gamma_0 - \Gamma_0 A_0 \bigg\} - A_0 \bigg] ~~ \text{at}~~ y=1. \label{eqn76}
	\end{eqnarray}
	Solving these equations with appropriate boundary conditions, the solutions can be written as
	\begin{eqnarray}
		& u_0 = 2 (y + \delta) h_0 ; ~~ p_0 = h_0 \cot(\xi); ~~ \Gamma_0 = \displaystyle \frac{\Gamma_{a0}}{R_a (1 - \Gamma_{a0})} \label{eqn77} 
		\\
		&\text{and}~~ \gamma_0(y) = \displaystyle \frac{A_0}{R_a (1 - \Gamma_{a0})^2}\label{eqn78}
	\end{eqnarray}
	From the zero$^{th}$ order expression of the velocity component in the flow direction, it can be observed that it behaves linearly in $y$ and depends on the slip parameter $\delta$ (when $h_0 \ne 0$). Also, the base concentration of the surfactant present in the bulk remains unaffected by the slip parameter $\delta$ and the externally imposed shear stress $\tau$. Moreover, the leading order surfactant concentration $\gamma_0(y)$ is an independent function of $y$ $i.e.,$ a constant quantity. Additionally, the expressions for $u_0(y)$, $p_0(y)$ and $\gamma_0(y)$ coincides with \citet{karapetsas2014role} in absence of slip parameter $\delta$, external imposed shear $\tau$ and $h_0 = 1$.
	
	\subsubsection{First-order analysis}
	
	The first-order equations are as follows
	
	\begin{eqnarray}
		& v^{'}_1 + \mathrm{i} u_0 = 0; \label{eqn79} 
		\\
		& u^{''}_1 - 2 \mathrm{i} p_0 = Re \biggl[ \biggl\{ U(y) - \lambdaup_0 \biggr\} \mathrm{i} u_0 + U^{'}(y) v_1 \biggr]; \label{eqn80} 
		\\
		& v^{''} - 2 p^{'}_1 = 0;\label{eqn81} 
		\\
		& u^{'}_1 - 2 h_1 = \displaystyle - 2 \mathrm{i} We \frac{E_0}{\Gamma_{a0}} A_0, ~~ \text{at}~~ y = 1; \label{eqn82} 
		\\
		& p_1 + \mathrm{i} u_0 + \mathrm{i} U^{'}(1) h_0 - h_1 \cot(\xi) = 0,~~ \text{at}~~ y=1; \label{eqn83} 
	\end{eqnarray}
	\begin{eqnarray}
		& v_1 = \biggl\{ U(1) - \lambdaup_0 \biggr\} \mathrm{i} h_0, ~~ \text{at}~~ y = 1; \label{eqn84} 
		\\
		& v_1 = 0~~ \text{at}~~ y = 0; \label{eqn85} 
		\\
		& u_1 = \delta u^{'}_1,~~ \text{at}~~ y = 0; \label{eqn86} 
		\\
		& \mathcal{J}_{ba,1} = \mathrm{i} \biggl\{ U(1) - \lambdaup_0 \biggr\} A_0 + \mathrm{i} \Gamma_{a0} \biggl\{ U^{'}(1) h_0 + u_0 \biggr\},~~ \text{at}~~ y = 1; \label{eqn87} 
		\\
		& \gamma^{''}_1 = \mathrm{i} Pe_b \biggl\{ U(y) - \lambdaup_0 \biggr\} \gamma_0; \label{eqn88} 
		\\
		& \gamma^{'}_1 + \beta_a Pe_b \mathcal{J}_{ba,1} = 0,~~ \text{at}~~ y = 1; \label{eqn89} 
		\\
		& \gamma^{'}_1 = 0 ~~\text{at}~~y=0; \label{eqn90}
		\\
		& \mathcal{J}_{ba,1} = k_a \biggl[ R_a \biggl\{ (1 - \Gamma_{a0}) \gamma_1 - \Gamma_0 A_1 \biggr\} - A_1 \biggr],~~ \text{at}~~ y = 1. \label{eqn91}
	\end{eqnarray}
	
	The above system of equations yields the zero$^{th}$ order eigenvalue 
	\begin{eqnarray}
		& \lambdaup_0 = (2 + \tau)(1 + \delta) + 2 \delta. \label{eqn92}
	\end{eqnarray}
	Again, it can be seen that the expression of $\lambdaup_0$ converges with that of \citet{karapetsas2014role} in the limit $\tau \rightarrow 0$, and $\delta \rightarrow 0$. This mode, designated as ``interface mode'' by \citet{wei2005effect}, agrees with the same in the absence of slip parameter $\delta$ ($i.e., \delta = 0$) and in the insoluble surfactant limit ($\xi_a = \beta_a R_a \rightarrow \infty$). It is of interest to point out that the eigenvalue $\lambdaup_0$ indicates the existence of surface mode when the zero$^{th}$ order surface deformation $h_0 \ne 0$. Here, the phase speed of the surface mode relies on the slip parameter $\delta$ as well as external shear stress $\tau$. It is noteworthy to provide the zero$^{th}$ order amplitude of the free surface surfactant concentration as
	\begin{eqnarray}
		& A_0 = \displaystyle \frac{6 \{ 2(1 + \delta) + \tau \} \beta_a R_a \Gamma_{a0} (1 - \Gamma_{a0})^2 h_0}{6 (2 \delta + 1) \beta_a R_a (1 - \Gamma_{a0})^2 + (8 + 3 \tau + 12 \delta )} \label{eqn93}
	\end{eqnarray}
	This expression coincides with the result of \citet{karapetsas2014role} in the limit $\tau, \delta \rightarrow 0$ and with \citet{bhat2018linear} for $\tau \rightarrow 0$, $\xi_a = \beta_a R_a \rightarrow \infty$, $\Gamma_{a0} \rightarrow 1$. On the other hand, if $h_0 = 0$ and $A_0 \ne 0$, then a different mode can be found by using the surfactant transport equation, which is predominantly generated due to the distortion of surfactant concentration. The expression of the mode is the following
	\begin{eqnarray}
		& \lambdaup_0 = \displaystyle \frac{6 U(1) \beta_a R_a (1 - \Gamma_{a0})^2 + ( 4 + 3 \tau + 12 \delta + 6 \tau \delta)}{6 \{ 1 + \beta_a R_a (1 - \Gamma_{a0})^2 \}} \label{eqn94}
	\end{eqnarray}
	This mode is termed surfactant mode \cite{wei2005effect} or Marangoni mode \cite{samanta2021effect} or concentration mode \cite{pereira2008dynamics}. It produces the same result as of \citet{wei2005effect} with the slip parameter set to zero and $\xi_a \rightarrow \infty$ (in the case of insoluble surfactant, $\xi_a = \beta_a R_a \rightarrow \infty$ \cite{karapetsas2013primary}). Moreover, when external shear force $\tau$ is ignored ($i.e., \tau = 0$) and $\xi_a \rightarrow \infty$ (for insoluble surfactant), the expression of $\lambdaup_0$ replicate the same result as of \citet{bhat2018linear}. The nomenclature of $\lambdaup_0$ for different modes is redefined as $\lambdaup_{0_{s}}$ and $\lambdaup_{0_{sf}}$ to address the surface and surfactant mode respectively. Here also, the phase speed corresponding to the surfactant mode depends on both the slip parameter $\delta$ as well as on the external shear force $\tau$ alike. Moreover, this phase speed relied on flux parameters and the base concentrations of the surfactant at the interfaces explicitly, which makes its behaviour complex to observe.
	\begin{figure}
		\begin{center}
			\subfigure[]{\label{fig_3_a}\includegraphics*[width=8cm]{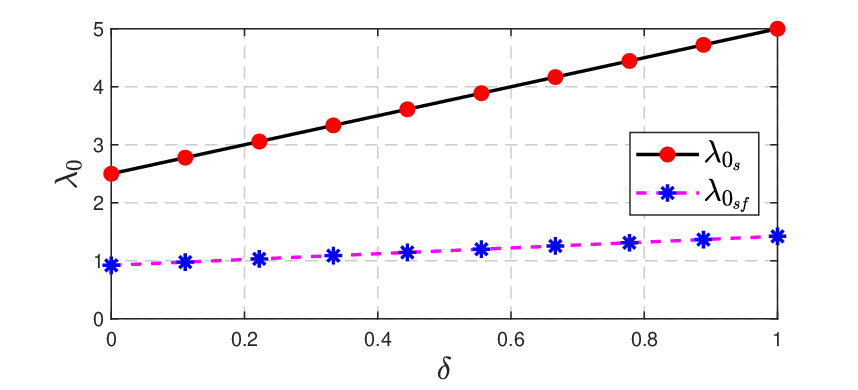}}
			\subfigure[]{\label{fig_3_b}\includegraphics*[width=8cm]{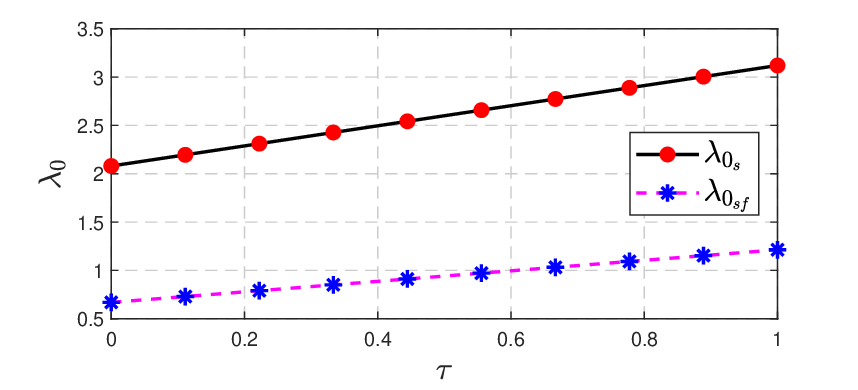}}
		\end{center}
		\caption{ The comparison of the phase speed between the surface mode and surfactant mode against (a) external shear force $\tau$, (b) slip parameter $\delta$. The other parameters are $\beta_a = 0.01$, $R_a = 1.0$, $k_a = 0.01$, $M_{tot} = 0.1$. } \label{fig_3}
	\end{figure}
	To get a more clear understanding, the phase speeds of zero$^{th}$ order of two modes are portrayed against the external shear force $\tau$ and the slip parameter $\delta$ in figure (\ref{fig_3}). It is evident that the slip parameter $\delta$ elevates the zero$^{th}$ phase speeds $\lambdaup_{0_{s}}$ of surface mode and $\lambdaup_{0_{sf}}$ of surfactant mode, but the phase speed for surface mode is elevated more profoundly compared to the phase speed $\lambdaup_{0_{sf}}$ of the surfactant mode (see figure \ref{fig_3_a}) and the zero$^{th}$ order phase speed is always more for the surface mode. Similarly, the phase speeds experience an increment for raising values of imposed shear, $\tau$ and in this case also the magnitude of the zero$^{th}$ order phase speed for the surface mode is more than that of the surfactant mode, as can be seen from the figure \ref{fig_3_b}.

	\subsubsection{Second-order analysis}
	
	Here, the second-order equations, i.e., O$(k^2)$ are 
	
	\begin{eqnarray}
		& v^{'}_2 + \mathrm{i} u_1 = 0, \label{eqn95}
		\\
		& Re \biggl[ \biggl\{ U(y) - \lambdaup_0 \biggr\} \mathrm{i} u_1 - \mathrm{i} u_0 \lambdaup_1 + U^{'}(y) v_2 \biggr] = - 2 \mathrm{i} p_1 + u^{''}_2 - u_0, \label{eqn96} 
		\\
		& Re \biggl[ U(y) - \lambdaup_0 \biggr] \mathrm{i} v_1 = - 2 p^{'}_2 + v^{''}_2, \label{eqn97} 
		\\
		& u^{'}_2(1) + \mathrm{i} v_1(1) - 2 h_2 = \displaystyle - 2 \mathrm{i} We \frac{E_0}{\Gamma_{a0}} A_1, ~~ \text{at}~~ y=1, \label{eqn98} 
		\\
		& p_2(1) + \mathrm{i} u_1(1) + \mathrm{i} U^{'}(1) h_1 - h_2 \cot(\xi) = We \sigma_0 h_0, ~~ \text{at}~~ y=1, \label{eqn99} 
		\\
		& v_2(1) = \{ U(1) - \lambdaup_0 \} \mathrm{i} h_1 - \mathrm{i} h_0 \lambdaup_1, ~~ \text{at}~~ y=1, \label{eqn100} 
		\\
		& v_2(0) = 0; ~~ u_2(0) = \delta u_2^{'}(0), ~~ \text{at}~~ y=0, \label{eqn101} 
		\\
		& \mathrm{i} \{ U(1) - \lambdaup_0 \} A_1 - \mathrm{i} \lambdaup_1 A_0 + \mathrm{i} \Gamma_{a0} \{ U^{'}(1) h_1 + u_1(1) \} = - \displaystyle \frac{A_0}{Pe_{ca}} + \mathcal{J}_{ba,2}, ~~ \text{at}~~ y=1, \label{eqn102} 
		\\
		& \mathrm{i} \{ U(y) - \lambdaup_0 \} \gamma_1(y) - \mathrm{i} \lambdaup_1 \gamma_0(y) = \displaystyle \frac{1}{Pe_b} \{ \gamma^{''}_2(y) - \gamma_0(y) \}, \label{eqn103} 
		\\
		& \gamma^{'}_2(1) = - \beta_a Pe_b \mathcal{J}_{ba,2}, ~~ \text{at}~~ y=1, \label{eqn104} 
		\\
		& \gamma^{'}_2(0) = 0, ~~ \text{at}~~ y=0, \label{eqn105} 
		\\
		& \mathcal{J}_{ba,2} = k_a [ R_a \{ (1 - \Gamma_{a0}) \gamma_2(1) - \Gamma_0 A_2 \} - A_2 ]. \label{eqn106}
	\end{eqnarray}
	
	\subsubsection[]{Surface mode}

	Solving the above equations, the first order eigenvalue for the surface mode $\lambdaup_{1{s}}$ can be expressed as follows
	
	\begin{eqnarray}
		& \nonumber \lambdaup_{1_{s}} = \displaystyle \mathrm{i} \biggl[ \frac{2}{15} Re (2 + 2 \delta + \tau) (2 + 10 \delta + 15 \delta^2) - \frac{2}{3}(3 \delta + 1) \cot(\xi) \\
		& \displaystyle - (2 \delta + 1) We E_0 \frac{6 \{ 2(1 + \delta) + \tau \} \beta_a R_a (1 - \Gamma_{a0})^2 }{6 (2 \delta + 1) \beta_a R_a (1 - \Gamma_{a0})^2 + (8 + 3 \tau + 12 \delta )} \biggr] \label{eqn107}
	\end{eqnarray}
	This term $\lambdaup_{1{s}}$ is purely imaginary and assists in the growth rate of the flow if greater than zero. The first term shows the inertial force destabilizes the flow, whereas the term for gravitational effect in the second has a stabilizing impact, and the last term is due to the soluble surfactant, which happens to have a stabilizing effect on the flow. Now, to derive the critical Reynolds number, the neutral stability condition $\mathcal{\Im{\lambdaup}} = 0$ is used, and the following expression is derived for surface mode as
	\begin{eqnarray}
		& \hspace{-8cm} \nonumber Re_{c_s} = \displaystyle \frac{5 (3 \delta + 1) \cot(\xi)}{(2 + 2 \delta + \tau) (2 + 10 \delta + 15 \delta^2)}\\
		& \hspace{2cm} +~~ \displaystyle \frac{45 (2 \delta + 1) We E_0 \beta_a R_a (1 - \Gamma_{a0})^2 }{ (2 + 10 \delta + 15 \delta^2)[6 (2 \delta + 1) \beta_a R_a (1 - \Gamma_{a0})^2 + (8 + 3 \tau + 12 \delta )]} \label{eqn108}
	\end{eqnarray}
	
	This result captures the exact form of the critical Reynolds number derived by \citet{karapetsas2014role, karapetsas2013primary} in the limit $\lambdaup \rightarrow 0,~~ \tau \rightarrow 0$. Also, it reduces to the result of \citet{samanta2011falling}for an uncontaminated flow over a slippery bottom, i.e., when external shear stress and surfactants are absent from the ongoing article. One extra term $(1 + 2 \delta)$ is present in the numerator of the critical Reynolds number derived in \citet{samanta2011falling} due to the fact that it chooses the velocity scale as $U = \displaystyle \frac{g H^2 \sin(\xi)}{2 \nu (1 + 2 \delta)}$ instead of the velocity scale chosen in this article. 
	
	\begin{figure} 
		\begin{center}
			\subfigure[]{\label{fig_4_a}\includegraphics*[width=8cm]{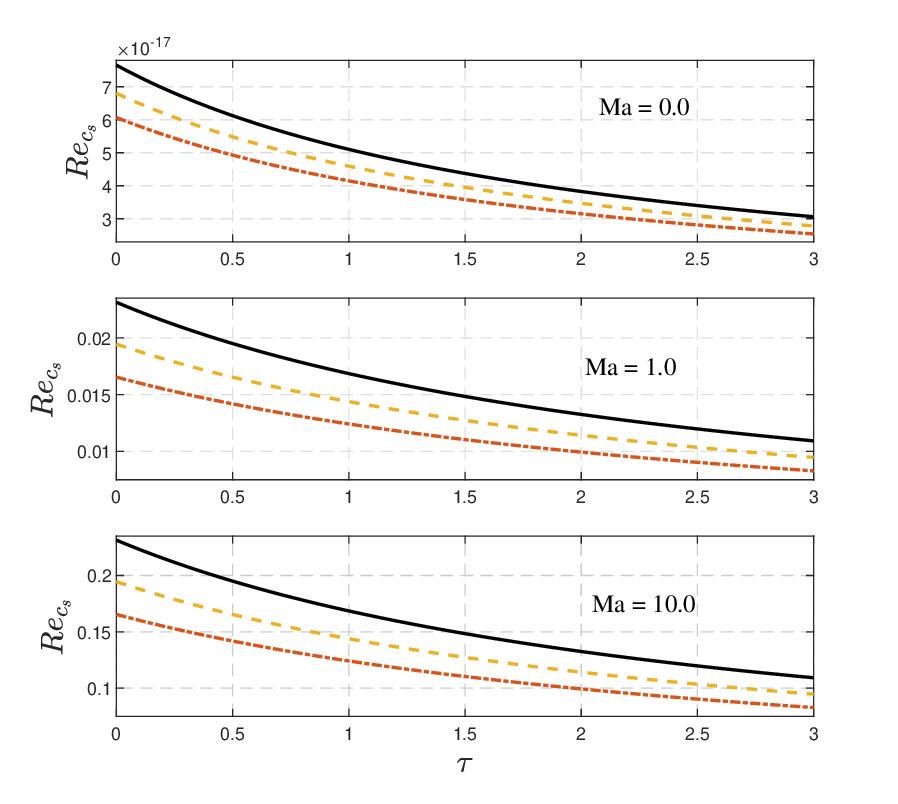}}
			\subfigure[]{\label{fig_4_b}\includegraphics*[width=8cm]{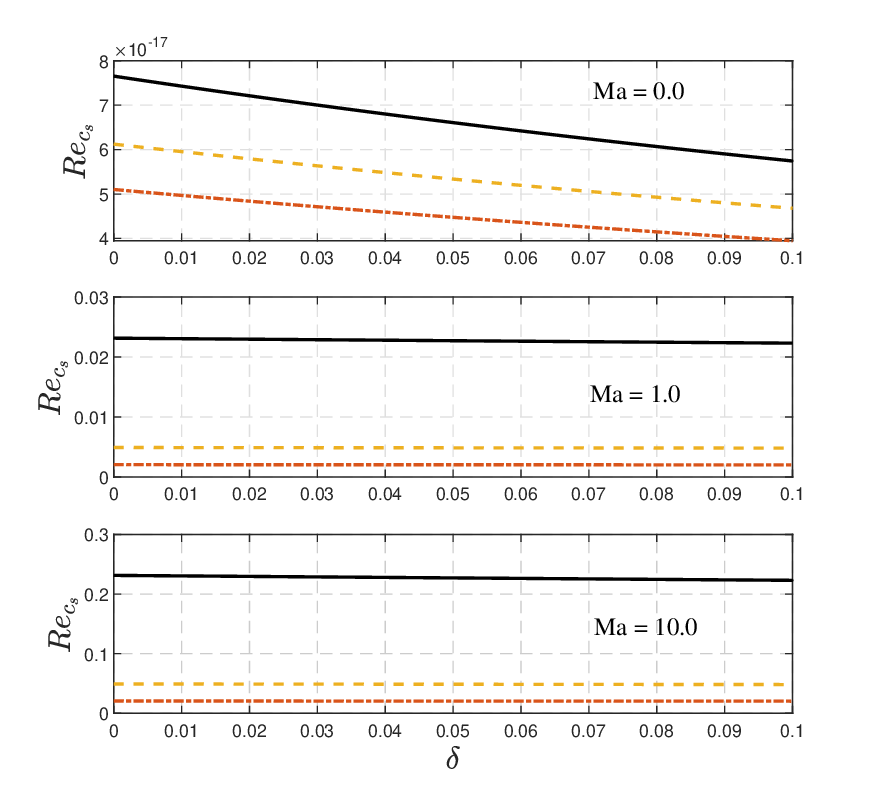}}
		\end{center} 
		\caption{ The variation of critical Reynolds number against (a) external shear force $\tau$, (b) slip parameter $\delta$. In the first set of figures, the solid lines, dashed lines, and dash-dot lines represent the values of external shear force $\tau = 0.0, 0.5, 1.0$ respectively. Similarly, in the second set of figures, the solid, dashed, and dash-dot lines represent the values of the slip parameter $\delta = 0.00, 0.04, 0.08$ respectively. Also, the Marangoni number $Ma$ for the corresponding figure is shown within the figure. The other parameters are $\beta_a = 0.01$, $R_a = 1.0$, $M_{tot} = 0.1$. } \label{fig_4}
	\end{figure}
	The characteristics of the critical Reynolds number $Re_{c_s}$ against external shear force $\tau$ is explored in figure \ref{fig_4_a}. It can be seen that in the absence of surfactant ($Ma = 0.0$), the critical Reynolds number converges to zero and varies for different values of slip parameters $\delta$ at $\order{10^{-17}}$. It can be seen that with an increment of the Marangoni number $Ma$, the critical Reynolds number grows considerably and maintains the same pattern of decreasing with an increment of slip parameter $\delta$. In figure \ref{fig_4_b}, the similar effect can be observed for the critical Reynolds number, and it decreases for increasing external imposed shear force $\tau$. Also, the critical modified Reynolds number $\chi_c$ can be derived from $Re_{c_s}$ using the relation
	\begin{eqnarray}
		\nonumber & \hspace{-4cm} \chi_c =  \displaystyle \frac{2 Re_{c_s}}{\sin(\xi)} = \frac{2}{\sin(\xi)}\bigg[ \displaystyle \frac{5 (3 \delta + 1) \cot(\xi)}{(2 + 2 \delta + \tau) (2 + 10 \delta + 15 \delta^2)}\\
		& \hspace{4.5cm} +~~ \displaystyle \frac{45 (2 \delta + 1) We E_0 \beta_a R_a (1 - \Gamma_{a0})^2 }{ (2 + 10 \delta + 15 \delta^2)[6 (2 \delta + 1) \beta_a R_a (1 - \Gamma_{a0})^2 + (8 + 3 \tau + 12 \delta )]} \bigg] \label{eqn109}
	\end{eqnarray}
	
	\subsubsection[]{Surfactant mode}
	\indent Assuming $h_0 = 0$ and using the first order equations, the first order interfacial deflection $h_1$ can be obtained as 
	\begin{eqnarray} 
		h_1 = \frac{ 6 \mathrm{i} (2 \delta +1) Ma A_0 (1 +\mu)}{\Gamma_{a0} [ (8 + 3 \tau + 12 \delta) + 6 (2 \delta + 1)\mu ]} \label{eqn110} 
	\end{eqnarray}
	where $Ma = We E_0$, $\mu = \beta_a R_a (1- \Gamma_{a0})^2$ as considered by \citet{karapetsas2013primary}. This expression is consistent with the result of \citet{wei2005effect} for the limiting case of $\xi_a = \beta_a R_a \rightarrow \infty$ (insoluble surfactant), $\delta \rightarrow 0$ except the fact that $Ma$ is there instead of $Ma/2$ because of the choosing of different scale of the non-dimensionalize parameters. Using equations \eqref{eqn102} and \eqref{eqn110}, the first-order eigenvalue corresponding to the surfactant mode can be derived as
	
	\begin{eqnarray} 
		&\nonumber \lambdaup_{1_{sf}} = \displaystyle \mathrm{i} \bigg[ \frac{6 \mu (2 + 2\delta + \tau) (2 \delta + 1) Ma }{8 + 3 \tau + 12 \delta + 6 (2 \delta + 1) \mu} - 2 (1 + \delta) Ma \frac{\mu}{1 + \mu} - \frac{\mu}{Pe_{ca}(1 + \mu)} - \frac{1}{(1 + \mu) Pe_b}\\
		& \displaystyle \nonumber + \frac{Pe_b}{1 + \mu}\biggl\{ -(2 + \tau)^2 \frac{(5 \delta + 1)}{120} + \frac{2 + \tau}{360} + \frac{(2 + \tau) \lambdaup_{0_{sf}}}{24} - \frac{(2 + \tau)(18 \delta + 5)}{180} + \frac{1}{84} + \frac{\lambdaup_{0_{sf}}}{10} \\
		&\nonumber \displaystyle + \frac{\delta (2 + \tau) (4 \delta + 1)}{12} + \frac{\delta}{30} + \frac{\delta \lambdaup_{0_{sf}}}{3} \biggr\} - \frac{(2 + 3 \tau) \mu}{6 (1 + \mu)^2} \biggl\{ \frac{(2 + 3 \tau) (1 - \Gamma_{a0})}{6 k_a (1 + \mu)^2}\\
		& \displaystyle + Pe_b \frac{(1 + \tau) + (3 + 4 \tau)\mu}{12 \mu (1 + \mu)} \biggr\} \bigg]\label{eqn111}
	\end{eqnarray}

	Here again, the first-order eigenvalue corresponding to the surfactant mode matches with that of \citet{bhat2019linear} when insoluble surfactant ($\xi_a \rightarrow \infty$) is considered with no-slip condition ($\delta \rightarrow 0$). It is apparent from equation \eqref{eqn111}, that the eigenvalue is purely imaginary, and it is contributing to the growth rate of the flow if $\Im{\lambdaup_{1_{sf}}} > 0$ and in turn which gives the critical value for the free surface P\'eclet Number as
	\begin{eqnarray}
		Pe_{ca_m} = \frac{72 k_a Pe_b \{ 8 + 3\tau + 12 \delta + 6 (2 \delta + 1) \mu \} \mu (1 + \mu)^2}{72 k_a (1 + \mu)^2 d_1 - d_2 \mu Pe_b}, \label{eqn112}
	\end{eqnarray} 
	where 
	\begin{eqnarray}
		d_1 &=& 6 (2 + 2 \delta + \tau) (2 \delta + 1) Ma \mu (1 + \mu) Pe_b  \nonumber \\
		&& - \bigg[ 2 \mu (1 + \delta) Ma Pe_b + 1 - Pe^2_b 
		\bigg\{ -(2 + \tau)^2 \frac{(5 \delta + 1)}{120} + \frac{2 + \tau}{360} 
		+ \frac{(2 + \tau) \lambdaup_{0_{sf}}}{24}  \nonumber \\
		&& - \frac{(2 + \tau)(18 \delta + 5)}{180} + \frac{1}{84} 
		+ \frac{\lambdaup_{0_{sf}}}{10} + \frac{\delta (2 + \tau) (4 \delta + 1)}{12} 
		+ \frac{\delta}{30} + \frac{\delta \lambdaup_{0_{sf}}}{3} 
		\bigg\} \bigg]  \nonumber \\
		&& \nonumber \times \{ 8 + 3\tau + 12 \delta + 6 (2 \delta + 1) \mu \}, \\
		d_2 &=& \bigg[ 2 (1 - \Gamma_{a0}) (2 + 3\tau)^2 
		+ (2 + 3 \tau) k_a Pe_b \{ (1 + \tau) + (3 + 4 \tau) \mu \} \bigg] \nonumber \\
		&& \nonumber \times \{ 8 + 3 \tau + 12 \delta + 6 \mu (2 \delta + 1) \}.
	\end{eqnarray}

\begin{figure}
		\begin{center}
			\subfigure[]{\label{fig_5_a}\includegraphics*[width=8cm]{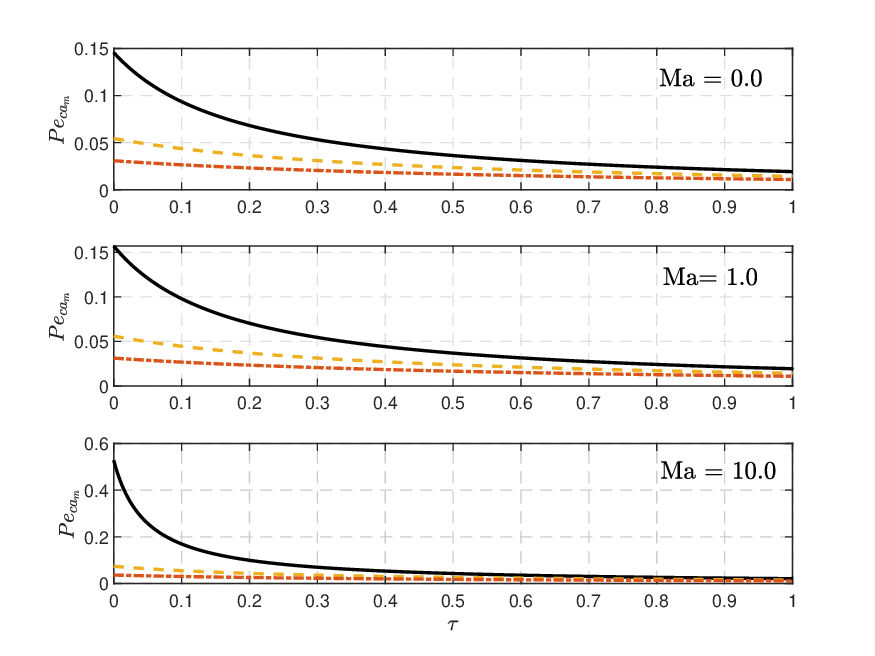}}
			\subfigure[]{\label{fig_5_b}\includegraphics*[width=8cm]{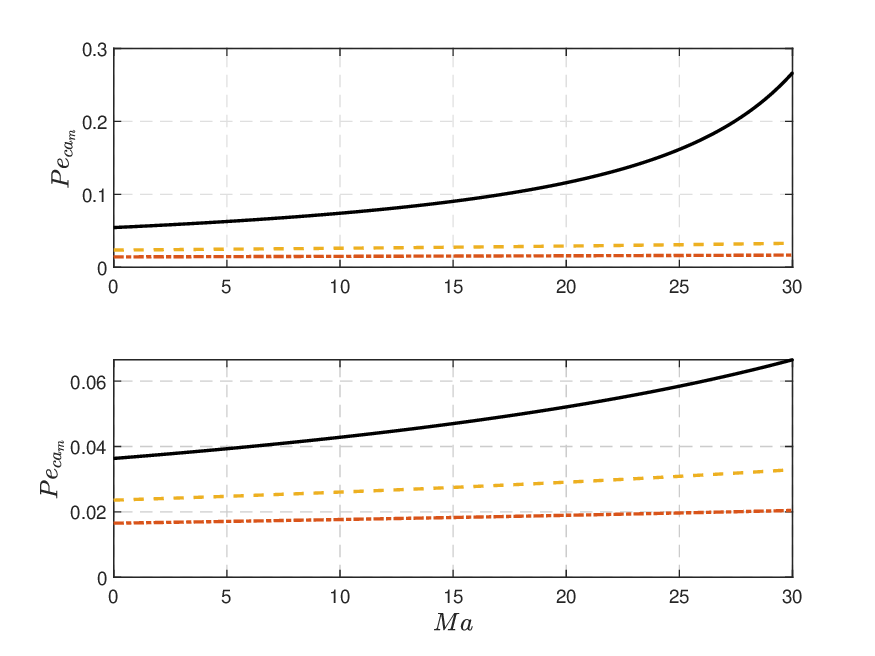}}
		\end{center} 
		\caption{ The variation of critical P\'eclet number against (a) external shear force $\tau$, (b) Marangoni number $Ma$. In the first set of figures, the solid lines, dashed lines, and dash-dot lines represent the values of the slip parameter $\delta = 0.00, 0.04, 0.08$ respectively. Similarly, in the second set of figures, the top figure represents the critical P\'eclet number against the Marangoni number $Ma$ for varying external shear force $\tau$ with $\delta = 0.04$ and the bottom one for varying slip parameter $\delta$ with $\tau = 0.5$. The other parameters are $\xi = 90^{\circ}$, $\chi = 1$, $\Sigma = 2.0$, $Sc_b = 10$,$M_{tot} = 0.1$, $\beta_a = 0.01$, $R_a = 1.0$ and $k_a = 1.0$ .}\label{fig_5}
	\end{figure}

	Here, the critical P\'eclet number depends on the soluble surfactant parameter as well as on the external shear force $\tau$ and slip parameter $\delta$. This dependence is explored in the figure (\ref{fig_5}).

	In figure \ref{fig_5_a}, the critical P\'eclet number is plotted against external shear $\tau$ for three different values of the Marangoni number $Ma$ in the case of varying slip parameter $\delta$. It is observed that the critical P\'eclet number reduces for increasing slip parameter $\delta$. The deviation of critical P\'eclet number  is negligible for $Ma \in [0.0, 1.0]$, but for $Ma = 10.0$, the value of the critical P\'eclet number increases for each value of the slip parameter $\delta$. It can be noticed from figure \ref{fig_5_b} that for large values of the Marangoni number $Ma$, the critical P\'eclet number accelerates sharply.

	\subsection{Linear stability analysis for arbitrary wavenumber}
	
\begin{figure} 
		\begin{center}
			\subfigure[]{\label{fig_6_a}\includegraphics*[width=8cm]{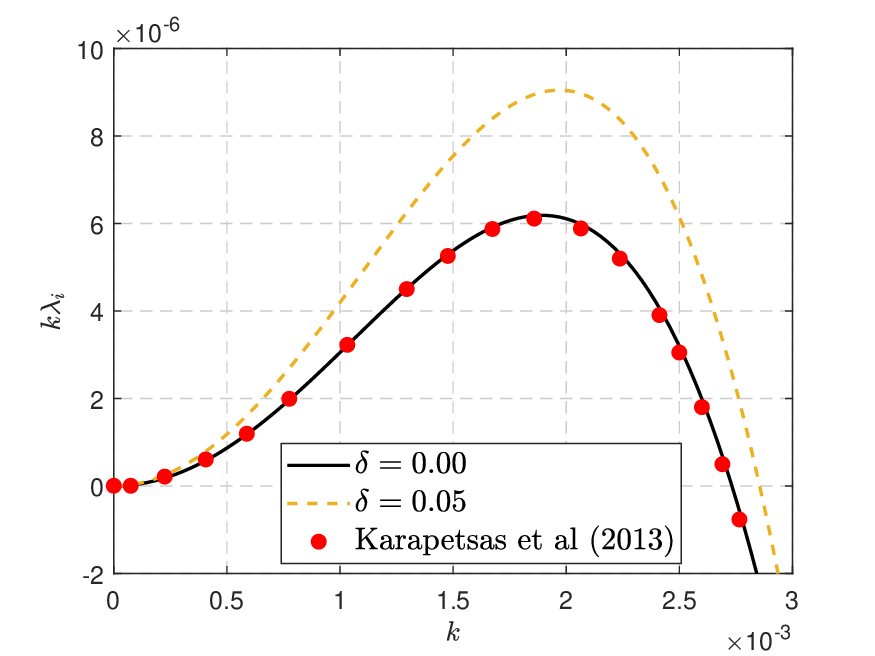}}
			\subfigure[]{\label{fig_6_b}\includegraphics*[width=8cm]{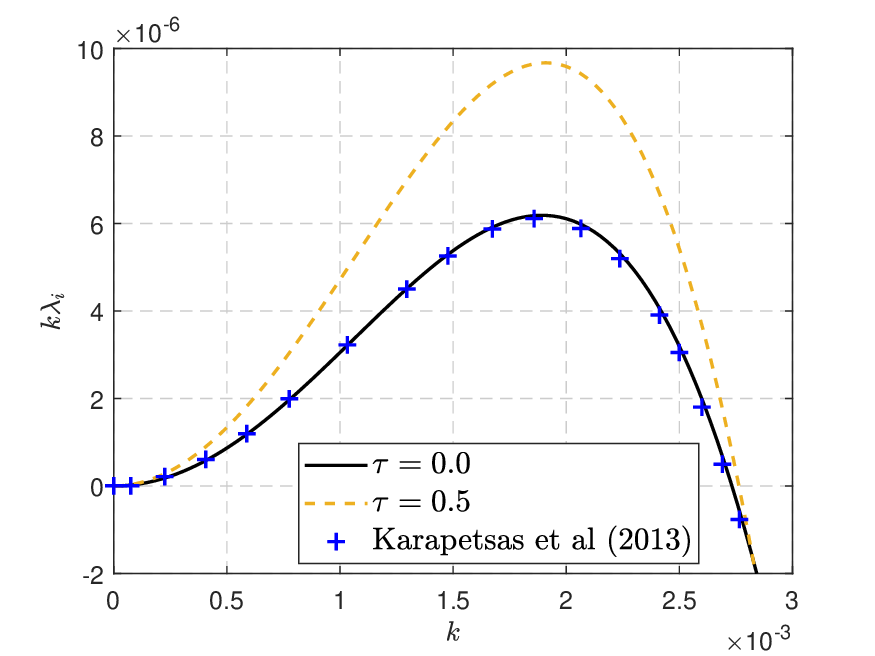}}
		\end{center} 
		\caption{ The temporal growth rate of instability against wavenumber $k$ for varying (a) $\delta$ with $\tau = 0.0$ (b) $\tau$ with $\delta = 0.0$. The other parameters are $\xi = 90^{\circ}$, $\chi = 15.0$, $Sc_a = 100$, $Sc_b = 10$, $Ka = 3000$, $\Sigma = 2$, $\beta_a = k_a = 0.01$, $R_a = 1$, $M_{tot} = 0.1$. The red bullets and blue addition signs are the digitally scanned data points from figure $(4)$ of \citet{karapetsas2013primary}.} \label{fig_6}
	\end{figure}

	This section is dedicated to investigate the linear stability analysis for arbitrary wavenumber for infinitesimal perturbation of the flow using the Orr-Sommerfeld boundary value problem (BVP). To construct the BVP, the perturbed velocity components are defined as 
	\begin{eqnarray}
		\widehat{u}(y) = \psi^{'}(y) ~;~~ \widehat{v}(y) = - \psi(y) \label{eqn113}
	\end{eqnarray}
	where $\psi(y)$ is the streamfunction of the flow. Now using the aforementioned substitutions, the equations \eqref{eqn47}-\eqref{eqn58} take form as 
	
	\begin{eqnarray}
		& \nonumber \psi^{''''}(y) - 2 k^2 \psi^{''}(y) + k^4 \psi(y) - \mathrm{i} k Re \biggl[ \bigg\{ U(y) - \lambdaup\bigg\}\bigg\{ \psi^{''}(y) - k^2 \psi(y) \bigg\} \\
		& \hspace{7cm}- U^{''}(y) \psi (y) \biggr] = 0~,0~\leq y \leq 1,\label{eqn114} 
	\end{eqnarray}
	\begin{eqnarray}
		& \Gamma^{''}(y) - k^2 \Gamma(y) - \mathrm{i} k Pe_b \biggl\{ U(y) - \lambdaup \biggr\} \Gamma(y) = 0~, ~ 0~\leq y \leq 1, \label{eqn115} \\
		& \displaystyle \psi^{''}(1) + k^2 \psi(1) - 2 h + 2 \mathrm{i} k We \frac{E_0}{\Gamma_{a0}} \Gamma_a = 0,~~\text{at}~~ y = 1, \label{eqn116} \\
		& \nonumber \psi^{'''}(1) - 3 k^2 \psi^{'}(1) - \mathrm{i} k Re \biggl[ \biggl\{ U(1) - \lambdaup \biggr\} \psi^{'}(1) - U^{'}(1) \psi(1) \biggr] - \mathrm{i} k \biggl\{ 2 \cot(\xi) \\
		& \hspace{5cm} - 2 \mathrm{i} k U^{'}(1) + 2 k^2 We \sigma_0 \biggr\} h = 0, ~~\text{at}~~ y = 1, \label{eqn117} \\
		& \psi(1) + \biggl\{ U(1) - \lambdaup \biggr\} h = 0, ~~\text{at}~~ y = 1, \label{eqn118} \\
		& \nonumber \biggl[ \mathrm{i} k \biggl\{ U(1) - \lambdaup \biggr\} + \displaystyle \frac{k^2}{Pe_{ca}} + k_a (R_a \Gamma_0 +1) \biggr] \Gamma_a - k_a R_a (1 - \Gamma_{a0}) \Gamma(1) \\
		& \hspace{5cm} + \mathrm{i} k \Gamma_{a0} \psi^{'} (1) + \mathrm{i} k \Gamma_{a0} U^{'}(1) h = 0,~~\text{at}~~ y = 1, \label{eqn119} \\
		& \psi(0) = 0, ~~\text{at}~~ y = 0, \label{eqn120} \\
		& \delta \psi^{''}(0) - \psi^{'}(0)= 0 ,~~\text{at}~~ y = 0, \label{eqn121} \\
		& \displaystyle \frac{1}{Pe_b} \Gamma^{'}(1) + \beta_a k_a \biggl[ R_a \biggl\{ \Gamma(1) (1 - \Gamma_{a0} ) - \Gamma_0 \Gamma_a\biggr\} - \Gamma_a \biggr] = 0, ~~\text{at}~~ y = 1, \label{eqn122}\\
		& \Gamma^{'} (0) = 0, ~~\text{at}~~ y = 0, \label{eqn123}
	\end{eqnarray}
	This system of equations is the Orr-Sommerfeld Boundary value system corresponding to the flow problem. In order to solve this problem, the numerical Chebyshev spectral collocation method is implemented. To this end, the streamfunction is expanded up to $(N+1)$ terms as 
	\begin{eqnarray}
		\psi(y) = \sum_{i=0}^{N} \breve{\psi}_i \mathscr{T}_i(y)~,~ \Gamma(y) = \sum_{i=0}^{N} \breve{\Gamma}_i \mathscr{T}_i(y) \label{eqn124}
	\end{eqnarray}
	where $\mathscr{T}_i(x)$ are the Chebyshev polynomials of the first kind, $\breve{\psi}_i~,~\breve{\Gamma}_i$ are the unknown constants to be determined. As the Chebyshev polynomials are ranged over $[-1,1]$, so the domain of the flow is converted from $[0,1]$ to $[-1,1]$ using the transformation $y = (x+1)/2$. Thereafter, substituting equation \eqref{eqn124} into the system of the equations \eqref{eqn114}-\eqref{eqn123}, and hence the polynomials are evaluated at Gauss-Lobatto points $x_i = \cos(\displaystyle \frac{\pi i}{N})$, $j = 0,1,2,\cdots,N$, which results in the system of linear eigenvalue problem as 
	\begin{eqnarray} 
		\mathscr{A Y} =\lambdaup \mathscr{B Y} \label{eqn125}
	\end{eqnarray}
	where $\mathscr{A}$, $\mathscr{B}$ are matrices of order $(2N + 4)$ and $\lambdaup$ is the eigenvalue. The precision of the code is shown in figure (\ref{fig_6}) by verifying the temporal growth rate of instability with the results of \citet{karapetsas2013primary} for $\chi = 15$ and varying $\delta$ and $ \tau $ (Figure 4 of their paper). In the first figure \ref{fig_6_a}, slip parameter is varied with externally imposed shear $\tau$ set to zero. Similarly, external imposed shear $\tau$ is varied in the second figure with slip parameter $\delta = 0$ (see figure \ref{fig_6_b}) and in both cases, it can be seen that the results are in very good agreement when parameters ($\delta = 0$ in first figure and $\tau = 0$ in second figure). The red bullets and blue addition signs are the data points extracted from the figure 4 of \citet{karapetsas2013primary}.
	
\begin{figure} 
		\begin{center}
			\subfigure[]{\label{fig_7_a}\includegraphics*[width=5.4cm]{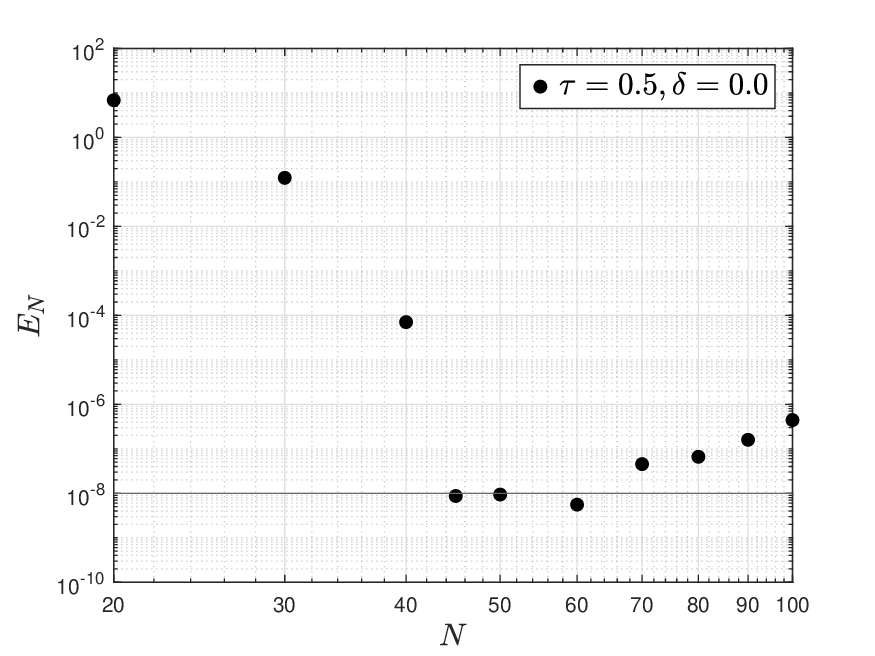}}
			\subfigure[]{\label{fig_7_b}\includegraphics*[width=5.4cm]{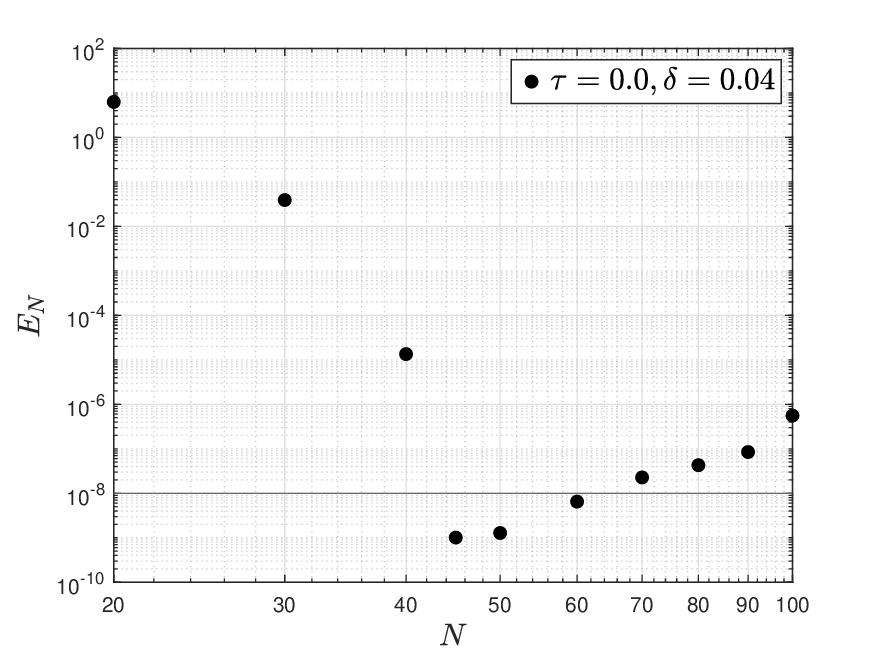}}
			\subfigure[]{\label{fig_7_c}\includegraphics*[width=5.4cm]{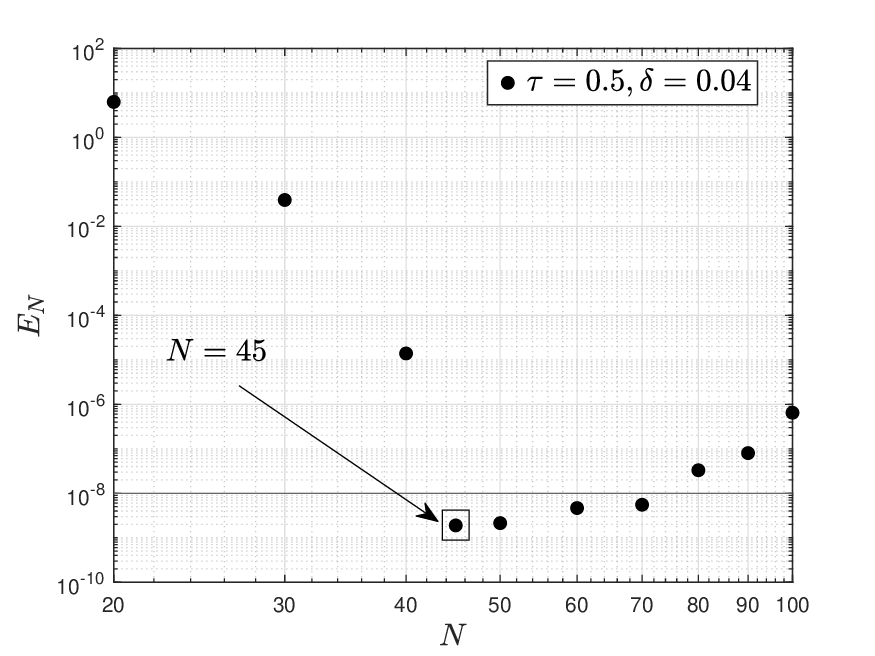}}
		\end{center} 
		\caption{ The variation of relative error $E_N$ against the number of polynomials $N$ for (a) $\tau = 0.5$, $\delta = 0.0$, (b) $\tau = 0.0$, $\delta = 0.04$ and (c) $\tau = 0.5$, $\delta = 0.04$. The other relevant parameters are $\xi = 90^{\circ}$, $\chi = 15$, $Ka = 3000$, $\Sigma = 2.0$, $Sc_a = 100$, $Sc_s = 100$, $Sc_b = 10$, $M_{tot} = 1.0$, $\beta_a = 0.01$, $R_a = 1.0$, $k_a = 0.01$.} \label{fig_7}
	\end{figure}

	In order to determine the number of sufficient polynomials required to solve the problem \eqref{eqn125} most accurately, the convergence test following \citet{tilton2008linear}, is being carried out using the relation
	\begin{eqnarray} 
		E_N = \frac{||e_{N+1} - e_N||_2}{||e_N||_2} \label{eqn126}
	\end{eqnarray}
	where $||.||_2$ denotes the $L_2$ norm. Also, the $e_{N+1}$ and $e_N$ are column vectors containing the twenty most unstable eigenvalues for polynomials of order $(N+1)$ and $N$, respectively, and the relative error is portrayed in figure (\ref{fig_7}). Three cases have been considered where the individual effects of imposed shear $\tau$ and slip parameter $\delta$ were considered during error analysis along with the combined effect of them is also tested. It can be seen that the relative errors are of $\order{10^{-8}}$ or below when $45 \le N \le 60$ and for $N \ge 60$, the error shows a steady increment, and hence the number of collocation point is fixed between $45 \le N \le 60$ so that the corresponding error will be of $\order{10^{-8}}$. 
	
	\begin{figure} 
		\begin{center}
			\subfigure[]{\label{fig_8_a}\includegraphics*[width=8cm]{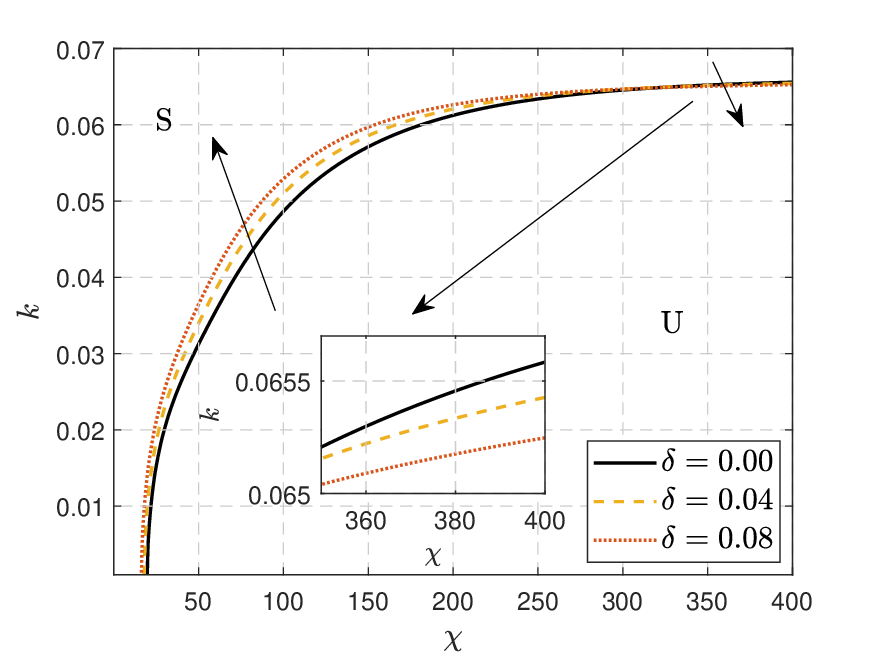}}
			\subfigure[]{\label{fig_8_b}\includegraphics*[width=8cm]{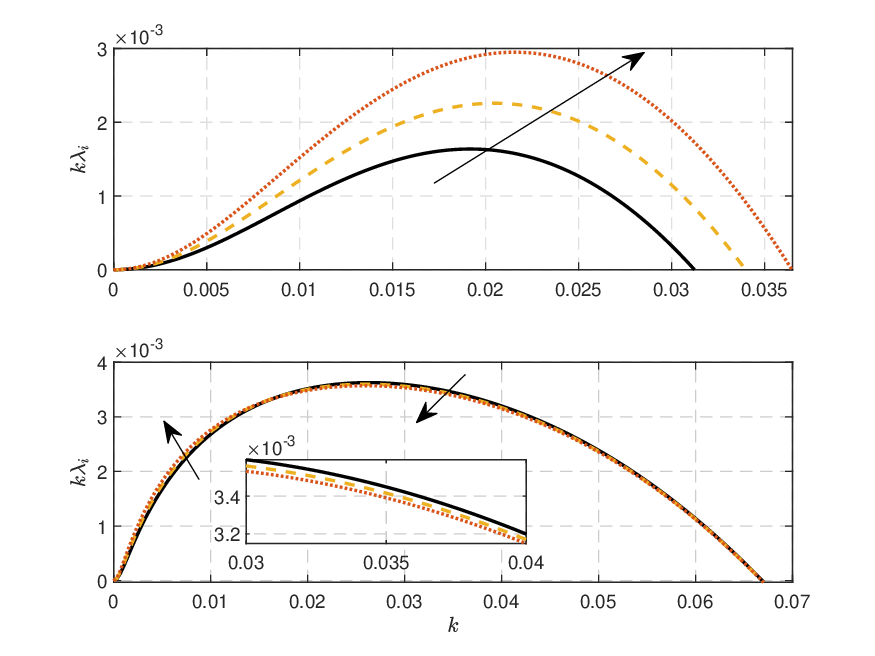}}
		\end{center}
		\caption{ (a) The variation of the marginal stability curve in the ($\chi-k$) plane for surface mode. The arrows are pointed from lower to higher values of the slip parameter $\delta$. The ``$S"$ and ``$U"$ indicate the stable and unstable zones. (b) The temporal growth rate variations against wavenumber $k$ for a surface mode where the solid line is for $\delta = 0.00$, the dashed line for $\delta = 0.04$, and the dotted line for $\delta = 0.08$. The values of the modified Reynolds number $\chi$ for the temporal growth rates are $\chi = 50$(top) and $\chi = 1000$(bottom). Here again, the arrow indicates the lower to higher value of the slip parameter $\delta$. The other parameters are $\xi = 90^{\circ}$, $Ka = 5000$, $\Sigma = 2.0$, $Sc_a = 100$, $Sc_b = 10$, $M_{tot} =0.1$, $\beta_a = 0.01$, $R_a = 10.0$, $k_a = 1.0$, $\tau= 0.5$.} \label{fig_8}
	\end{figure}
	
	In order to observe the influence of the slip parameter for the surface mode, the marginal stability curve in the $(\chi, k)$ plane is numerically computed and depicted in figure \ref{fig_8_a}. It can be observed apparently that the slip parameter $\delta$ helps to reduce the critical modified Reynolds number ($\chi_c$) near the onset of instability and thus increases the unstable region, suggesting a destabilizing effect. At the same time, for larger modified Reynolds number $\chi$, the slip parameter changes its behaviour by decreasing the unstable wavenumber region and thus helping stabilize the flow as can be seen in the inset figure of figure \ref{fig_8_a}. This phenomenon was previously reported by earlier studies \citep{samanta2011falling, bhat2018linear, davalos2023marangoni, selim2024gravity}. The slippery bottom chosen here is a hydrophobic surface that has a contact angle of more than $90^{\circ}$ for repelling action. When the modified Reynolds number is small, the inertial force is less then possibly the hydrophobicity of the bottom is dominant and repels the fluid due to the greater contact angle, but as soon as the inertial force gains supremacy and crosses some threshold value, the contact angle becomes less than $90^{\circ}$ which allows wetting, and this might be the reason behind the dual nature of the slip parameter \cite{jeevahan2018superhydrophobic}. The associated temporal growth rates are depicted in figure \ref{fig_8_b}. The modified Reynolds number preferred for the growth rate of the top figure is $\chi = 50$, and for the bottom figure, it is $\chi = 1000$. It can be observed that the figure at the top displays the destabilizing nature of the slip parameter at the onset of instability, and the bottom figure shows that it has a dual nature in different regions of wavenumbers.
	
	\begin{figure} 
		\begin{center}
			\subfigure[]{\label{fig_9_a}\includegraphics*[width=8cm]{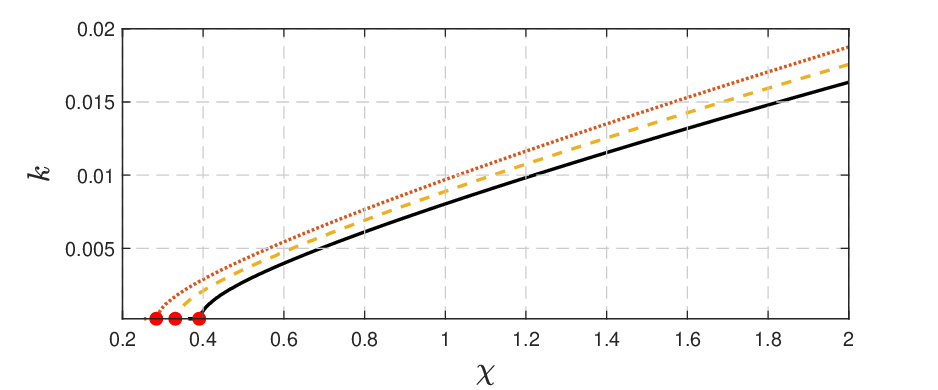}}
			\subfigure[]{\label{fig_9_b}\includegraphics*[width=8cm]{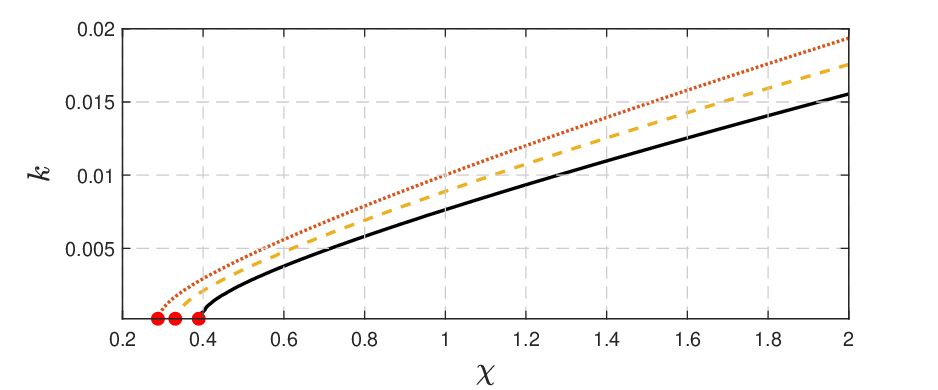}}
		\end{center} 
		\caption{ The marginal stability curve for varying (a) slip parameter $\delta$, (b) imposed shear $\tau$ for surface mode. In the first figure, the solid curve stands for $\delta = 0.00$, dashed for $\delta = 0.04$, and dotted for $\delta = 0.08$, Similarly, in the second figure, the marginal stability curve for $\tau = 0.0$ is shown in solid, $\tau = 0.5$ in dashed, and $\tau = 1.0$ in dotted line. The red dots in both figures are critical modified Reynolds numbers obtained from the analytical longwave approximation method.The other parameters are $\xi = 90^{\circ}$, $Ka = 5000$, $\Sigma = 2.0$, $Sc_a = 100$, $Sc_b = 10$, $M_{tot} =0.1$, $\beta_a = 0.01$, $R_a = 1.0$, $k_a = 1.0$, $Ma = 10.0$, $\delta= 0.04$ and $\tau = 0.5$.} \label{fig_9}
	\end{figure}
	The comparison between the analytical calculation and numerical computation for the surface mode has been shown in figure \ref{fig_9}. The marginal stability curves are plotted for different values of slip parameter $\delta$ in case of surface mode in the figure \ref{fig_9_a}, and the red dots are analytically computed data points for the critical modified Reynolds number, and an excellent agreement can be observed. Similarly, the marginal stability curves are plotted for different values of imposed shear $\tau$ in figure \ref{fig_9_b}, and again, the red dots are data points corresponding to the critical modified Reynolds number for varying imposed shear, and it matches perfectly with the numerical result.\\
	\begin{figure}
		\begin{center}
			\subfigure[]{\label{fig_10_a}\includegraphics*[width=8cm]{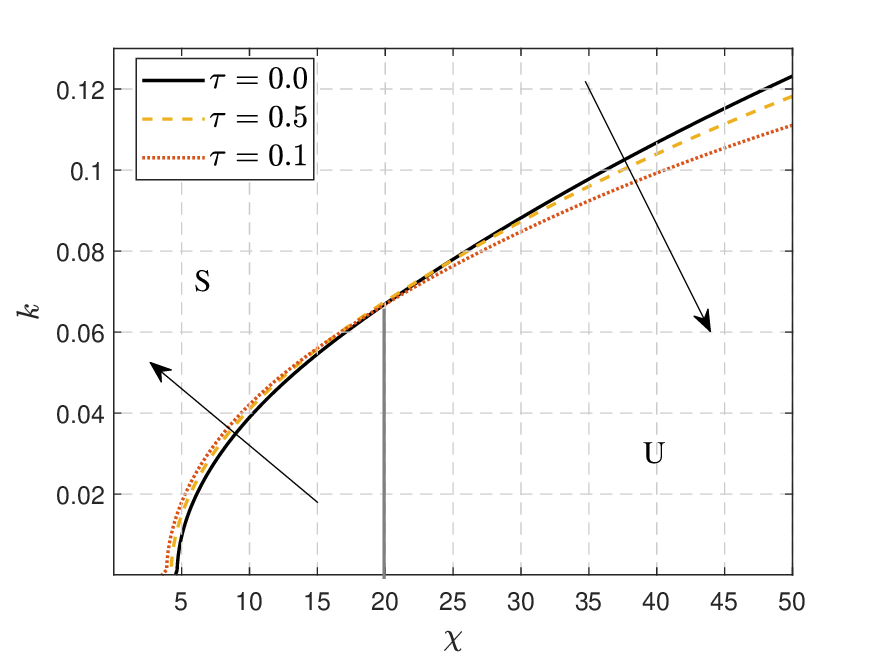}}
			\subfigure[]{\label{fig_10_b}\includegraphics*[width=8cm]{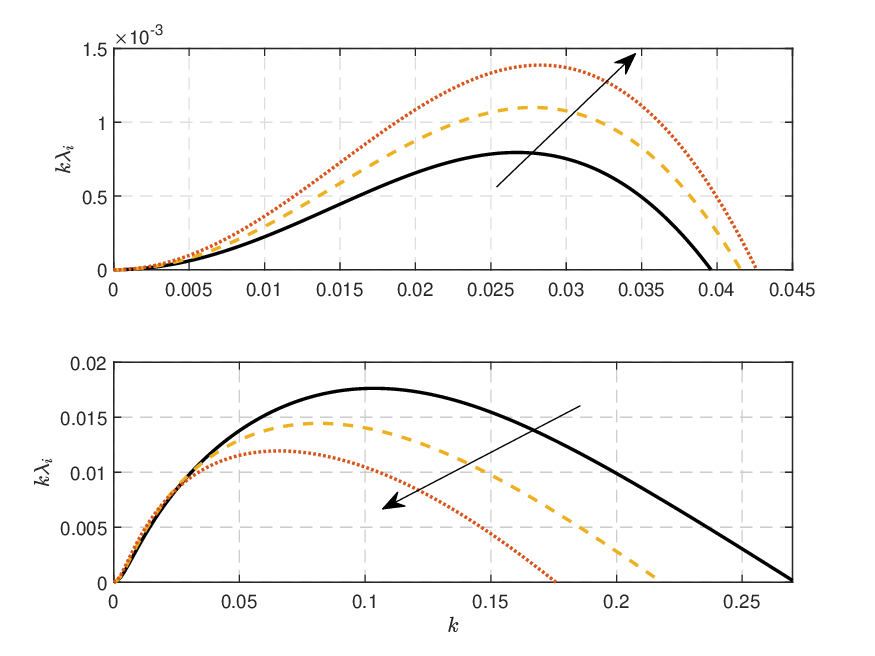}}	
		\end{center} 
		\caption{(a) The variation of the marginal stability curve in the ($\chi-k$) plane for surface mode. The arrows are pointed from lower to higher values of external shear force $\tau$. The ``$S"$ and ``$U"$ indicate the stable and unstable zones. (b) The temporal growth rate variations against wavenumber $k$ for a surface mode where the solid line is for $\tau = 0.0$, the dashed line for $\tau = 0.5$, and the dotted line for $\tau = 1.0$. The values of the modified Reynolds number $\chi$ for the temporal growth rates are $\chi = 10$(top) and $\chi = 200$(bottom). Here again, the arrow indicates the lower to the higher value of external shear force $\tau$. The other parameters are $\xi = 90^{\circ}$, $Ka = 5000$, $\Sigma = 2.0$, $Sc_a = 100$, $Sc_b = 10$, $M_{tot} =0.1$, $\beta_a = 0.01$, $R_a = 1.0$, $k_a = 1.0$, $\delta= 0.04$.} \label{fig_10}
	\end{figure}
	\indent On the other hand, the marginal stability curve in case of surface mode for different external shear force $\tau$ in figure \ref{fig_10_a} conveys that external shear $\tau$ attenuates the critical modified Reynolds number ($\chi_c$) and hence destabilizes the primary instability, but for higher modified Reynolds numbers, it stabilizes the flow by curtailing the unstable region which is consistent with the results of \citet{samanta2014shear}. The reason can be attributed to the fact that the inertial force dominates the shear-induced normal stress near the inception of instability, but when the modified Reynolds number reaches a threshold point, the normal stress gains prominence over the inertial force and stabilizes the flow. The temporal growth rate against the wavenumber $k$ associated with the above discussion is shown in figure \ref{fig_10_b}, where the modified Reynolds number for the top figure is chosen as $\chi = 10$ and that for the bottom is $\chi = 200$. The temporal growth rates in the case of surface mode clearly indicate the unstable nature of the external shear $\tau$ in the longwave regime, and the bottom figure shows that the externally imposed shear $\tau$ has a stabilizing effect on the flow away from the onset of instability.

	\begin{figure}
		\begin{center}
			\subfigure[]{\label{fig_11_a}\includegraphics*[width=5.4cm]{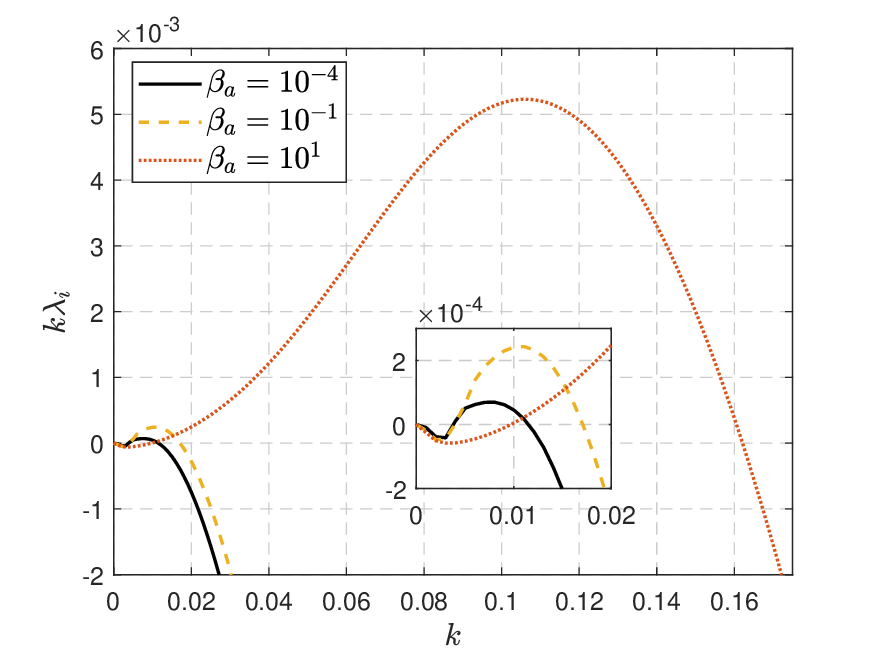}}
			\subfigure[]{\label{fig_11_b}\includegraphics*[width=5.4cm]{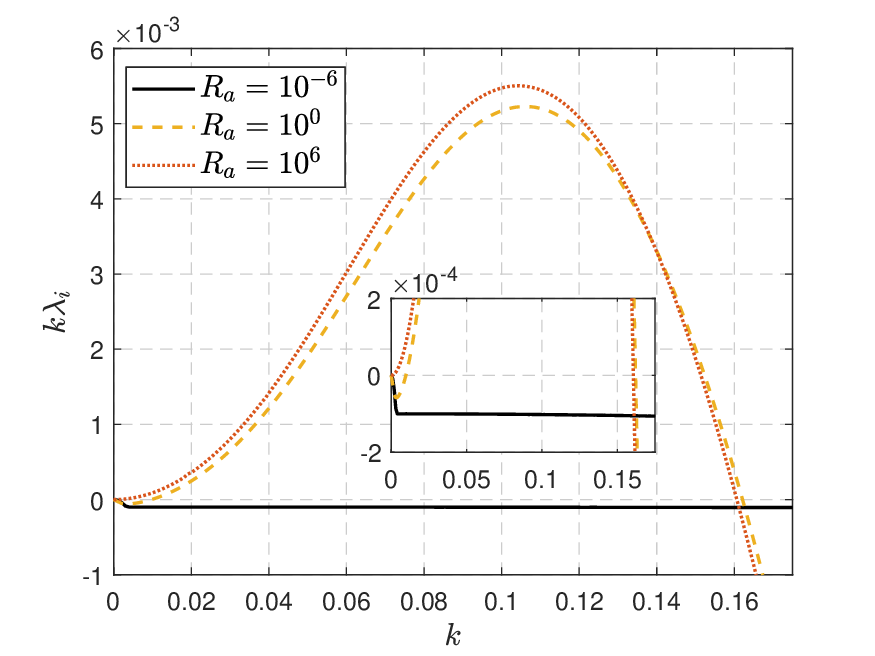}}
			\subfigure[]{\label{fig_11_c}\includegraphics*[width=5.4cm]{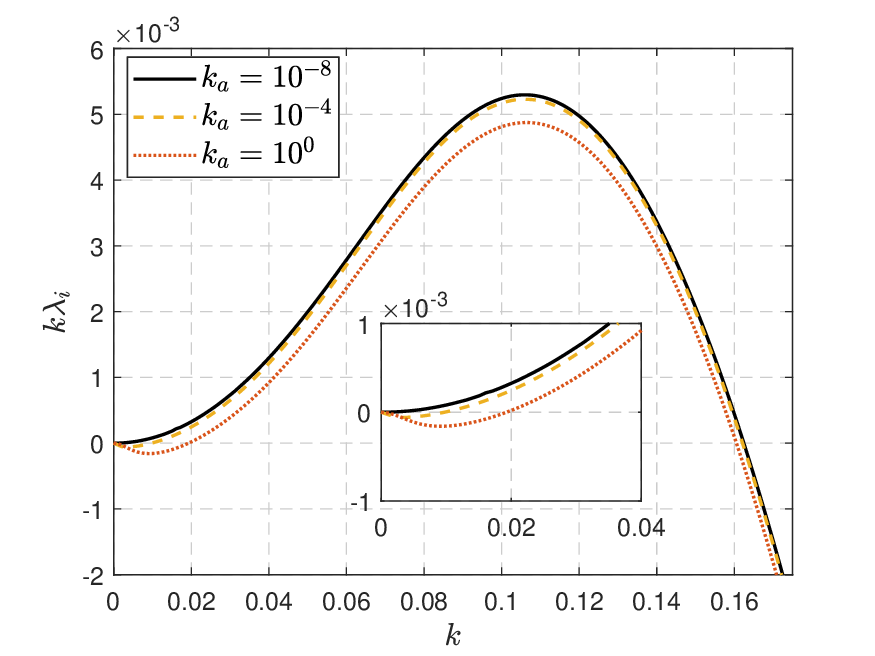}}
		\end{center}
		\caption{ The variation of the temporal growth rates against wavenumber $k$ when (a) surfactant parameter $\beta_a$, (b) solubility parameter $R_a$, (c) sorption kinetics rate $k_a$ varying for surfactant mode. During computing results for figures (a) $k_a = 10^{-4}$, $R_a = 1.0$, (b) $k_a = 10^{-4}$, $\beta_a = 10.0$, and (c) $\beta_a = 10.0$, $R_a = 1.0$ are kept fixed. The other parameters are $\xi = 90^{\circ}$, $\chi = 100$, $Ka = 5000$, $\Sigma = 2.0$, $Sc_a = 100$, $Sc_b = 10$, $M_{tot} = 0.1$, $\tau = 0.5$, $\delta = 0.04$. } \label{fig_11}
	\end{figure}
	
\begin{figure}
		\begin{center}
			\subfigure[]{\label{fig_12_a}\includegraphics*[width=8cm]{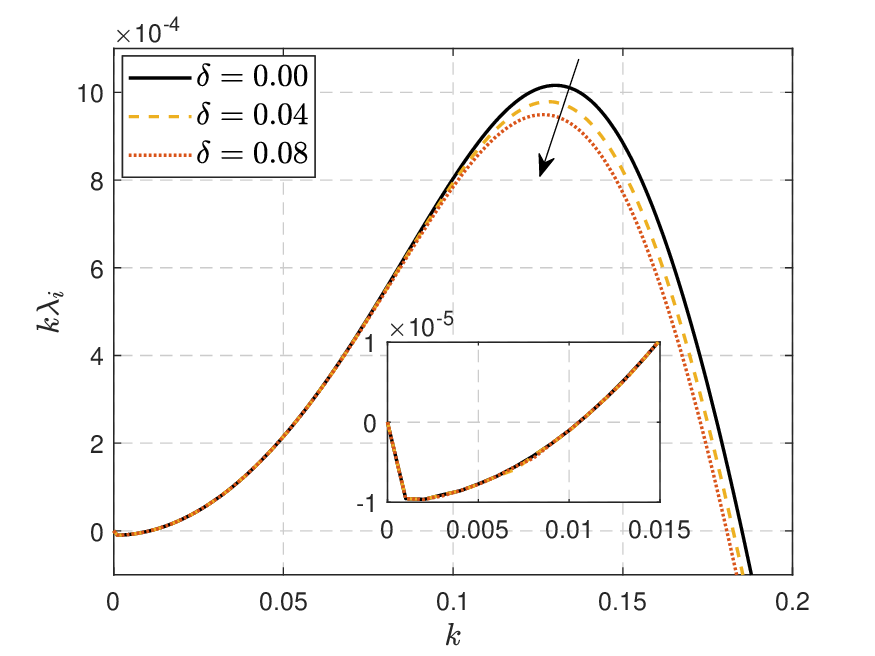}}
			\subfigure[]{\label{fig_12_b}\includegraphics*[width=8cm]{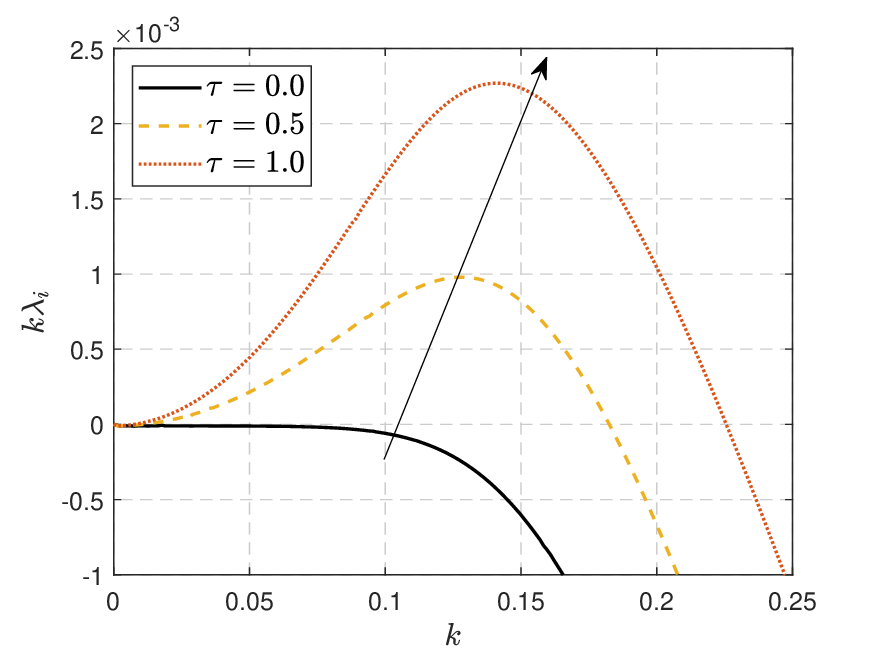}}
		\end{center}
		\caption{ The variation of temporal growth rates for surface mode against wavenumber $k$ when(a) $\chi = 100$ in the case of different values of slip length $\delta$, and (b) when $\chi = 1000$ in the case of different values of external imposed shear $\tau$. The other parameters are $\xi = 90^{\circ}$, $Ka = 5000$, $\Sigma = 2.0$, $Sc_a = 100$, $Sc_b = 10$, $M_{tot} = 0.1$, $\beta_a = 0.01$, $R_a = 0.01$, $k_a = 0.00001$, $\tau = 0.5$. } \label{fig_12}
	\end{figure}

	Moreover, the behaviour of the surfactant mode is explored in the figures (\ref{fig_11}) and (\ref{fig_12}). The surfactant parameter $\beta_a$ is varied and displayed in \ref{fig_11_a} in the case of surfactant mode. The flow remains stabilized for wavenumber $k \rightarrow 0$, and then for $k \ge 0.004$, the flow becomes unstable for a range of wavenumbers and then stabilizes again except for $\beta_a = 10.0$, which shows a strictly monotonic increasing nature (for $k \ge 0.01$) of the growth rate of instability for surfactant parameter $\beta_a$. The larger values of the surfactant parameter make the flow unstable, which happens to be the insoluble surfactant limit (given $R_a$ is a not-so-small fixed quantity as $\xi_a = \beta_a R_a \rightarrow \infty$ is the insoluble surfactant limit), as physical interpretation implies that higher values of $\beta_a$ mean that the ability to adsorb monomers at the free surface is high. Further, the solubility parameter $R_a$ indicates that for small values (precisely $R_a = 10^{-6}$) of it, the flow remains stable throughout but becomes unstable for moderate to high values (see figure \ref{fig_11_b}). The high values of $R_a$ indicate the surfactant present in the configuration is an insoluble surfactant, and flow behaviour with insoluble surfactant is well documented \citep{choudhury2024thermocapillary, samanta2021effect}. Figure \ref{fig_11_b} shows that moderate values such as $R_a = 10^0 = 1$ are good enough to observe instability in the surfactant mode for soluble surfactant.

	Furthermore, the variation of the sorption kinetics rate $k_a$ portrays that for moderate to large values of $k_a$ in the longwave regime, the flow remains stable, and only for very low values ($k_a = 10^{-8}$), the flow is unstable in the longwave zone. The flow in the surfactant mode eventually (for $k \ge 0.02$) gets unstable for all the considered values of the sorption rate $k_a$ for a sufficiently good range of unstable wavenumbers in the surfactant mode, suggesting a stabilizing effect as increment of $k_a$ implies a decrease in the growth rate of instability. For large values of $k_a$, the transport is governed by diffusion, and the exchange kinetics between the bulk and the interface remains in equilibrium \citep{kalogirou2019role}. All these help to conclude that the soluble surfactant parameters largely stabilize the flow in the surfactant mode, and the analytical computation certainly strengthens the claim with the presence of these parameters. This can be explained from the analytical calculation for the longwave regime as done by \citet{wei2005effect}. The first-order tangential stress condition \eqref{eqn82} can be recast into
	\begin{eqnarray} 
		h_1 = \frac{ 6 \mathrm{i} (2 \delta +1) Ma A_0 (1 +\mu)}{\Gamma_{a0} [ (8 + 3 \tau + 12 \delta) + 6 (2 \delta + 1)\mu ]} \label{eqn127}
	\end{eqnarray}
	which is equivalent to
	\begin{eqnarray} 
		& k \breve{h}_1 = \displaystyle \frac{ 6 (2 \delta +1) Ma \breve{A}_{0_x} (1 +\mu)}{\Gamma_{a0} [ (8 + 3 \tau + 12 \delta) + 6 (2 \delta + 1)\mu ]} \label{eqn128}
	\end{eqnarray}
	for opting the reference frame which moves with the speed of base velocity $U$ and transformation $\langle\breve{A}_0, \breve{h}_1 \rangle \longmapsto \langle A_0, h_1 \rangle exp\{ \mathrm{i} k (x - \lambdaup t) \}$. The equation \eqref{eqn128} indicates that first-order surface deformation is influenced by soluble surfactant parameters as well as imposed shear and slip parameters. Also, the zero$^{th}$ order concentration of surfactant observes a change due to the Marangoni stress as an effect of tangential stress balance. Now, if the solubility parameter $R_a$ or the surfactant parameter $\beta_a$ is very small then the equation \eqref{eqn128} converts into a much simpler form where, for fixed external shear force $\tau$, and slip length $\delta$, the slope of the zero$^{th}$-order surfactant concentration purely depends on the Marangoni number (inversely proportional) and the free surface surfactant concentration $\Gamma_{a0}$ (directly proportional).
	
	\begin{eqnarray} 
		\breve{A}_{0_x} = \frac{k \breve{h}_1 \Gamma_{a0} (8 + 3 \tau + 12 \delta)}{6 (2 \delta + 1) Ma} \label{eqn129}
	\end{eqnarray}
	At the same time, the definition of $\Gamma_{a0}$ reveals that $\Gamma_{a0}\rightarrow 0$ as $R_a \rightarrow 0$ and hence there won't be any change in $A_0$ (from \eqref{eqn129}) which ensures stability as there is no surfactant concentration change. Now, if surfactant parameter $\beta_a$ becomes too small, then the equation \eqref{eqn30} implies $\Gamma_0 \approx M_{tot}$ and the impact of surfactant at the free surface will tend to vanish and help to stabilize the flow in the surfactant mode.
	
	Additionally, the impact of the slip parameter $\delta$ on the surfactant mode is explored by demonstrating the temporal growth rate in figure \ref{fig_12_a}. It reveals that the flow remains stable for $k \rightarrow 0$ i.e., in the longwave regime, and subsequently transition to instability takes place and during this phase, growth rate curves displays minimal sensitivity for different values of $\delta$ upto $k \le 0.1$, it is only for higher values of k (more specifically $k \ge 0.1$), the variation of the curves are more prominent and where the growth rate reduces for increasing values of slip parameter $\delta$ highlighting a stabilizing effect of slip parameter in case of surfactant mode. However, the temporal growth rate curves for surfactant mode, when varied for imposed shear $\tau$, divulges (see figure \ref{fig_12_b}) the opposite phenomenon with growth rate increases for increasing imposed shear and also, it displays that for zero external imposed shear $\tau$, the system always remains stable in the surfactant mode. The reason can be attributed to the fact that the surfactant concentration in the presence of imposed shear moves faster \cite{bhat2019linear} and gets enhanced if the intensity of the imposed shear is greater.
	
	\begin{figure} 
		\begin{center}
			\subfigure[]{\label{fig_13_a}\includegraphics*[width=8cm]{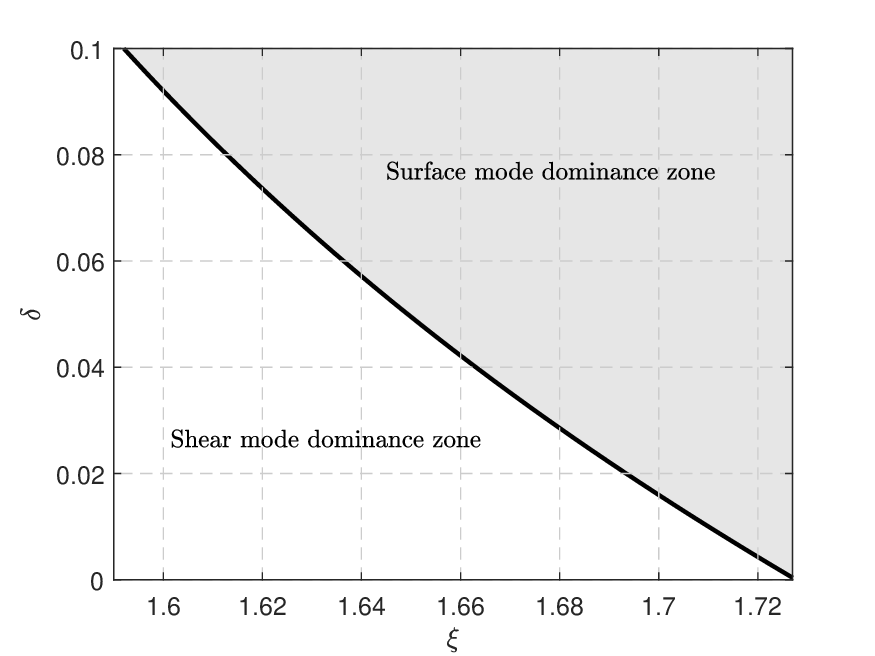}}
			\subfigure[]{\label{fig_13_b}\includegraphics*[width=8cm]{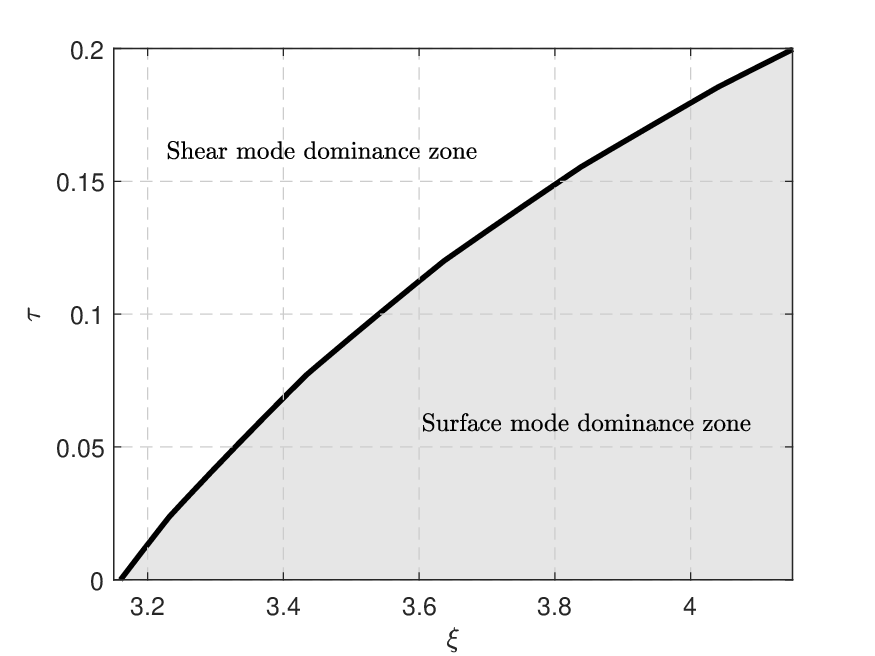}}
		\end{center} 
		\caption{ The phase boundary to segregate the surface mode and shear mode in the (a) $(\xi, \delta)$ , (b) $(\xi, \tau)$ plane. The other parameters chosen are $\chi = 10^4$, $Ka = 5000$, $\Sigma = 2.0$, $Sc_a = 100$, $Sc_b = 10.0$, $M_{tot} = 0.1$, $\beta_a = 0.01$, $R_a = 1.0$, $k_a = 1.0$, $\tau = 0.5$(when applicable), $\delta = 0.04$(when applicable). The wavenumber chosen for (a) is $k = 0.1$, and for (b) is $k = 0.2$.} \label{fig_13}
	\end{figure}
	
	\begin{figure}
		\begin{center}
			\subfigure[]{\label{fig_14_a}\includegraphics*[width=8cm]{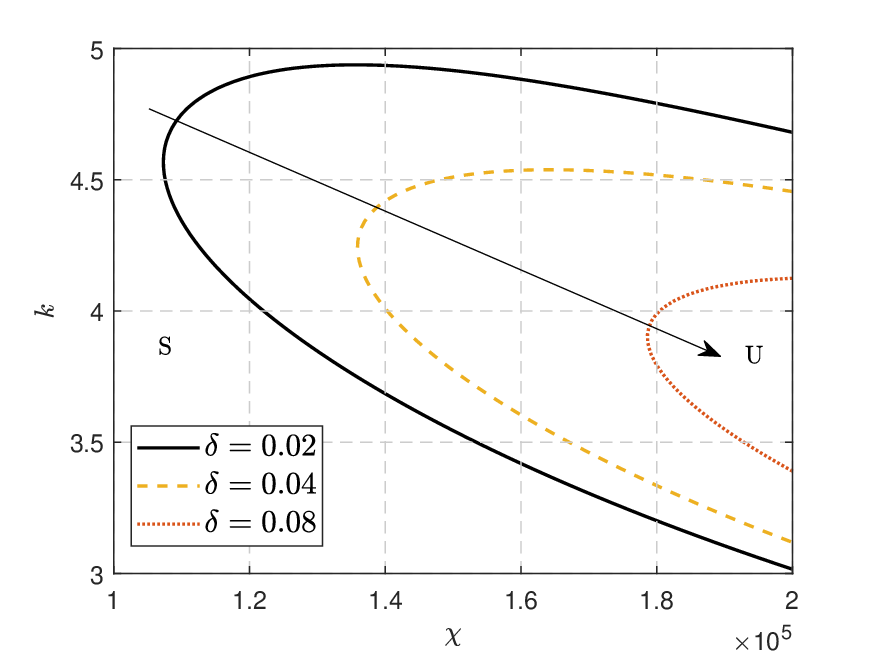}}
			\subfigure[]{\label{fig_14_b}\includegraphics*[width=8cm]{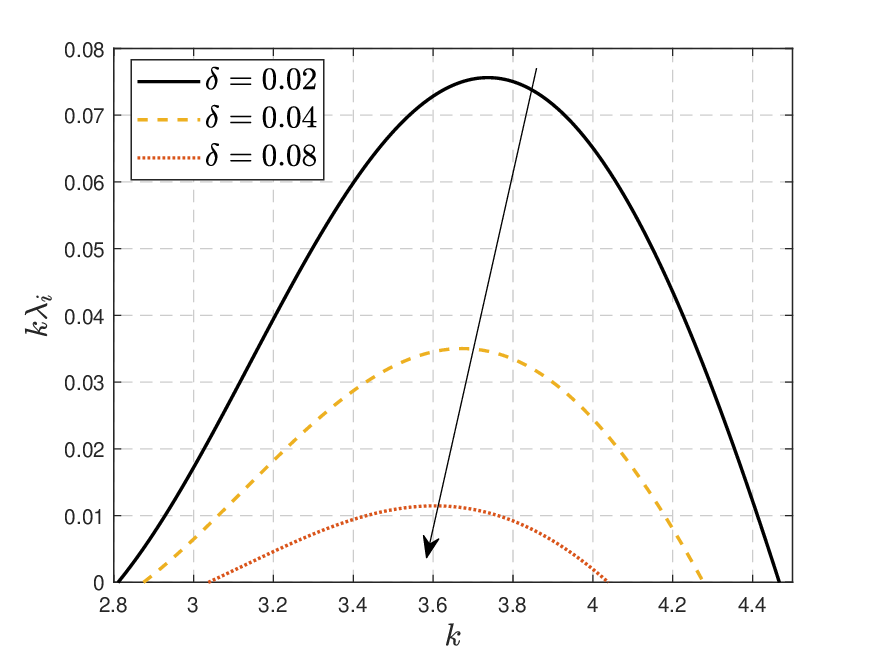}}
		\end{center}
		\caption{ (a) The variation of the marginal stability curve in the ($\chi-k$) plane for shear mode when the slip parameter $\delta$ varies. The arrow is pointed from lower to higher values of the slip parameter $\delta$. The ``S'' and ``U'' indicate the stable and unstable regions. (b) The temporal growth rates against wavenumber $k$ when the modified Reynolds number $\chi = 2 \times 10^5$ in the case of different values of $\delta$. Here, again, the arrow indicates lower to higher values of the slip parameter. The other parameters are $\xi = 1^{\circ}$, $Ka = 50000$, $\Sigma = 2.0$, $Sc_a = 100$, $Sc_b = 10$, $M_{tot} = 0.1$, $\beta_a = 0.01$, $R_a = 1.0$, $k_a = 1.0$, $Ma = 0.01$, $\tau = 0.5$. } \label{fig_14}
	\end{figure}
	
	With the aim to recognize the significance of external shear $\tau$ and slip length $\delta$ on the shear mode, which can be observed for high modified Reynolds number at small inclination angle \citep{chin1986gravity}, the phase boundary that distinguishes the surfactant mode and the shear mode, is plotted numerically in figure (\ref{fig_13}). The plane for the phase boundary is chosen as $(\xi, \tau)$ and $ ( \xi, \delta ) $ for the range of the imposed shear $\tau$ and slip parameter $\delta$ respectively, as well as the angle of inclination. The modified Reynolds number $\chi = 10^{4}$, the wavenumber $ k = 0.1$ for $\delta$ and $k = 0.2$ for $\tau$ are chosen during these plots. Both of the plots permit the choice of $\xi = 1^{\circ}$ for the range of imposed shear $\tau$ and the slip length $\delta$ preferred throughout the article .\\
	\indent Using these parametric values, the marginal stability curves are displayed in the figures (\ref{fig_14_a}) for varying slip parameter $\delta$. It is evident that the shear mode arises in the finite wavenumber regime and that the unstable region for the shear mode undergoes a contraction, and the critical modified Reynolds number surges on the scale of $\order{10^5}$ which highlights the stabilizing nature of the slip parameter $\delta$ in case of the shear mode. This very fact is strengthened by the corresponding temporal growth rate curves in figure \ref{fig_14_b} against wavenumber $k$, whose peaks drop drastically to provide the information that the slip parameter plays a major role in stabilizing the flow in the shear mode regime. Here, the modified Reynolds number $\chi = 2 \times 10^{5}$ is selected by noticing the marginal stability curve in the figure \ref{fig_14_a}.
	
	\begin{figure}
		\begin{center}
			\subfigure[]{\label{fig_15_a}\includegraphics*[width=8cm]{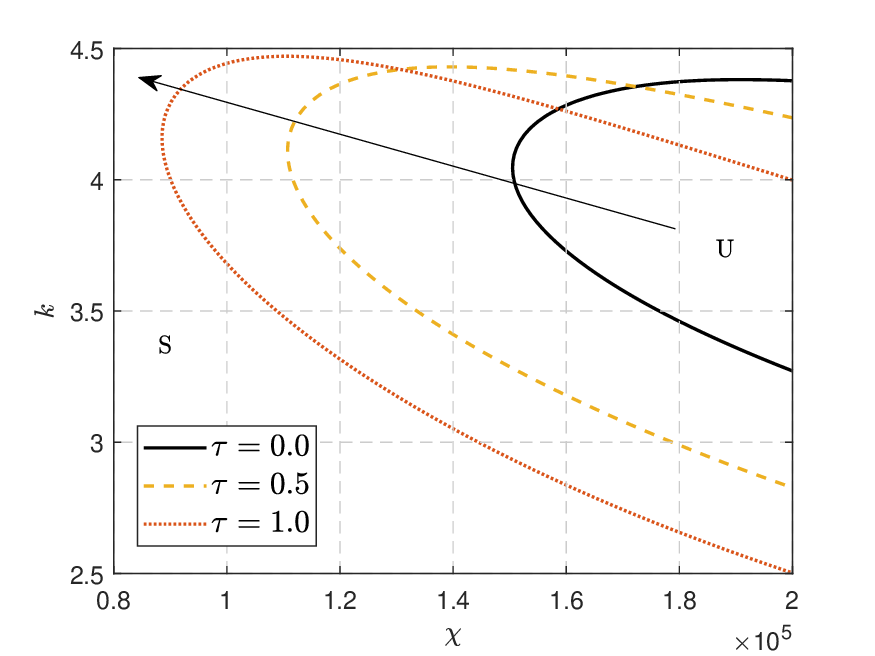}}
			\subfigure[]{\label{fig_15_b}\includegraphics*[width=8cm]{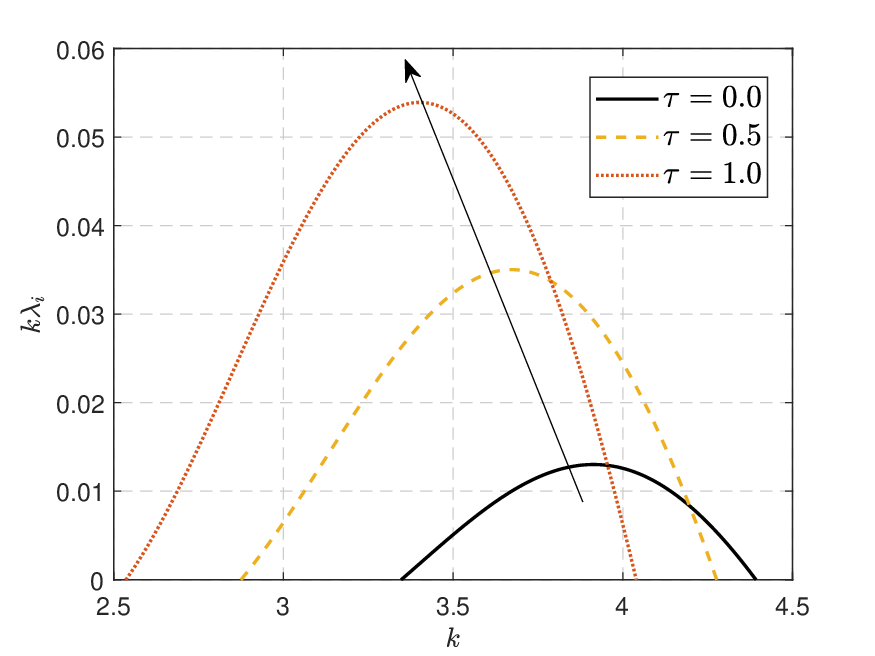}}
		\end{center}
		\caption{ (a) The variation of the marginal stability curve in the ($\chi-k$) plane for shear mode when external shear force $\tau$ varies. The arrow is pointed from lower to higher values of external shear force $\tau$. The ``S'' and ``U'' indicate the stable and unstable regions. (b) The temporal growth rates against wavenumber $k$ when the modified Reynolds number $\chi = 2 \times 10^5$ in the case of different values of $\tau$. Here, again, the arrow indicates lower to higher values of external shear force. The other parameters are $\xi = 1^{\circ}$, $Ka = 50000$, $\Sigma = 2.0$, $Sc_a = 100$, $Sc_b = 10$, $M_{tot} = 0.1$, $\beta_a = 0.01$, $R_a = 1.0$, $k_a = 1.0$, $Ma = 0.01$, $\delta = 0.04$. } \label{fig_15}
	\end{figure}
	
	On the other hand, the marginal stability curves (see figure \ref{fig_15_a}) in the $\chi-k$ plane for the imposed shear $\tau$ variation in the shear mode regime amplify as the unstable region gets enhanced and the corresponding critical modified Reynolds number increases considerably on the scale of $\order{10^5}$. That means more imposed shear helps the onset of instability for less modified Reynolds numbers in the shear mode regime. The corresponding temporal growth rate curves against wavenumber $k$ help to fortify the earlier claim. Along with that, the maximum value of growth rate increases with external shear stress $\tau$ (figure \ref{fig_15_b}). It is noteworthy to mention that during plotting the growth rate for varied imposed shear in the shear mode regime, the modified Reynolds number is chosen as $\chi = 2 \times 10^{5}$ by observing the marginal stability curve. This clearly dictates a destabilizing effect of the imposed shear $\tau$ in the shear mode regime. 
	
	\begin{figure} 
		\begin{center}
			\subfigure[]{\label{fig_16_a}\includegraphics*[width=5.4cm]{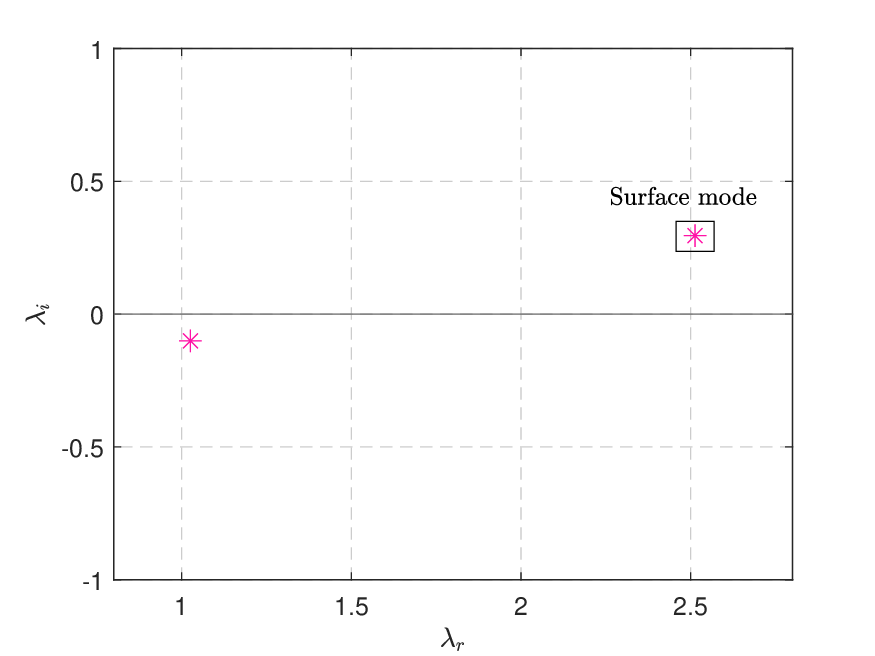}}
			\subfigure[]{\label{fig_16_b}\includegraphics*[width=5.4cm]{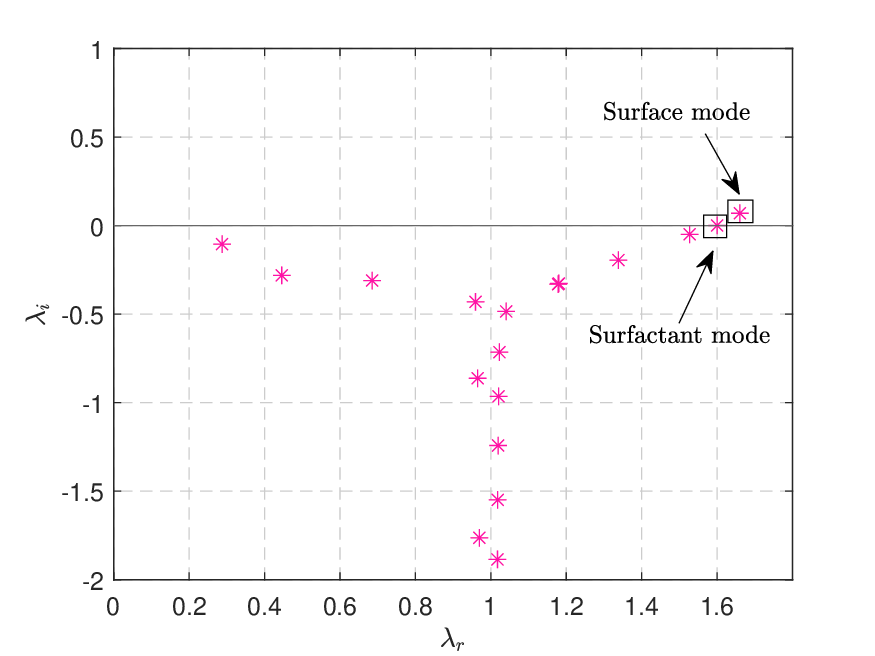}}
			\subfigure[]{\label{fig_16_c}\includegraphics*[width=5.4cm]{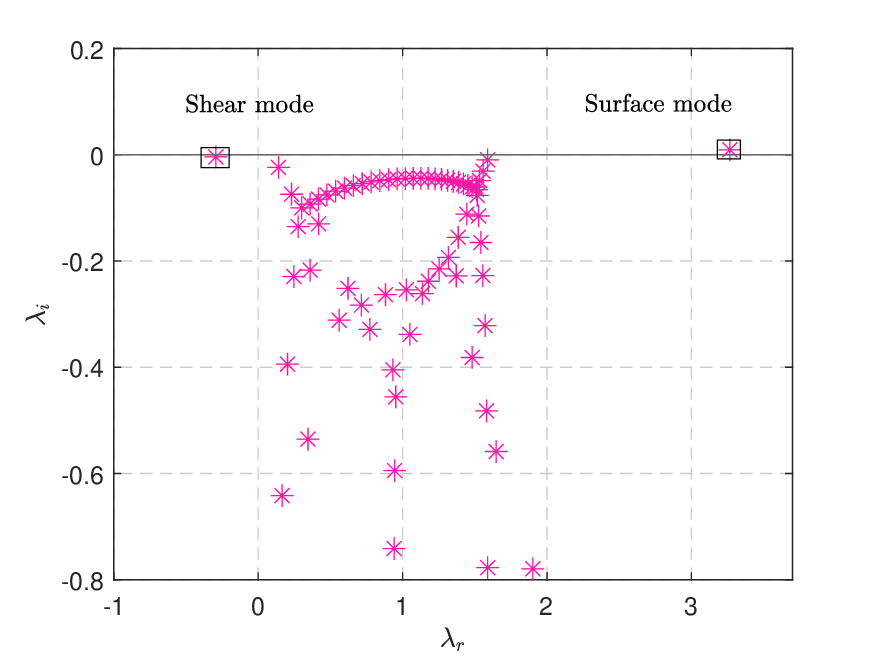}}
		\end{center}
		\caption{The eigenspectrum of the Orr-Sommerfeld eigenvalue problem \eqref{eqn114} when (a) $\chi = 50$, $k = 0.02$, $R_a = 1.0$, $k_a = 1.0$, $Ka = 5000$, $\xi = 90^{\circ}$ (b) $\chi = 1000$, $k = 0.125$, $\xi = 90^{\circ}$, $R_a = 0.01$, $k_a = 0.00001$, $Ka = 5000$, (c) $\xi = 1^{\circ}$, $\chi = 2 \times 10^5$, $k = 3.5$, $R_a = 1.0$, $k_a = 1.0$, $Ka = 50000$. The other relevant parameters are $\Sigma = 2.0$, $Sc_a = 100$, $Sc_b = 10$, $M_{tot} = 0.1$, $\beta_a = 0.01$, and $Ma = 0.01$.} \label{fig_16}
	\end{figure}
	The eigenspectrum of the eigenvalue problem is demonstrated in the figure (\ref{fig_16}) for different unstable regimes on the $(\lambdaup_r, \lambdaup_i)$ plane. It can be seen that in figure \ref{fig_16_a}, the surface mode is clearly the most unstable in the regime. In the following figure \ref{fig_16_b}, the surfactant mode is shown along with the surface mode, and they are distinguishable because of their different phase speed $\lambdaup_r$, with the phase speed of the surfactant mode being less than that of the surface mode \citep{hossain2024odd}. It can be seen that the surfactant mode is damped, as discussed in the earlier section, and the reason can be attributed to the fact that the soluble surfactant parameters behave as stabilizing agents in the surface mode. The shear mode is shown in the last figure \ref{fig_16_c} with the surface mode. Here, they are discernible because they appear in different branches of the eigenspectrum, and along with that, the shear mode moves with a lower phase speed than that of the surface mode.
	
	\begin{figure}
		\begin{center}
			\subfigure[]{\label{fig_17_a}\includegraphics*[width=8cm]{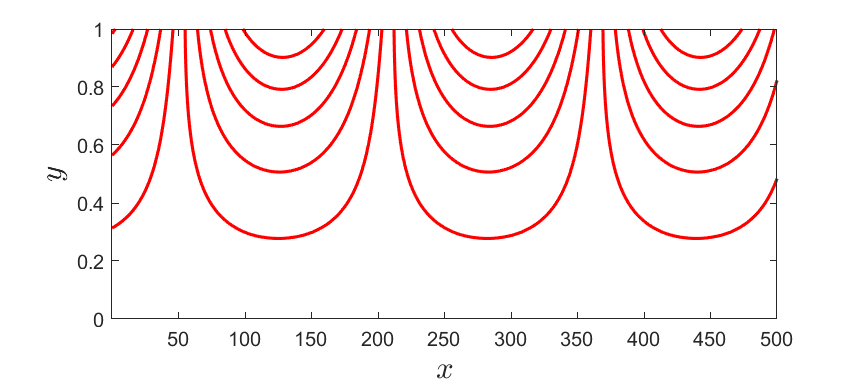}}
			\subfigure[]{\label{fig_17_b}\includegraphics*[width=8cm]{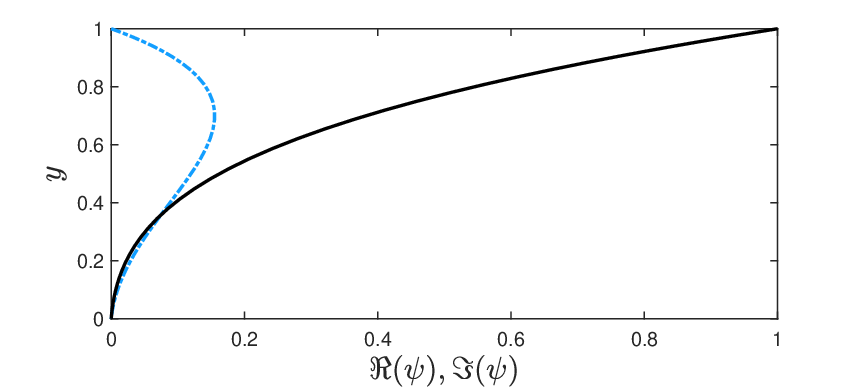}}
			\subfigure[]{\label{fig_17_c}\includegraphics*[width=8cm]{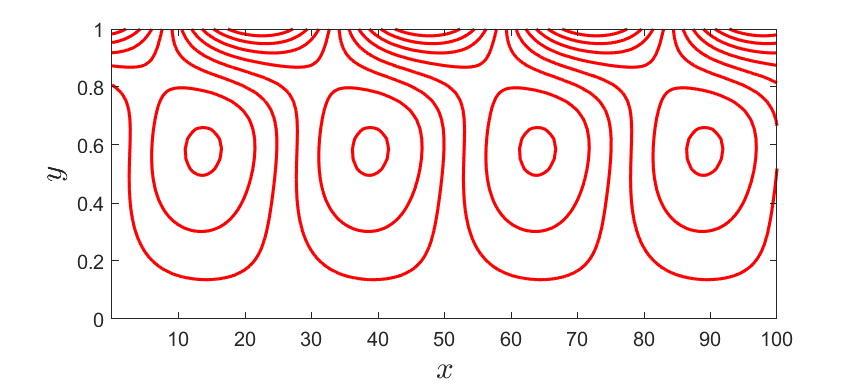}}
			\subfigure[]{\label{fig_17_d}\includegraphics*[width=8cm]{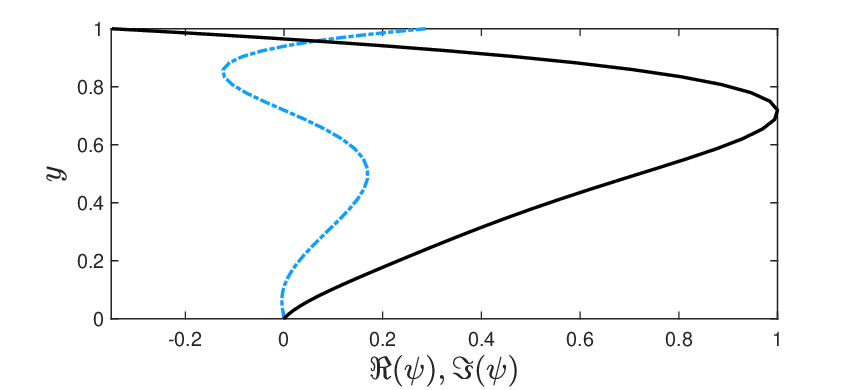}}
			\subfigure[]{\label{fig_17_e}\includegraphics*[width=8cm]{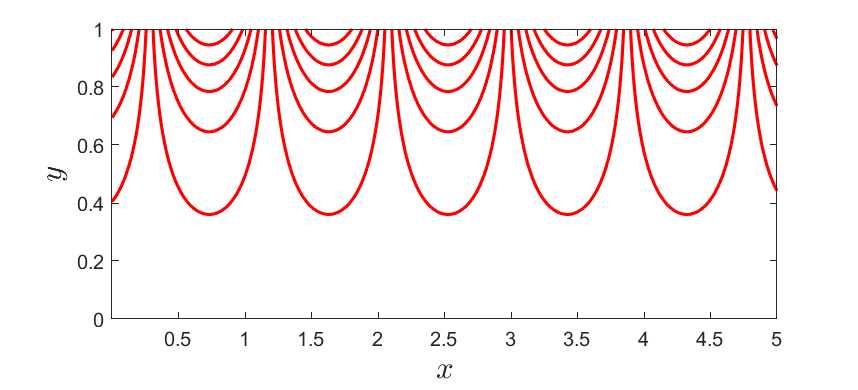}}
			\subfigure[]{\label{fig_17_f}\includegraphics*[width=8cm]{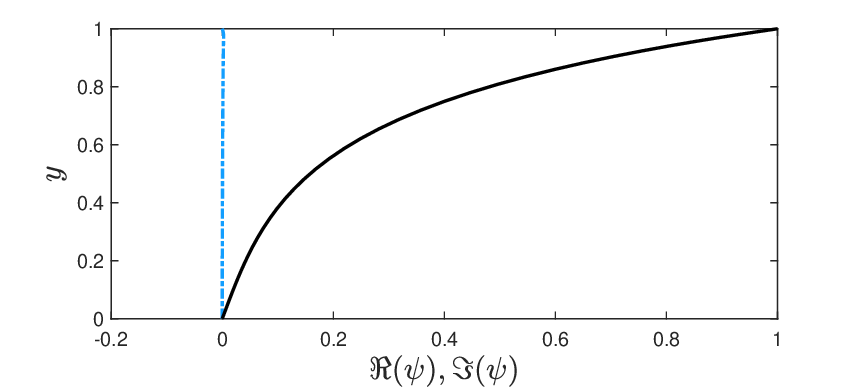}}
		\end{center}
		\caption{ (a) The contour plot of the streamfunction in the $(x,y)$ plane for surface mode with $\chi = 50$, $k = 0.02$, $\xi = 90^{\circ}, Ka = 5000$, $R_a = 1.0$, $k_a = 1.0$, (b) corresponding normalized eigenfunction against $y$ direction for surface mode, (c) contour plot of the streamfunction in the $(x, y)$ plane for surfactant mode with $\chi = 1000$, $k = 0.125$, $\xi = 90^{\circ}, Ka = 5000$, $R_a = 0.01$, $k_a = 0.00001$ (d) corresponding normalized eigenfunction against $y$ direction for surfactant mode, (e) contour plot of the streamfunction in the $(x, y)$ plane for shear mode with $\chi = 2 \times 10^5$, $k = 3.5$, $\xi = 1^{\circ}, Ka = 50000$, $R_a = 1.0$, $k_a = 1.0$ (f) corresponding normalized eigenfunction against $y$ direction for shear mode. The black curve in the eigenfunction plot denotes the real part of the function and the sky blue curve represents the imaginary part of the same. The other relevant parameters are  $\Sigma = 2.0$, $Sc_a = 100$, $Sc_b = 10$, $M_{tot} = 0.1$, $\beta_a = 0.01$, $\tau = 0.5$, and $\delta = 0.04$. } \label{fig_17}
	\end{figure}
	
	The contour plots of the streamfunctions in the $(x, y)$ plane and the real and imaginary parts of the corresponding eigenfunctions are shown in figure (\ref{fig_17}) to decipher the flow features in the different mode regimes. In the surface mode, the wavenumber is chosen as $k = 0.02$ and modified Reynolds number $\chi = 50$ and it is found that open vortices formed near the free surface (see figure \ref{fig_17_a}) which is further confirmed by the eigenfunctions, where a significant change between the real and imaginary parts of the eigenvectors can be observed near the free surface (\ref{fig_17_b}). However, the closed vortices can be seen near the middle of the bulk flow (\ref{fig_17_c}) in the surfactant mode regime with $\chi = 1000$, $k = 0.125$ chosen for the surfactant mode zone. The normalized eigenvectors also authenticate this behaviour by showing the substantial change between real and imaginary parts of the eigenvector in the center part of the flow (\ref{fig_17_d}). However, the contour plots in the case of shear mode divulge that more open vortices formed near the free surface (see figure \ref{fig_17_e}) compared to the surface mode, and the analogous normalized eigenvectors display this behaviour by showing a considerable difference between the real and imaginary parts of the eigenvectors(\ref{fig_17_f}) at the free surface of the flow.
	
	\begin{figure}
		\begin{center}
			\subfigure[]{\label{fig_18_a}\includegraphics*[width=8cm]{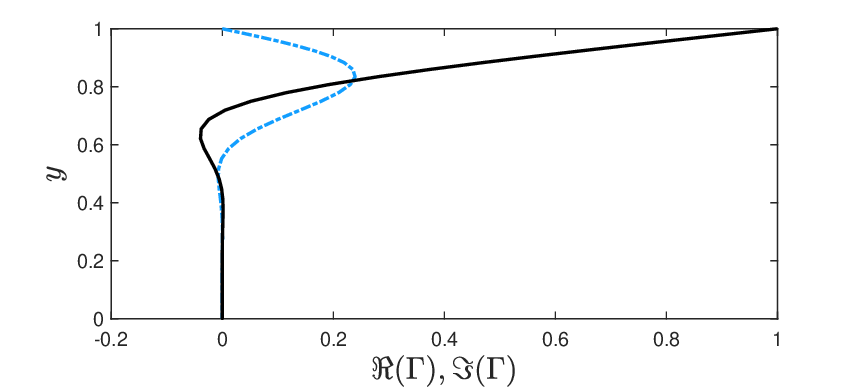}}
			\subfigure[]{\label{fig_18_b}\includegraphics*[width=8cm]{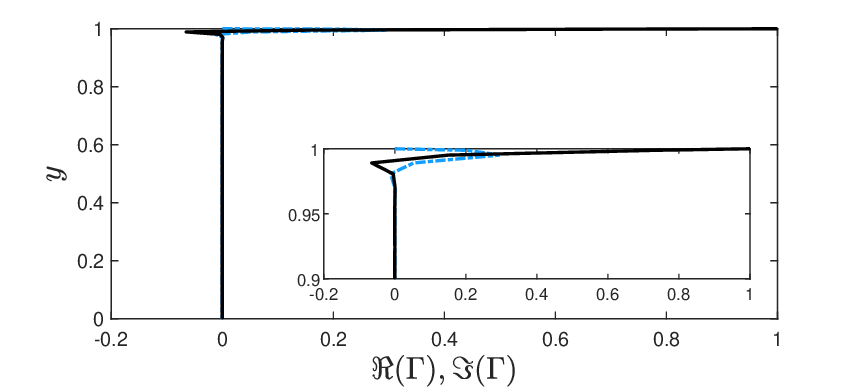}}
		\end{center}
		\caption{ (a) The normalized eigenfunction of perturbed bulk surfactant concentration against $y$ direction for surfactant mode with $\chi = 1000$, $k = 0.125$, $\xi = 90^{\circ}, Ka = 5000$, $R_a = 0.01$, $k_a = 0.00001$, and (b)  the normalized eigenfunction of perturbed bulk surfactant concentration against $y$ direction for shear mode with $\chi = 2 \times 10^5, ~\text{and}~ k = 3.5$, $Ka = 50000$, $R_a = 1.0$, $k_a = 1.0$, $Ma = 0.01$. The black curve in the eigenfunction plot denotes the real part of the function and the sky blue curve represents the imaginary part of the same. The other relevant parameters are $\Sigma = 2.0$, $Sc_a = 100$, $Sc_b = 10$, $M_{tot} = 0.1$, $\beta_a = 0.01$, $\tau = 0.5$, and $\delta = 0.04$. } \label{fig_18}
	\end{figure}
	
	Further, the eigenfunction corresponding to the perturbed bulk surfactant concentration is plotted for different modes in the figures \ref{fig_18}. The figure \ref{fig_18_a} shows the real and imaginary parts of the aforementioned eigenfunction, and it is evident that up to the center there is no deviation in between them, and the difference intensifies near the free surface, and the real part becomes negative for a certain range of y before turning positive. However, the eigenfunction of the perturbed bulk surfactant concentration for shear mode displays that the real and imaginary parts are overlapping and near the free surface, those parts changes rapidly, satisfying the boundary condition \eqref{eqn40} (see figure \ref{fig_18_b}).

	\section{Energy budget analysis} \label{eba}
	\indent The hydrodynamic instability of a flow onsets when the kinetic energy of the whole system grows with time, and this energy can be supplied from multiple sources, primarily from the infinitesimal disturbances given to the base flow. However, the exact mechanism for this energy growth is not clear, and the factors contributing to this energy transfer are ambiguous based on the above analysis. To deal with this, the energy budget analysis is one of the most favoured tools, used by many studies \citep{kelly1989mechanism, hossain2023instability, samanta2023elliptic, kalogirou2021instabilities} in the literature. This method helps to identify the dominant source of energy that moves the flow towards stabilization/destabilization. On that note, the procedure adopted by \citet{kelly1989mechanism} served as a reference during the investigation. The modified Reynolds-Orr energy balance equation takes the following form, \citep{kelly1989mechanism, samanta2023elliptic}
	\begin{eqnarray} 
		& \nonumber \displaystyle \frac{Re}{2 \zetaup} \int_{0}^{1} \int_{0}^{\zetaup} \frac{\partial}{\partial t} ( u^2 + v^2) dx dy = - \frac{Re}{\zetaup} \int_{0}^{1} \int_{0}^{\zetaup} U^{'}(y) u v dx dy~ + \frac{2 We \sigma_0}{\zetaup} \int_{0}^{\zetaup} v\bigg|_{y = 1} h_{xx} dx \\
		& \nonumber \displaystyle - \frac{2 \cot(\xi)}{\zetaup} \int_{0}^{\zetaup} v \bigg|_{y = 1} h dx~ - \frac{2}{\zetaup} \int_{0}^{\zetaup} v v_y dx~ + \frac{2 \tau}{\zetaup} \int_{0}^{\zetaup} v\bigg|_{y = 1} h_x dx~ + \frac{1}{\zetaup} \int_{0}^{\zetaup} u(u_y + v_x) \bigg|_{y = 1} dx\\
		& - \displaystyle \frac{\delta}{\zetaup} \int_{0}^{\zetaup} u_y (u_y + v_x)\bigg|_{y = 0} dx - \frac{1}{\zetaup} \int_{0}^{1} \int_{0}^{\zetaup} \biggl\{ 2 u^2_x + 2 v^2_y + (u_y + v_x)^2 \biggr\} dx dy \label{eqn130}
	\end{eqnarray}
	\noindent where $\zetaup = 2 \piup / k$ is the wavelength of the infinitesimal disturbance. Similarly, the perturbed bulk surfactant concentration equation, with the help of appropriate boundary conditions listed above (equations \eqref{eqn38}, \eqref{eqn39}, $\&$ \eqref{eqn43}), can be transformed into the following form \citep{kalogirou2021instabilities},
	\begin{eqnarray} 
		& \nonumber \displaystyle \frac{1}{2 \zetaup} \int_{0}^{1} \int_{0}^{\zetaup} \frac{\partial}{\partial t} (\Gamma^2) dx dy = - \frac{1}{\zetaup} \int_{0}^{\zetaup} k_a \beta_a \Gamma\bigg|_{y = 1} \biggl[ R_a \biggl\{ (1 - \Gamma_{a0} ) \Gamma\bigg|_{y = 1} - \Gamma_0 ~ \Gamma_a \biggr\} - \Gamma_a \biggr] dx \\
		& \displaystyle - \frac{1}{\zetaup Pe_b} \int_{0}^{1} \int_{0}^{\zetaup} (\Gamma^2_{xx} + \Gamma^2_{yy}) dx dy \label{eqn131}
	\end{eqnarray}
	
	\noindent and finally, the free surface surfactant concentration equation is transformed into
	
	\begin{eqnarray} 
		& \nonumber \displaystyle \frac{1}{2 \zetaup} \int_{0}^{\zetaup} \frac{\partial}{\partial t} (\Gamma_a) dx =- \frac{\Gamma_{a0}}{\zetaup} \bigg[ \int_{0}^{\zetaup} u_x \Gamma_a dx + U^{'}(1) \int_{0}^{\zetaup} h_x \Gamma_a dx \bigg] - \frac{1}{\zetaup Pe_{ca}} \int_{0}^{\zetaup} \Gamma^2_{a, x} dx\\
		& \displaystyle + \frac{k_a}{\zetaup} \bigg[ R_a (1-\Gamma_{a0}) \int_{0}^{\zetaup} \Gamma(1) \Gamma_a dx - \frac{1}{1 - \Gamma_{a0}} \int_{0}^{\zetaup} \Gamma^2_a dx \bigg] \label{eqn132}
	\end{eqnarray}
	conforming to the preceding technique. In order to achieve the total kinetic energy of the flow, the equations \eqref{eqn130}, \eqref{eqn131}, and \eqref{eqn132} are summed up, and then the equation for the energy budget is as follows, 
	\begin{eqnarray} 
		& \nonumber Kin + Kinc + Kinfs = Rys + Ste + Hyd + Ishe + She + Per + Diss \\
		& + Flxc + Diffc + Vlgf + Bsh + Difff + Uflx \label{eqn133}
	\end{eqnarray}
	where 
	\begin{eqnarray}
		& \nonumber Kin = \displaystyle \frac{Re}{2 \zetaup} \int_{0}^{1} \int_{0}^{\zetaup} \frac{\partial}{\partial t} ( u^2 + v^2) dx dy;~ Kinc = \frac{1}{2 \zetaup} \int_{0}^{1} \int_{0}^{\zetaup} \frac{\partial}{\partial t} (\Gamma^2) dx dy, 
		\\
		& \nonumber Kinfs = \displaystyle \frac{1}{2 \zetaup} \int_{0}^{\zetaup} \frac{\partial}{\partial t} (\Gamma_a) dx;~	Rys = - \frac{Re}{\zetaup} \int_{0}^{1} \int_{0}^{\zetaup} U^{'}(y) u v dx dy, 
	\end{eqnarray}
	\begin{eqnarray}
		& \displaystyle \nonumber Ste = \frac{2 We \sigma_0}{\zetaup} \int_{0}^{\zetaup} v\bigg|_{y = 1} h_{xx} dx;~	Hyd = - \frac{2 \cot(\xi)}{\zetaup} \int_{0}^{\zetaup} v \bigg|_{y = 1} h dx, \\
		& \displaystyle \nonumber Ishe = \frac{2 \tau}{\zetaup} \int_{0}^{\zetaup} v\bigg|_{y = 1} h_x dx;~ She = \frac{1}{\zetaup} \int_{0}^{\zetaup} u(u_y + v_x) \bigg|_{y = 1} dx, \\
		& \nonumber Per = - \displaystyle \frac{\delta}{\zetaup} \int_{0}^{\zetaup} u_y (u_y + v_x)\bigg|_{y = 0} dx;~	Diss = - \frac{1}{\zetaup} \int_{0}^{1} \int_{0}^{\zetaup} \biggl\{ 2 u^2_x + 2 v^2_y + (u_y + v_x)^2 \biggr\} dx dy, \\
		& \displaystyle \nonumber Flxc = - \frac{1}{\zetaup} \int_{0}^{\zetaup} \beta_a \Gamma\bigg|_{y = 1} k_a \biggl[ R_a \biggl\{ (1 - \Gamma_{a0} ) \Gamma\bigg|_{y = 1} - \bar{\Gamma}~ \Gamma_a \biggr\} - \Gamma_a \biggr] dx;\\
		& \nonumber \displaystyle Diffc = - \frac{1}{\zetaup Pe_b} \int_{0}^{1} \int_{0}^{\zetaup} (\Gamma^2_{xx} + \Gamma^2_{yy}) dx dy, \\
		& \displaystyle \nonumber Vlgf = - \frac{\Gamma_{a0}}{\zetaup} \int_{0}^{\zetaup} u_x \Gamma_a dx; Bsh = - \frac{\Gamma_{a0}}{\zetaup} U^{'}(1) \int_{0}^{\zetaup} h_x \Gamma_a dx, \\
		& \displaystyle \nonumber Difff = - \frac{1}{\zetaup Pe_{ca}} \int_{0}^{\zetaup} \Gamma^2_{a, x} dx;~	Uflx = \frac{k_a}{\zetaup} \bigg[ R_a (1-\Gamma_{a0}) \int_{0}^{\zetaup} \Gamma(1) \Gamma_a dx - \frac{1}{1 - \Gamma_{a0}} \int_{0}^{\zetaup} \Gamma^2_a dx \bigg].
	\end{eqnarray}
	\noindent `$Kin$' represents the growth or decay of the kinetic energy with respect to time based on its positive or negative sign, respectively. `$Kinc$' and `$Kinfs$' symbolize the energy supply due to the presence of surfactant present in the bulk and on the free surface, respectively. `$Rys$' stand for the energy transport to the perturbation due to the Reynolds stress, `$Ste$' depicts the amount of work done for the infinitesimal perturbation against surface tension, `$Hyd$' conveys the rate of work done against hydrostatic pressure, `$Ishe$' expresses the energy contribution due to the externally imposed shear force, `$She$' measures the shear work done in unit time by the infinitesimal disturbances on the frees surface, `$Per$' demonstrates the energy dispersion due to the slippery bottom, `$Diss$' signifies the total amount of viscous energy dissipation of the perturbed flow, `$Flxc$' indicates the energy term due to the flux of monomers between the bulk and free surface, `$Diffc$' advocates the diffusion of energy due to the surfactants present in the bulk, `$Vlgf$' describes the energy transportation due the horizontal velocity gradient at the perturbed free surface, `$Bsh$' accounts for the energy contribution via background shear, `$Difff$' serves as the diffusion of energy due to the presence of free surface surfactant, and `$Uflx$' is a associated with the flux of surfactant between the free surface and the bulk.
	
	\begin{figure}
		\begin{center}
			\subfigure[]{\label{fig_19_a}\includegraphics*[width=8cm]{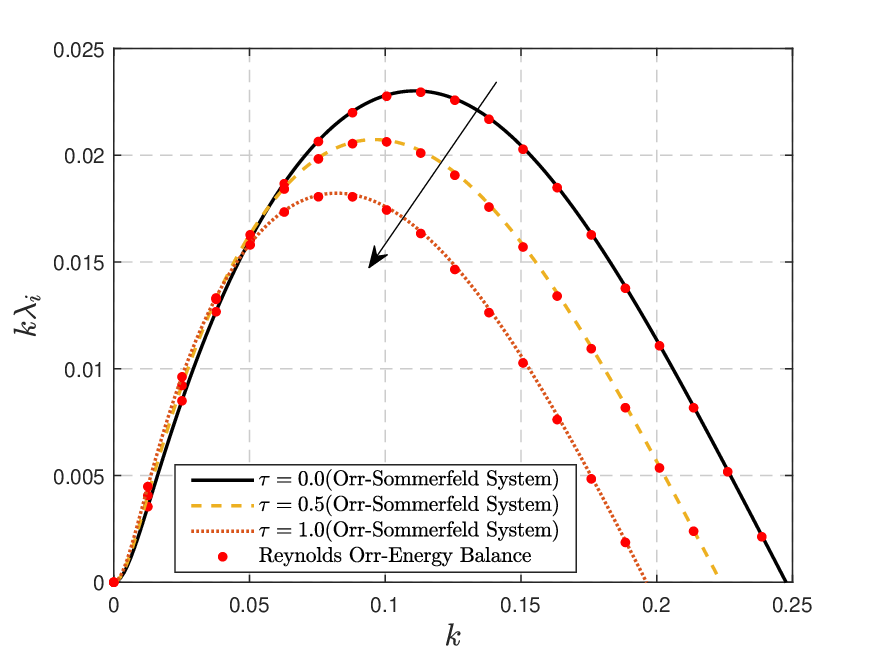}}
			\subfigure[]{\label{fig_19_b}\includegraphics*[width=8cm]{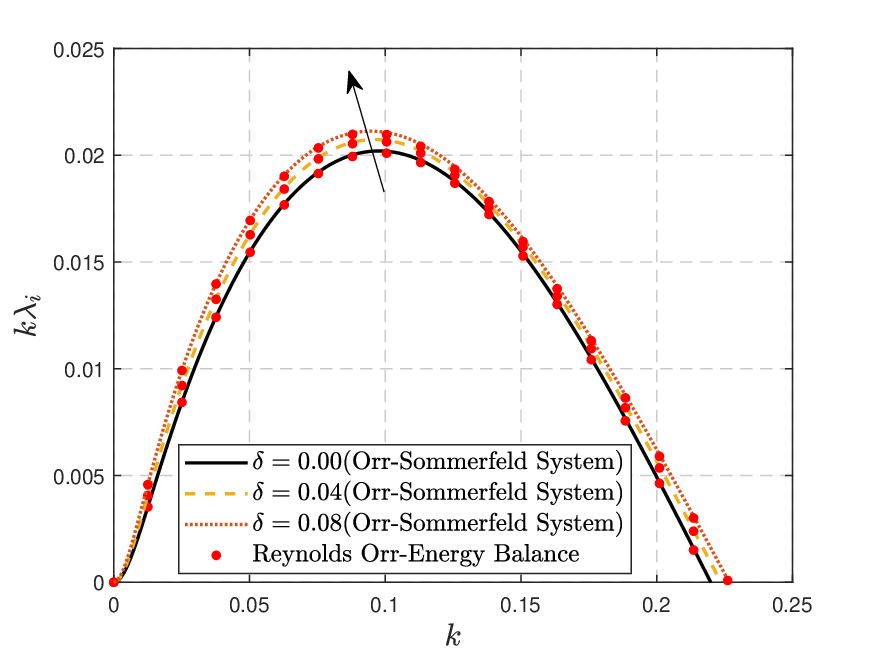}}
		\end{center}
		\caption{ (a) The comparison of growth rate variations against wavenumber $k$ for different external shear force $\tau$. The solid, dashed, and dotted lines are curves corresponding to $\tau = 0.0$, $\tau = 0.5$, and $\tau = 1.0$, respectively. (b) The comparison of growth rate variations against wavenumber $k$ for different slip parameters $\delta$. The solid, dashed, and dotted lines are curves corresponding to $\delta = 0.00$, $\delta = 0.04$, and $\delta = 0.08,$ respectively. The arrow indicates the direction of the growth rates from lower parametric values (of $\tau, \delta$) to higher. The red dots represent the data obtained from solving the Reynolds-Orr energy balance equation. The other relevant parameters are $\xi = 90^{\circ}, \chi = 100, Ka = 3000$, $\Sigma = 2.0$, $Sc_a = 100$, $Sc_b = 10$, $M_{tot} = 0.1$, $\beta_a = 0.01$, $R_a = 1.0$, $k_a = 1.0$, $\tau = 0.5$, and $\delta = 0.04$. } \label{fig_19}
	\end{figure}
	The equation \eqref{eqn133} is normalized by the left-hand side of the equation in order to get the temporal growth rate of the system following \citet{hossain2023instability}. The temporal growth rates for surface mode from energy budget calculations are compared with those of the numerical solution of the equation \eqref{eqn125} for the same values of the parameters with $\chi = 100$ in the figure (\ref{fig_19}), and an excellent agreement can be observed. After normalization, the energy components are denoted in capital letters, such as, `$KIN$' for `$Kin$' and so on. In figure \ref{fig_19_a}, the temporal growth rates are plotted for different values of external shear force $\tau$, and in figure \ref{fig_19_b}, they are plotted for different slip parameters $\delta$. It is evident that temporal growth rates corresponding to the surface mode against wavenumber $k$ follow the same results as reported earlier during numerical solutions in both scenarios. The growth rate results in figures \ref{fig_19_a}, \ref{fig_19_b} are a testament to the fact that the code related to the energy budget analysis is correct, and based on that, the energy components are illustrated below under various values of the parameters for the surface mode.\\
	\begin{figure} 
		\begin{center}
			\subfigure[]{\label{fig_20_a}\includegraphics*[width=5.4cm]{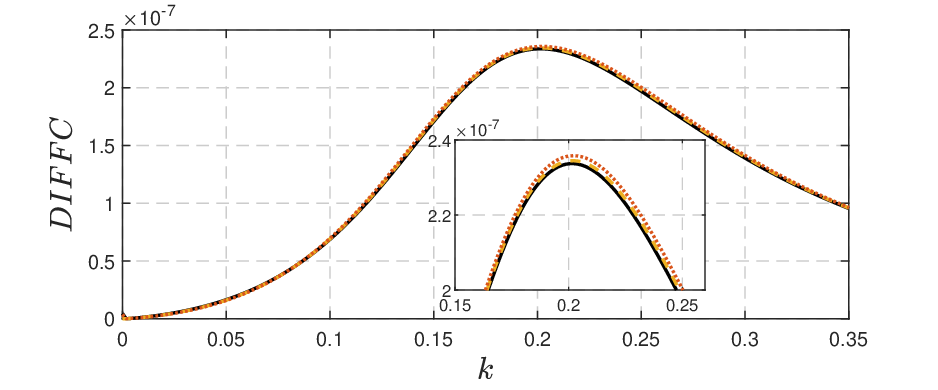}}
			\subfigure[]{\label{fig_20_b}\includegraphics*[width=5.4cm]{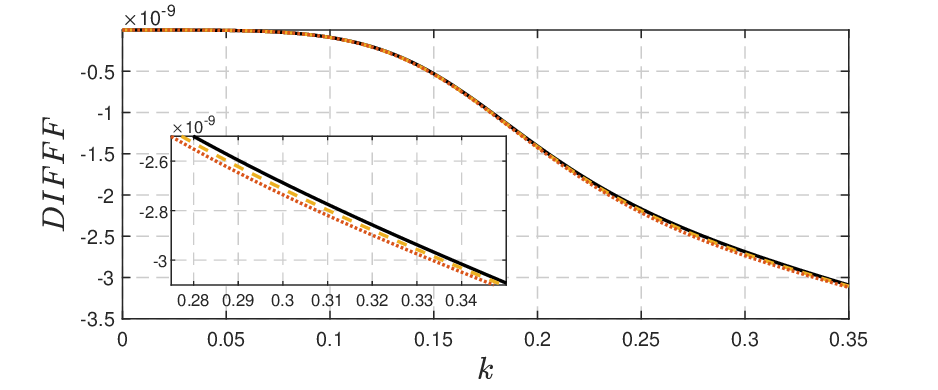}}
			\subfigure[]{\label{fig_20_c}\includegraphics*[width=5.4cm]{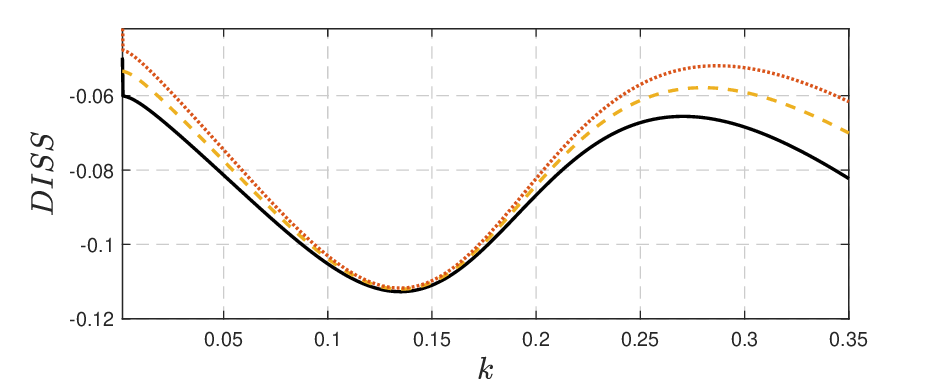}}
			\subfigure[]{\label{fig_20_d}\includegraphics*[width=5.4cm]{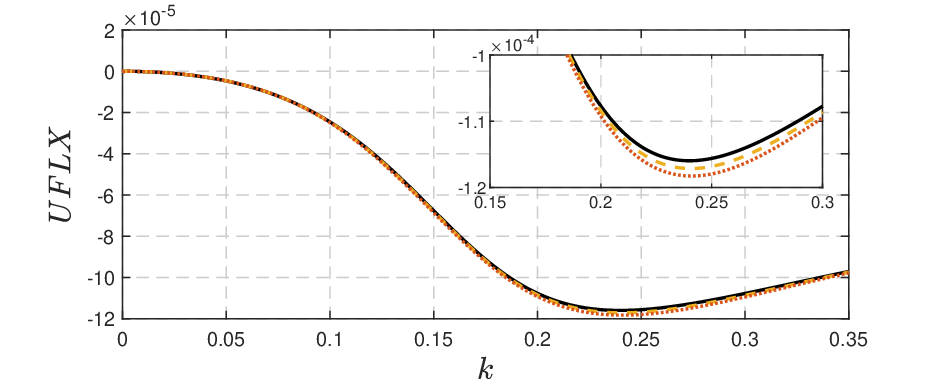}}
			\subfigure[]{\label{fig_20_e}\includegraphics*[width=5.4cm]{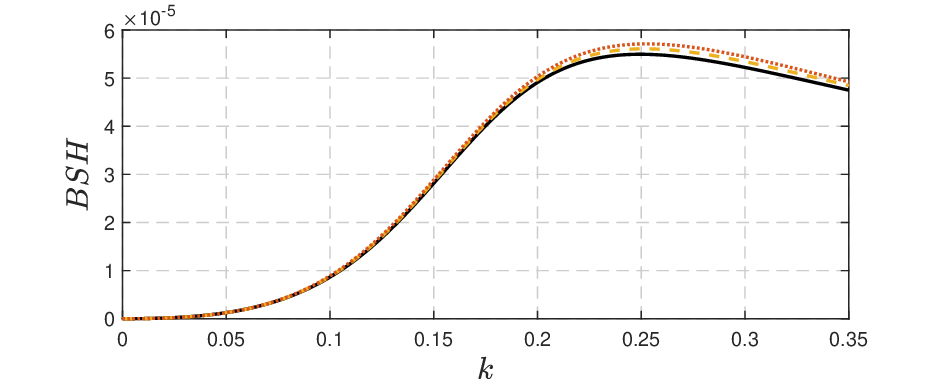}}
			\subfigure[]{\label{fig_20_f}\includegraphics*[width=5.4cm]{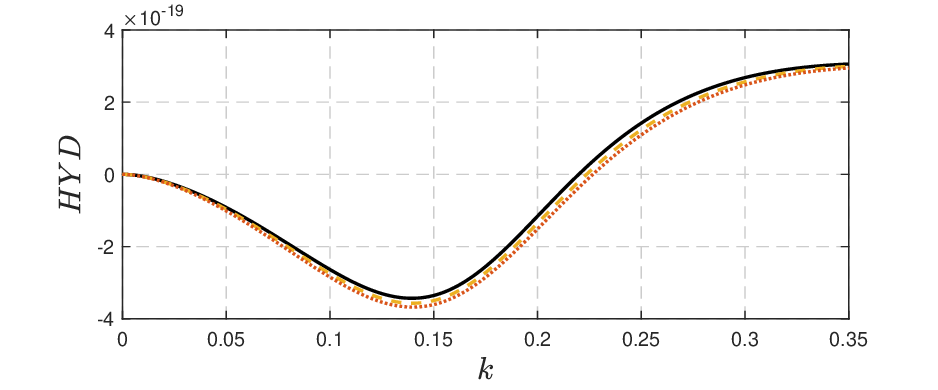}}
			\subfigure[]{\label{fig_20_g}\includegraphics*[width=5.4cm]{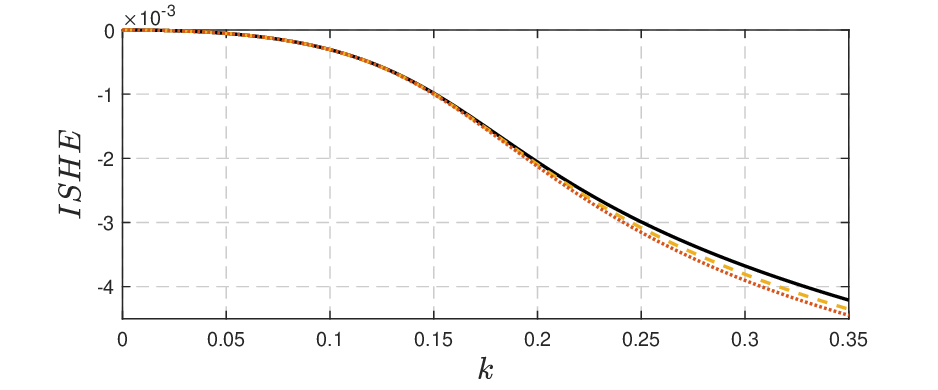}}
			\subfigure[]{\label{fig_20_h}\includegraphics*[width=5.4cm]{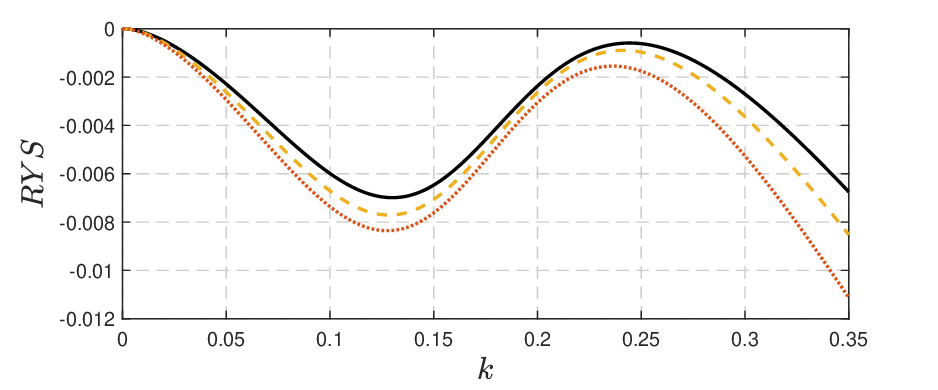}}
			\subfigure[]{\label{fig_20_i}\includegraphics*[width=5.4cm]{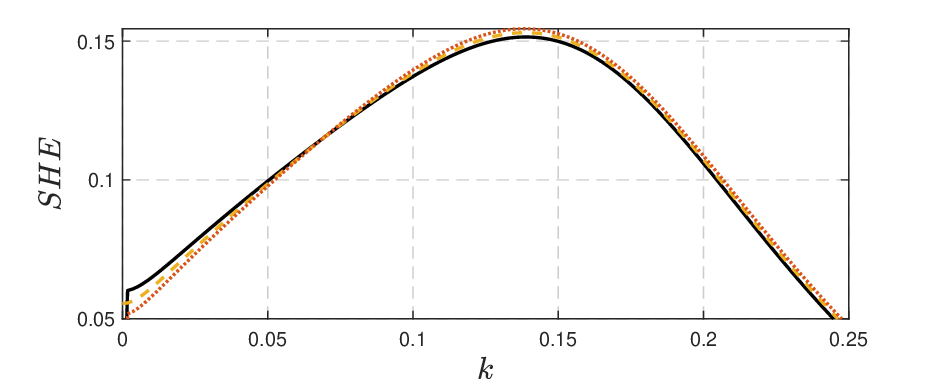}}
			\subfigure[]{\label{fig_20_j}\includegraphics*[width=5.4cm]{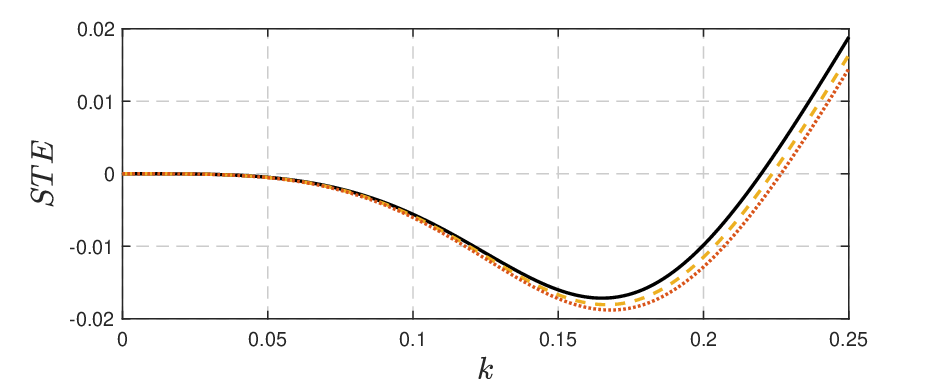}}
		\end{center}
		\caption{The variation of normalized energy components (a) DIFFC, (b) DIFFF, (c) DISS, (d) UFLX, (e) BSH, (f) HYD, (g) ISHE, (h) RYS, (i) SHE, and (j) STE against wavenumber $k$ for different slip parameter $\delta$ in the case of surface mode. The solid, dashed, and dotted lines represent the energy component curves for slip parameters $\delta = 0.00$, $\delta = 0.04$, and $\delta = 0.08$, respectively. The other relevant parameters are $\xi = 90^{\circ}, \chi = 100, Ka = 3000$, $\Sigma = 2.0$, $Sc_a = 100$, $Sc_b = 10$, $M_{tot} = 0.1$, $\beta_a = 0.01$, $R_a = 1.0$, $k_a = 1.0$, and $\tau = 0.5$. } \label{fig_20}
	\end{figure}
	\indent In the first place, the energy components are varied against wavenumber $k$ for various values of slip parameter $\delta$ in the figures (\ref{fig_20}). It can be observed from the figures that the most prominent energy contributor to the accumulated kinetic energy source is the work done against shear stress ($SHE$) with work done against surface tension ($STE$), along with the effect for background shear ($BSH$), diffusion of energy due to the presence of surfactant in the bulk ($DIFFC$) and hydrostatic pressure ($HYD$) being are other four contributors, as can be seen from the figure (\ref{fig_20}). In fact, for $k \rightarrow 0$, $STE$ is damped, and only for higher values of wavenumber ($k \ge 0.221106$) does it contribute to the kinetic energy of the system. Moreover, the contribution of $HYD$ to the kinetic energy of the system is very minimal (of order $\order{10^{-19}}$), and that too for $k \ge 0.221106$ and for longwave $i.e., k \rightarrow 0$, $HYD$ is having a damped effect. The rest of the energy components contribute to dampening the onset of instability. Also, the diffusion term for the bulk surfactant contributes to the positive energy and drives the kinetic energy. Furthermore, figures \ref{fig_20_c} and \ref{fig_20_i} clearly indicate that the energy dissipation term $DISS$ due to the bulk fluid flow is completely balanced by the shear stress perturbation term $SHE$ in the $k \rightarrow 0$ limit, as observed in the literature\citep{kelly1989mechanism, samanta2023elliptic}. The rest of the terms, such as $DIFFF$, $DISS$, $UFLX$, $BSH$, $ISHE$, and $RYS$, promote stabilization by reducing the kinetic energy of the flow, with $DISS$ being the most pronounced in having a damping effect on the flow. Also, the effect of the slip parameter is portrayed in those figures. The considerable changes due to the slip parameter can be observed in the $DISS$, $HYD$, $RYS$, $SHE$, $STE$ terms with the magnitude of the $DISS$ getting decreased and that of the $HYD$, $RYS$, $STE$ get increased with extra slip length ($\delta$). The exception among them is $SHE$, which experiences a dual effect due to the slip parameter $\delta$ as it decreases with increasing slip parameter till $k \approx 0.069$, and then increases with larger wavenumbers.
	
	\begin{figure} 
		\begin{center}
			\subfigure[]{\label{fig_21_a}\includegraphics*[width=5.4cm]{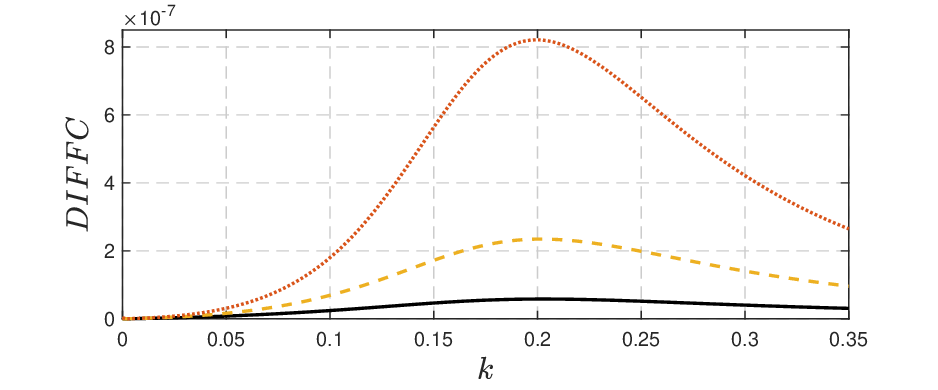}}
			\subfigure[]{\label{fig_21_b}\includegraphics*[width=5.4cm]{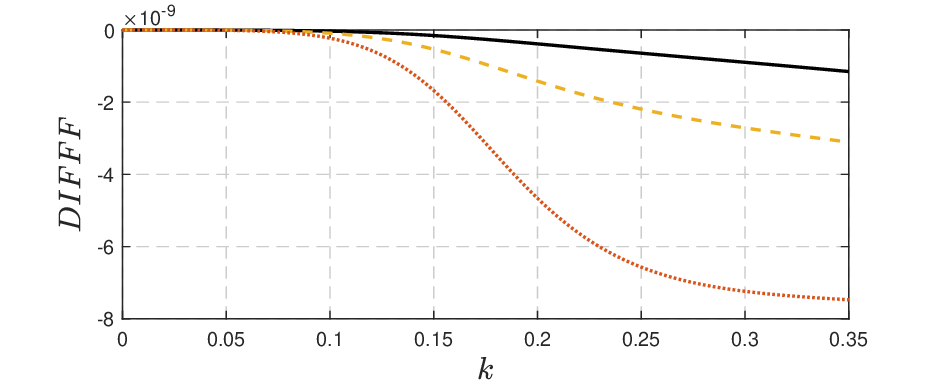}}
			\subfigure[]{\label{fig_21_c}\includegraphics*[width=5.4cm]{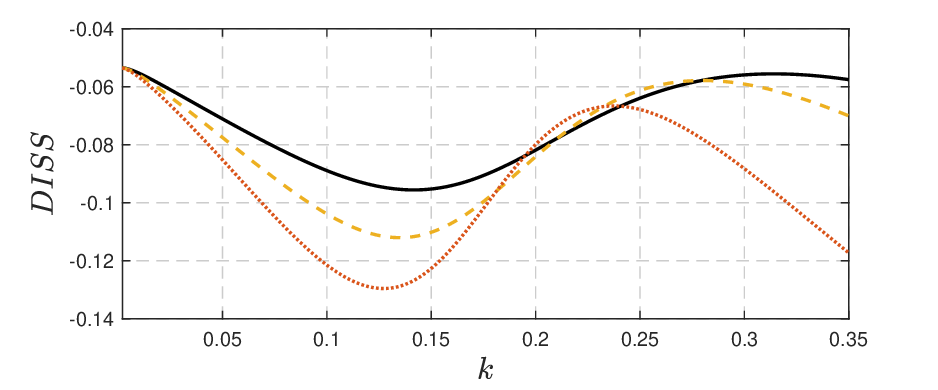}}
			\subfigure[]{\label{fig_21_d}\includegraphics*[width=5.4cm]{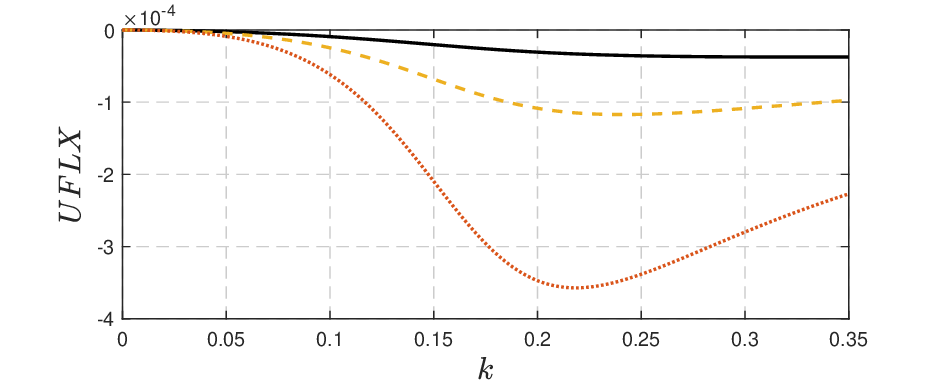}}
			\subfigure[]{\label{fig_21_e}\includegraphics*[width=5.4cm]{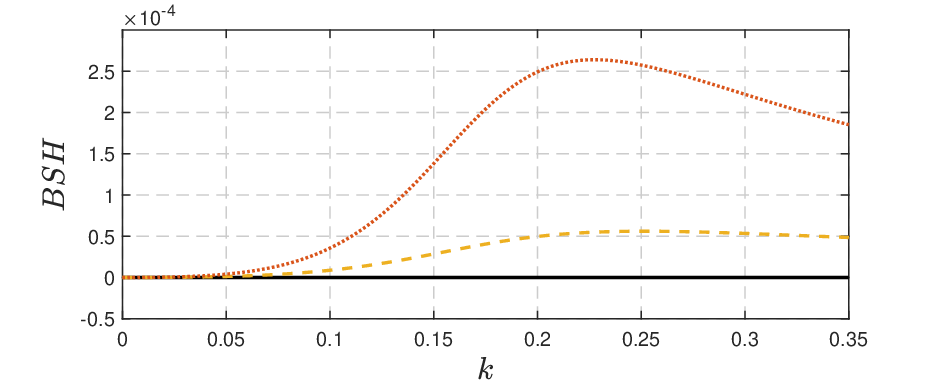}}
			\subfigure[]{\label{fig_21_f}\includegraphics*[width=5.4cm]{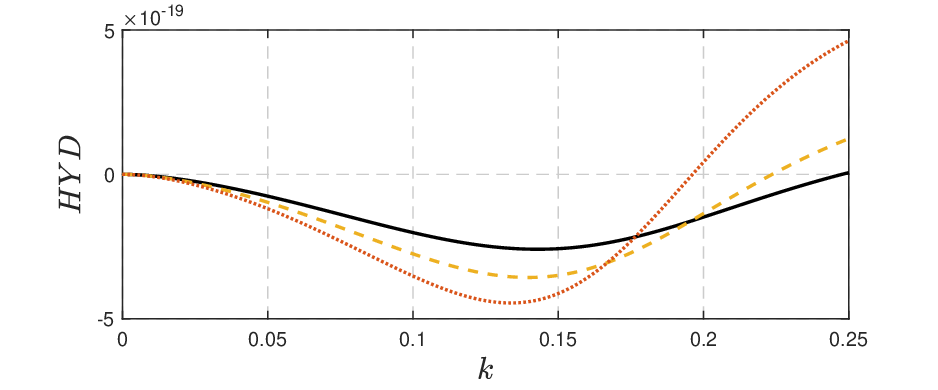}}
			\subfigure[]{\label{fig_21_g}\includegraphics*[width=5.4cm]{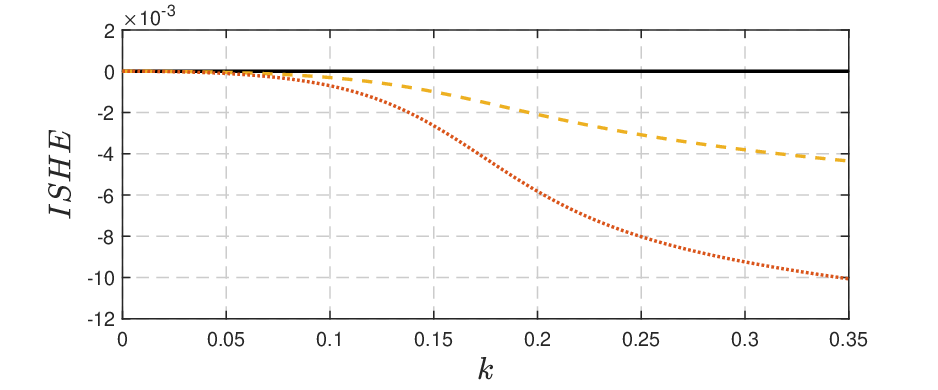}}
			\subfigure[]{\label{fig_21_h}\includegraphics*[width=5.4cm]{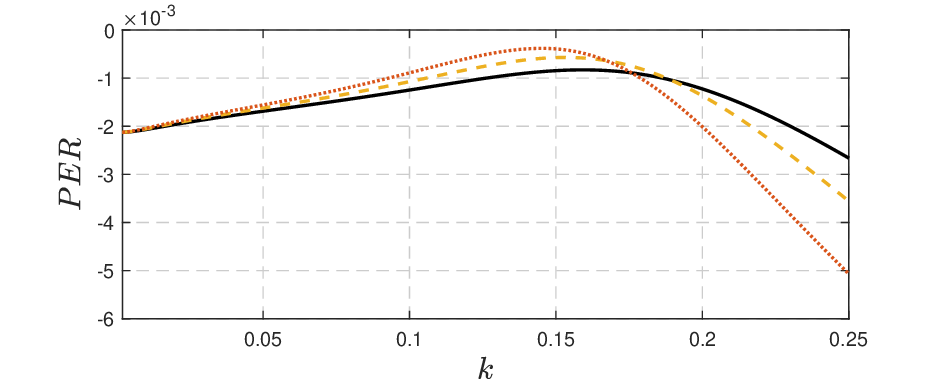}}
			\subfigure[]{\label{fig_21_i}\includegraphics*[width=5.4cm]{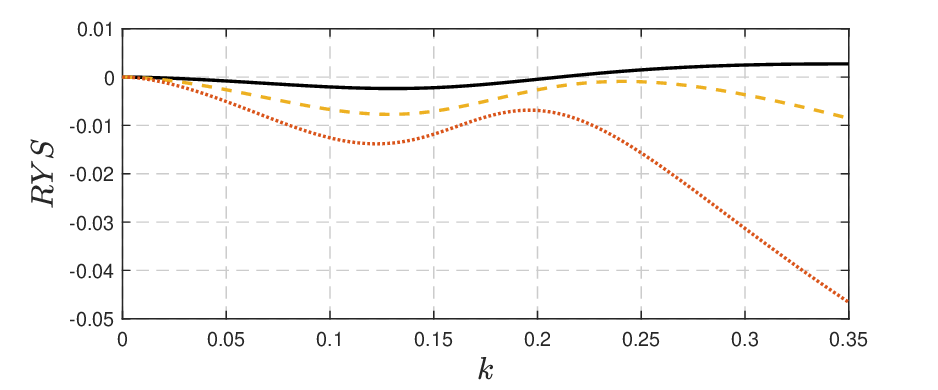}}
			\subfigure[]{\label{fig_21_j}\includegraphics*[width=5.4cm]{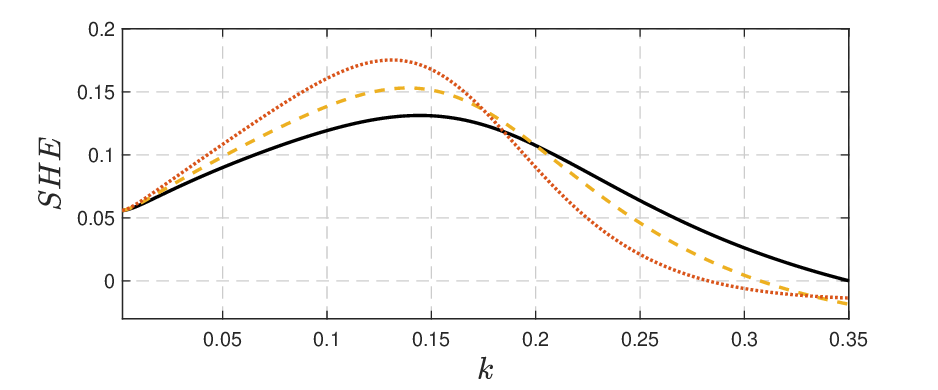}}
			\subfigure[]{\label{fig_21_k}\includegraphics*[width=5.4cm]{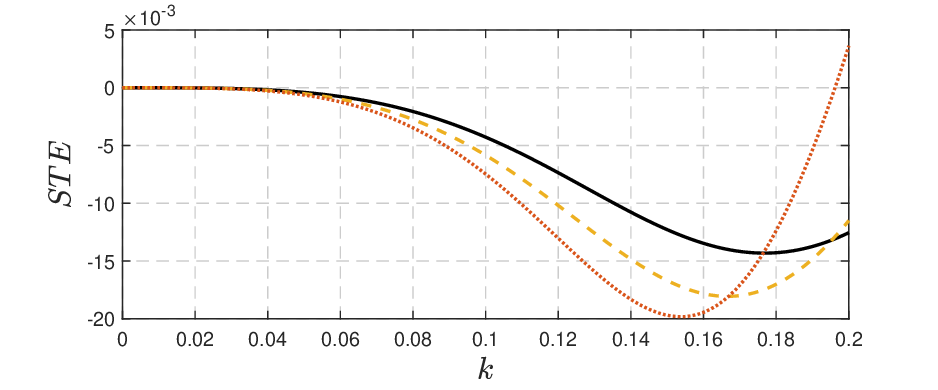}}
		\end{center}
		\caption{The variation of normalized energy components (a) DIFFC, (b) DIFFF, (c) DISS, (d) UFLX, (e) BSH, (f) HYD, (g) ISHE, (h) PER, (i) RYS, (j) SHE, and (k) STE against wavenumber $k$ for different imposed shear force $\tau$ in the case of surface mode. The solid, dashed, and dotted lines represent the energy component curves for imposed shear force $\tau = 0.0$, $\tau = 0.5$, and $\tau = 1.0$, respectively. The other relevant parameters are $\xi = 90^{\circ}, \chi = 100, Ka = 3000$, $\Sigma = 2.0$, $Sc_a = 100$, $Sc_b = 10$, $M_{tot} = 0.1$, $\beta_a = 0.01$, $R_a = 1.0$, $k_a = 1.0$, and $\delta = 0.04$.} \label{fig_21}
	\end{figure}
	On the other hand, when the imposed shear force ($\tau$) is varied, figures (\ref{fig_21}) suggest that $DIFFC$, $BSH$, and $SHE$ are the ones that contribute to increasing the accumulated kinetic energy of the whole system. Also, figure \ref{fig_21_f} suggests that $HYD$ provides energy support to the instability, but for large values of wavenumber ($k \ge 0.249146 $ for the values of the imposed shear chosen in this article). The rest of the energy components weaken the instability of the flow by reducing the magnitude of the kinetic energy. It can be observed that figure $SHE$ is one of the major contributors to the kinetic energy, followed by $DIFFC$ and $BSH$. At the longwave regime ($k \rightarrow 0$), perturbed shear stress $SHE$ is balanced by the energy dissipation $DISS$ of the bulk flow \citep{bhat2019linear, kelly1989mechanism}, as can be noticed from figures \ref{fig_21_c} and \ref{fig_21_j}. Moreover, the work done against hydrostatic pressure $HYD$ is negligible, as can be seen in the figure \ref{fig_21_f} (as it is scaled in the $\order{10^{-19}}$). But the energy transfer from the base state to the perturbed state through the Reynolds stress $RYS$, dissipation of energy by viscous effect $DISS$, energy term associated with slip parameter $PER$, work done against surface tension $STE$, work done by the imposed shear against infinitesimal perturbation $ISHE$ are the terms having a greater impact to reduce the perturbed kinetic energy. The other terms that create an alleviating effect on the energy of the whole system are $UFLX$ and $DIFFF$. Also, the increasing imposed shear $\tau$ helps the diffusion of energy due to the surfactant present in the bulk $DIFFC$, and the term due to the background shear $BSH$ to increase the kinetic energy source strictly, whereas, $DIFFF$, $ISHE$, $RYS$, $STE$, increases in magnitude but effects in damping. Further, the other energy terms, such as $DISS$, $HYD$, $PER$, and $SHE$, all have dual effects as far as increasing imposed shear is concerned.

	\section{\bf{Conclusions}}\label{con}
	The current study focuses on linear stability analysis for soluble surfactant-contaminated thin film streaming over a slippery inclined surface in the presence of externally imposed shear. The Orr-Sommerfeld eigenvalue system corresponding to the problem is obtained and solved analytically by the longwave expansion method and numerically using the Chebyshev spectral collocation method. The advection-diffusion equation for soluble surfactant in the bulk and the kinetics laws are considered as \citet{karapetsas2014role}. The longwave analysis reveals that there exist two longwave modes, viz, surface mode and surfactant mode. It has been found that both slippery bottom and imposed shear maintain a dual nature for the surface mode. The results obtained for surfactant mode reveal that the surfactant parameter $\beta_a$ and the solubility parameter $R_a$ behave as stabilizing agents. The lower value of the product of the aforementioned parameters $(\xi_a = \beta_a R_a)$ makes the surfactant highly soluble, and higher values $(\xi_a \rightarrow \infty)$ make it insoluble, and here it is observed that for lower values of these parameters, the flow stabilizes. Also, it has been found that for moderate values of $\beta_a$ and $R_a$, their effect can be better observed. Similar results were observed by \citet{karapetsas2013primary}. For the moderate values of these parameters, the effect of the slip parameter $\delta$ and imposed shear $\tau$ in the surfactant mode regime is shown, and it is witnessed that the slip parameter stabilizes the flow, and the imposed shear $\tau$ follows the exact opposite route. The numerical analysis fortifies the claim of having two modes and divulges that there is another finite wavelength mode that exists, named shear mode. The phase boundary between the surface mode and shear mode reveals that $\xi = 1^{'}$ can be considered to observe the effect of shear mode, and based on this inclination angle, the slippery bottom parameter $\delta$ displays a stabilizing effect on the shear mode by increasing the critical modified Reynolds number. However, the externally imposed shear $\tau$ helps to enhance the unstable zone and hence contributes to the instability of the flow. In order to comprehend the mechanism for the surface mode regime more intricately, the method of energy budget is implemented. It has been found that the work done against shear stress $SHE$, energy term for background shear $BSH$, and the diffusion term due to the bulk surfactant $DIFFC$ are the major contributors to the kinetic energy of the system and there are terms such as the energy term for hydrostatic pressure $HYD$ and surface tension $STE$ also contribute to the instability but in the finite wavenumber regime. The rest of the energy terms actually help to stabilize the system.
	
	\section*{Acknowledgment}
	The first author would like to acknowledge the help he received from Dr. George Karapetsas, (Aristotle University of Thessaloniki, Greece) during the preparation of this manuscript. Also, he acknowledge the financial support from SERB, Department of Science and Technology, Government of India, through CRG project, Award No. CRG/2022/006698.
	
	\textbf{Declaration of interests:} The authors report no conflict of interest.
	
	\textbf{Data Availability:} The data that supports the findings of this study are available within the article, highlighted in each of the figure captions and corresponding discussions. 
	\section*{References}
	\bibliographystyle{unsrtnat}
	\bibliography{REF}

\begin{thebibliography}{88}
\providecommand{\natexlab}[1]{#1}
\providecommand{\url}[1]{\texttt{#1}}
\expandafter\ifx\csname urlstyle\endcsname\relax
  \providecommand{\doi}[1]{doi: #1}\else
  \providecommand{\doi}{doi: \begingroup \urlstyle{rm}\Url}\fi

\bibitem[Karapetsas and Bontozoglou(2014)]{karapetsas2014role}
George Karapetsas and Vasilis Bontozoglou.
\newblock The role of surfactants on the mechanism of the long-wave instability
  in liquid film flows.
\newblock \emph{Journal of fluid mechanics}, 741:\penalty0 139--155, 2014.

\bibitem[GREEK(1991)]{greek1991sales}
BRUCE~F GREEK.
\newblock Sales of detergents growing despite recession.
\newblock \emph{Chemical \& Engineering News Archive}, 69\penalty0
  (4):\penalty0 25--52, 1991.

\bibitem[Desai and Banat(1997)]{desai1997microbial}
Jitendra~D Desai and Ibrahim~M Banat.
\newblock Microbial production of surfactants and their commercial potential.
\newblock \emph{Microbiology and Molecular biology reviews}, 61\penalty0
  (1):\penalty0 47--64, 1997.

\bibitem[De et~al.(2015)De, Malik, Ghosh, Saha, and Saha]{de2015review}
Sourav De, Susanta Malik, Aniruddha Ghosh, Rumpa Saha, and Bidyut Saha.
\newblock A review on natural surfactants.
\newblock \emph{RSC advances}, 5\penalty0 (81):\penalty0 65757--65767, 2015.

\bibitem[Halpern and Grotberg(1993)]{halpern1993surfactant}
D~Halpern and JB~Grotberg.
\newblock Surfactant effects on fluid-elastic instabilities of liquid-lined
  flexible tubes: a model of airway closure.
\newblock 1993.

\bibitem[Guerra-Santos et~al.(1984)Guerra-Santos, K{\"a}ppeli, and
  Fiechter]{guerra1984pseudomonas}
Luis Guerra-Santos, Othmar K{\"a}ppeli, and Armin Fiechter.
\newblock Pseudomonas aeruginosa biosurfactant production in continuous culture
  with glucose as carbon source.
\newblock \emph{Applied and environmental microbiology}, 48\penalty0
  (2):\penalty0 301--305, 1984.

\bibitem[Craster and Matar(2009)]{craster2009dynamics}
Richard~V Craster and Omar~K Matar.
\newblock Dynamics and stability of thin liquid films.
\newblock \emph{Reviews of modern physics}, 81\penalty0 (3):\penalty0 1131,
  2009.

\bibitem[Grotberg(1994)]{grotberg1994pulmonary}
James~B Grotberg.
\newblock Pulmonary flow and transport phenomena.
\newblock \emph{Annual review of fluid mechanics}, 26\penalty0 (1):\penalty0
  529--571, 1994.

\bibitem[Halliday(2008)]{halliday2008surfactants}
HL~Halliday.
\newblock Surfactants: past, present and future.
\newblock \emph{Journal of perinatology}, 28\penalty0 (1):\penalty0 S47--S56,
  2008.

\bibitem[Banat(1995{\natexlab{a}})]{banat1995biosurfactants}
IBRAHIM~M Banat.
\newblock Biosurfactants production and possible uses in microbial enhanced oil
  recovery and oil pollution remediation: a review.
\newblock \emph{Bioresource technology}, 51\penalty0 (1):\penalty0 1--12,
  1995{\natexlab{a}}.

\bibitem[Banat(1995{\natexlab{b}})]{banat1995characterization}
IM~Banat.
\newblock Characterization of biosurfactants and their use in pollution
  removal--state of the art.
\newblock \emph{Acta Biotechnologica}, 15\penalty0 (3):\penalty0 251--267,
  1995{\natexlab{b}}.

\bibitem[Fiechter(1992)]{fiechter1992biosurfactants}
Armin Fiechter.
\newblock Biosurfactants: moving towards industrial application.
\newblock \emph{Trends in biotechnology}, 10:\penalty0 208--217, 1992.

\bibitem[Santos et~al.(2016)Santos, Rufino, Luna, Santos, and
  Sarubbo]{santos2016biosurfactants}
Danyelle Khadydja~F Santos, Raquel~D Rufino, Juliana~M Luna, Valdemir~A Santos,
  and Leonie~A Sarubbo.
\newblock Biosurfactants: multifunctional biomolecules of the 21st century.
\newblock \emph{International journal of molecular sciences}, 17\penalty0
  (3):\penalty0 401, 2016.

\bibitem[Diez and Kondic(2001)]{diez2001contact}
Javier~A Diez and L~Kondic.
\newblock Contact line instabilities of thin liquid films.
\newblock \emph{Physical Review Letters}, 86\penalty0 (4):\penalty0 632, 2001.

\bibitem[Benjamin(1964)]{benjamin1964effects}
T~Brooke Benjamin.
\newblock Effects of surface contamination on wave formation in falling liquid
  films(stabilizing effect of surface active agents on wave formation in
  contaminated falling liquid film).
\newblock \emph{Archiwum Mechaniki Stosowanej}, 16\penalty0 (3):\penalty0
  615--626, 1964.

\bibitem[Whitaker(1964)]{whitaker1964effect}
Stephen Whitaker.
\newblock Effect of surface active agents on the stability of falling liquid
  films.
\newblock \emph{Industrial \& Engineering Chemistry Fundamentals}, 3\penalty0
  (2):\penalty0 132--142, 1964.

\bibitem[Emmert(1954)]{emmert1954study}
RE~Emmert.
\newblock A study of gas absorption in falling liquid films.
\newblock \emph{Chem. Eng. Prog.}, 50\penalty0 (2):\penalty0 87--93, 1954.

\bibitem[Stirba and Hurt(1955)]{stirba1955turbulence}
Clifford Stirba and Ds~M Hurt.
\newblock Turbulence in falling liquid films.
\newblock \emph{AIChE Journal}, 1\penalty0 (2):\penalty0 178--184, 1955.

\bibitem[Fulford(1964)]{fulford1964flow}
George~D Fulford.
\newblock The flow of liquids in thin films.
\newblock In \emph{Advances in Chemical Engineering}, volume~5, pages 151--236.
  Elsevier, 1964.

\bibitem[Skelland and Caenepeel(1972)]{skelland1972effects}
AHP Skelland and CL~Caenepeel.
\newblock Effects of surface active agents on mass transfer during droplet
  formation, fall, and coalescence.
\newblock \emph{AIChE Journal}, 18\penalty0 (6):\penalty0 1154--1163, 1972.

\bibitem[Roch{\'e} et~al.(2014)Roch{\'e}, Li, Griffiths, Le~Roux, Cantat,
  Saint-Jalmes, and Stone]{roche2014marangoni}
Matthieu Roch{\'e}, Zhenzhen Li, Ian~M Griffiths, S{\'e}bastien Le~Roux,
  Isabelle Cantat, Arnaud Saint-Jalmes, and Howard~A Stone.
\newblock Marangoni flow of soluble amphiphiles.
\newblock \emph{Physical review letters}, 112\penalty0 (20):\penalty0 208302,
  2014.

\bibitem[Kim et~al.(2017)Kim, Muller, Shardt, Afkhami, and
  Stone]{kim2017solutal}
Hyoungsoo Kim, Koen Muller, Orest Shardt, Shahriar Afkhami, and Howard~A Stone.
\newblock Solutal marangoni flows of miscible liquids drive transport without
  surface contamination.
\newblock \emph{Nature Physics}, 13\penalty0 (11):\penalty0 1105--1110, 2017.

\bibitem[Ji and Setterwall(1994)]{ji1994instabilities}
Wei Ji and Fredrik Setterwall.
\newblock On the instabilities of vertical falling liquid films in the presence
  of surface-active solute.
\newblock \emph{Journal of Fluid Mechanics}, 278:\penalty0 297--323, 1994.

\bibitem[Blyth and Pozrikidis(2004)]{blyth2004effect}
MG~Blyth and C~Pozrikidis.
\newblock Effect of surfactant on the stability of film flow down an inclined
  plane.
\newblock \emph{Journal of Fluid Mechanics}, 521:\penalty0 241--250, 2004.

\bibitem[Pereira and Kalliadasis(2008)]{pereira2008dynamics}
Antonio Pereira and Serafim Kalliadasis.
\newblock Dynamics of a falling film with solutal marangoni effect.
\newblock \emph{Physical Review E—Statistical, Nonlinear, and Soft Matter
  Physics}, 78\penalty0 (3):\penalty0 036312, 2008.

\bibitem[Hu et~al.(2020)Hu, Fu, and Yang]{hu2020falling}
Tao Hu, Qingfei Fu, and Lijun Yang.
\newblock Falling film with insoluble surfactants: effects of surface
  elasticity and surface viscosities.
\newblock \emph{Journal of Fluid Mechanics}, 889:\penalty0 A16, 2020.

\bibitem[Matar and Kumar(2004)]{matar2004rupture}
Omar~K Matar and Satish Kumar.
\newblock Rupture of a surfactant-covered thin liquid film on a flexible wall.
\newblock \emph{SIAM Journal on Applied Mathematics}, 64\penalty0 (6):\penalty0
  2144--2166, 2004.

\bibitem[Pal and Samanta(2021)]{pal2021linear}
Subham Pal and Arghya Samanta.
\newblock Linear stability of a surfactant-laden viscoelastic liquid flowing
  down a slippery inclined plane.
\newblock \emph{Physics of Fluids}, 33\penalty0 (5), 2021.

\bibitem[Hossain et~al.(2022)Hossain, Ghosh, and Behera]{hossain2022linear}
Md~Mouzakkir Hossain, Sukhendu Ghosh, and Harekrushna Behera.
\newblock Linear instability of a surfactant-laden shear imposed falling film
  over an inclined porous bed.
\newblock \emph{Physics of Fluids}, 34\penalty0 (8), 2022.

\bibitem[Wei(2005)]{wei2005effect}
Hsien-Hung Wei.
\newblock Effect of surfactant on the long-wave instability of a shear-imposed
  liquid flow down an inclined plane.
\newblock \emph{Physics of Fluids}, 17\penalty0 (1), 2005.

\bibitem[Batchvarov et~al.(2021)Batchvarov, Kahouadji, Constante-Amores,
  Gon{\c{c}}alves, Shin, Chergui, Juric, Craster, and
  Matar]{batchvarov2021three}
Assen Batchvarov, Lyes Kahouadji, Cristian~R Constante-Amores, Gabriel
  Farah~Nor{\~o}es Gon{\c{c}}alves, Seungwon Shin, Jalel Chergui, Damir Juric,
  Richard~V Craster, and Omar~K Matar.
\newblock Three-dimensional dynamics of falling films in the presence of
  insoluble surfactants.
\newblock \emph{Journal of Fluid Mechanics}, 906:\penalty0 A16, 2021.

\bibitem[Zivkov and Pascal(2025)]{zivkov2025thin}
Eugene Zivkov and Jean-Paul Pascal.
\newblock Thin liquid film stability in the presence of bottom topography and
  surfactant.
\newblock \emph{International Journal of Multiphase Flow}, 182:\penalty0
  105043, 2025.

\bibitem[Manikantan and Squires(2020)]{manikantan2020surfactant}
Harishankar Manikantan and Todd~M Squires.
\newblock Surfactant dynamics: hidden variables controlling fluid flows.
\newblock \emph{Journal of fluid mechanics}, 892:\penalty0 P1, 2020.

\bibitem[Shkadov et~al.(2004)Shkadov, Velarde, and
  Shkadova]{shkadov2004falling}
V~Ya Shkadov, MG~Velarde, and VP~Shkadova.
\newblock Falling films and the marangoni effect.
\newblock \emph{Physical Review E—Statistical, Nonlinear, and Soft Matter
  Physics}, 69\penalty0 (5):\penalty0 056310, 2004.

\bibitem[Georgantaki et~al.(2012)Georgantaki, Vlachogiannis, and
  Bontozoglou]{georgantaki2012effect}
A~Georgantaki, M~Vlachogiannis, and V~Bontozoglou.
\newblock The effect of soluble surfactants on liquid film flow.
\newblock In \emph{Journal of Physics: Conference Series}, volume 395, page
  012165. IOP Publishing, 2012.

\bibitem[Karapetsas and Bontozoglou(2013)]{karapetsas2013primary}
George Karapetsas and Vasilis Bontozoglou.
\newblock The primary instability of falling films in the presence of soluble
  surfactants.
\newblock \emph{Journal of Fluid Mechanics}, 729:\penalty0 123--150, 2013.

\bibitem[Samanta(2025{\natexlab{a}})]{samanta2025role}
Arghya Samanta.
\newblock Role of soluble surfactant in linear stability of a liquid film
  flowing down a compliant substrate.
\newblock \emph{Journal of Fluid Mechanics}, 1011:\penalty0 A6,
  2025{\natexlab{a}}.

\bibitem[D’Alessio et~al.(2020)D’Alessio, Pascal, Ellaban, and
  Ruyer-Quil]{d2020marangoni}
SJD D’Alessio, JP~Pascal, E~Ellaban, and C~Ruyer-Quil.
\newblock Marangoni instabilities associated with heated surfactant-laden
  falling films.
\newblock \emph{Journal of Fluid Mechanics}, 887:\penalty0 A20, 2020.

\bibitem[Samanta(2025{\natexlab{b}})]{samanta2025effect}
Arghya Samanta.
\newblock Effect of soluble surfactant on thermocapillary instability in
  falling film.
\newblock \emph{Physical Review Fluids}, 10\penalty0 (6):\penalty0 064001,
  2025{\natexlab{b}}.

\bibitem[Smith(1990)]{smith1990mechanism}
Marc~K Smith.
\newblock The mechanism for the long-wave instability in thin liquid films.
\newblock \emph{Journal of Fluid Mechanics}, 217:\penalty0 469--485, 1990.

\bibitem[Samanta(2014)]{samanta2014shear}
Arghya Samanta.
\newblock Shear-imposed falling film.
\newblock \emph{Journal of fluid mechanics}, 753:\penalty0 131--149, 2014.

\bibitem[Choudhury and Samanta(2024)]{choudhury2024thermocapillary}
Arnab Choudhury and Arghya Samanta.
\newblock Thermocapillary instability of a surfactant-laden shear-imposed film
  flow.
\newblock \emph{Physical Review Fluids}, 9\penalty0 (8):\penalty0 084002, 2024.

\bibitem[Espinosa and Kamm(1999)]{espinosa1999bolus}
FF~Espinosa and RD~Kamm.
\newblock Bolus dispersal through the lungs in surfactant replacement therapy.
\newblock \emph{Journal of Applied Physiology}, 86\penalty0 (1):\penalty0
  391--410, 1999.

\bibitem[Bhat and Samanta(2019)]{bhat2019linear}
Farooq~Ahmad Bhat and Arghya Samanta.
\newblock Linear stability analysis of a surfactant-laden shear-imposed falling
  film.
\newblock \emph{Physics of Fluids}, 31\penalty0 (5), 2019.

\bibitem[Hossain et~al.(2024{\natexlab{a}})Hossain, Saha, Behera, and
  Ghosh]{hossain2024instability}
Md~Mouzakkir Hossain, Mrityunjoy Saha, Harekrushna Behera, and Sukhendu Ghosh.
\newblock Instability behavior of odd viscosity-induced viscous fluid over a
  vibrating bed.
\newblock \emph{arXiv preprint arXiv:2409.20482}, 2024{\natexlab{a}}.

\bibitem[Hossain et~al.(2024{\natexlab{b}})Hossain, Ghosh, Behera, and
  Sekhar]{hossain2024hydrodynamic}
Md~Mouzakkir Hossain, Sukhendu Ghosh, Harekrushna Behera, and GP~Sekhar.
\newblock Hydrodynamic instability of shear imposed falling film over a
  uniformly heated inclined undulated substrate.
\newblock \emph{Physics of Fluids}, 36\penalty0 (8), 2024{\natexlab{b}}.

\bibitem[Miladinova et~al.(2002)Miladinova, Slavtchev, Lebon, and
  Legros]{miladinova2002long}
Svetla Miladinova, Slavtcho Slavtchev, Georgy Lebon, and Jean-Claude Legros.
\newblock Long-wave instabilities of non-uniformly heated falling films.
\newblock \emph{Journal of Fluid Mechanics}, 453:\penalty0 153--175, 2002.

\bibitem[D'alessio et~al.(2010)D'alessio, Pascal, Jasmine, and
  Ogden]{d2010film}
SJD D'alessio, JP~Pascal, HA~Jasmine, and KA~Ogden.
\newblock Film flow over heated wavy inclined surfaces.
\newblock \emph{Journal of fluid mechanics}, 665:\penalty0 418--456, 2010.

\bibitem[Usha et~al.(2005)Usha, Ravindran, and Uma]{usha2005dynamics}
R~Usha, R~Ravindran, and B~Uma.
\newblock Dynamics and stability of a thin liquid film on a heated rotating
  disk film with variable viscosity.
\newblock \emph{Physics of Fluids}, 17\penalty0 (10), 2005.

\bibitem[Samanta et~al.(2011)Samanta, Ruyer-Quil, and
  Goyeau]{samanta2011falling}
Arghya Samanta, Christian Ruyer-Quil, and Benoit Goyeau.
\newblock A falling film down a slippery inclined plane.
\newblock \emph{Journal of Fluid Mechanics}, 684:\penalty0 353--383, 2011.

\bibitem[Samanta(2021)]{samanta2021effect}
Arghya Samanta.
\newblock Effect of surfactant on the linear stability of a shear-imposed fluid
  flowing down a compliant substrate.
\newblock \emph{Journal of Fluid Mechanics}, 920:\penalty0 A23, 2021.

\bibitem[Choudhury and Samanta(2022)]{choudhury2022linear}
Arnab Choudhury and Arghya Samanta.
\newblock Linear stability of a falling film over a heated slippery plane.
\newblock \emph{Physical Review E}, 105\penalty0 (6):\penalty0 065112, 2022.

\bibitem[Neto et~al.(2005)Neto, Evans, Bonaccurso, Butt, and
  Craig]{neto2005boundary}
Chiara Neto, Drew~R Evans, Elmar Bonaccurso, Hans-J{\"u}rgen Butt, and
  Vincent~SJ Craig.
\newblock Boundary slip in newtonian liquids: a review of experimental studies.
\newblock \emph{Reports on progress in physics}, 68\penalty0 (12):\penalty0
  2859, 2005.

\bibitem[Lefevre et~al.(2004)Lefevre, Saugey, Barrat, Bocquet, Charlaix, Gobin,
  and Vigier]{lefevre2004intrusion}
Beno{\^\i}t Lefevre, Anthony Saugey, Jean-Louis Barrat, L~Bocquet, E~Charlaix,
  Pierre-Fran{\c{c}}ois Gobin, and G{\'e}rard Vigier.
\newblock Intrusion and extrusion of water in hydrophobic mesopores.
\newblock \emph{The Journal of chemical physics}, 120\penalty0 (10):\penalty0
  4927--4938, 2004.

\bibitem[Denn(2001)]{denn2001extrusion}
Morton~M Denn.
\newblock Extrusion instabilities and wall slip.
\newblock \emph{Annual Review of Fluid Mechanics}, 33\penalty0 (1):\penalty0
  265--287, 2001.

\bibitem[Cottin-Bizonne et~al.(2003)Cottin-Bizonne, Barrat, Bocquet, and
  Charlaix]{cottin2003low}
C{\'e}cile Cottin-Bizonne, Jean-Louis Barrat, Lyd{\'e}ric Bocquet, and
  Elisabeth Charlaix.
\newblock Low-friction flows of liquid at nanopatterned interfaces.
\newblock \emph{Nature materials}, 2\penalty0 (4):\penalty0 237--240, 2003.

\bibitem[Choi and Kim(2006)]{choi2006large}
Chang-Hwan Choi and Chang-Jin Kim.
\newblock Large slip of aqueous liquid flow over a nanoengineered
  superhydrophobic surface.
\newblock \emph{Physical review letters}, 96\penalty0 (6):\penalty0 066001,
  2006.

\bibitem[Lauga and Stone(2003)]{lauga2003effective}
Eric Lauga and Howard~A Stone.
\newblock Effective slip in pressure-driven stokes flow.
\newblock \emph{Journal of Fluid Mechanics}, 489:\penalty0 55--77, 2003.

\bibitem[Cho et~al.(2004)Cho, Law, and Rieutord]{cho2004dipole}
Jae-Hie~J Cho, Bruce~M Law, and Fran{\c{c}}ois Rieutord.
\newblock Dipole-dependent slip of newtonian liquids at smooth solid
  hydrophobic surfaces.
\newblock \emph{Physical review letters}, 92\penalty0 (16):\penalty0 166102,
  2004.

\bibitem[Blake(1990)]{blake1990slip}
Terence~D Blake.
\newblock Slip between a liquid and a solid: Dm tolstoi's (1952) theory
  reconsidered.
\newblock \emph{Colloids and surfaces}, 47:\penalty0 135--145, 1990.

\bibitem[Beavers and Joseph(1967)]{beavers1967boundary}
Gordon~S Beavers and Daniel~D Joseph.
\newblock Boundary conditions at a naturally permeable wall.
\newblock \emph{Journal of fluid mechanics}, 30\penalty0 (1):\penalty0
  197--207, 1967.

\bibitem[Pascal(1999)]{pascal1999linear}
JP~Pascal.
\newblock Linear stability of fluid flow down a porous inclined plane.
\newblock \emph{Journal of Physics D: Applied Physics}, 32\penalty0
  (4):\penalty0 417, 1999.

\bibitem[Ma et~al.(2020)Ma, Liu, Shao, Li, Li, and Xue]{ma2020effect}
Chicheng Ma, Jianlin Liu, Mingyu Shao, Bo~Li, Lei Li, and Zhangna Xue.
\newblock Effect of slip on the contact-line instability of a thin liquid film
  flowing down a cylinder.
\newblock \emph{Physical Review E}, 101\penalty0 (5):\penalty0 053108, 2020.

\bibitem[Chakraborty et~al.(2019)Chakraborty, Sheu, and
  Ghosh]{chakraborty2019dynamics}
Symphony Chakraborty, Tony Wen-Hann Sheu, and Sukhendu Ghosh.
\newblock Dynamics and stability of a power-law film flowing down a slippery
  slope.
\newblock \emph{Physics of Fluids}, 31\penalty0 (1), 2019.

\bibitem[Chattopadhyay et~al.(2021)Chattopadhyay, Desai, Gaonkar, Barua, and
  Mukhopadhyay]{chattopadhyay2021weakly}
Souradip Chattopadhyay, Akshay~S Desai, Amar~K Gaonkar, Amlan~K Barua, and
  Anandamoy Mukhopadhyay.
\newblock Weakly viscoelastic film on a slippery slope.
\newblock \emph{Physics of Fluids}, 33\penalty0 (11), 2021.

\bibitem[Ghosh and Usha(2016)]{ghosh2016stability}
Sukhendu Ghosh and R~Usha.
\newblock Stability of viscosity stratified flows down an incline: Role of
  miscibility and wall slip.
\newblock \emph{Physics of Fluids}, 28\penalty0 (10), 2016.

\bibitem[Mukhopadhyay and Pal(2024)]{mukhopadhyay2024modeling}
Anandamoy Mukhopadhyay and Subham Pal.
\newblock Modeling the stability of thin liquid film flows on a uniformly
  heated slippery inclined substrate: A realistic approach.
\newblock \emph{Physics of Fluids}, 36\penalty0 (3), 2024.

\bibitem[Bhat and Samanta(2018)]{bhat2018linear}
Farooq~Ahmad Bhat and Arghya Samanta.
\newblock Linear stability of a contaminated fluid flow down a slippery
  inclined plane.
\newblock \emph{Physical Review E}, 98\penalty0 (3):\penalty0 033108, 2018.

\bibitem[Agrawal et~al.(2024)Agrawal, Deepu, Usha, and
  Chattopadhyay]{agrawal2024impact}
Ankur Agrawal, P~Deepu, R~Usha, and Geetanjali Chattopadhyay.
\newblock Impact of wall-slip on the soluto-marangoni instability in a
  two-fluid system in a channel--creeping flow scenario.
\newblock \emph{International Journal of Multiphase Flow}, 175:\penalty0
  104813, 2024.

\bibitem[Sheludko(1967)]{sheludko1967thin}
A~Sheludko.
\newblock Thin liquid films.
\newblock \emph{Advances in Colloid and Interface Science}, 1\penalty0
  (4):\penalty0 391--464, 1967.

\bibitem[Gaver and Grotberg(1990)]{gaver1990dynamics}
Donald~P Gaver and James~B Grotberg.
\newblock The dynamics of a localized surfactant on a thin film.
\newblock \emph{Journal of Fluid Mechanics}, 213:\penalty0 127--148, 1990.

\bibitem[Edmonstone et~al.(2006)Edmonstone, Craster, and
  Matar]{edmonstone2006surfactant}
BD~Edmonstone, RV~Craster, and OK~Matar.
\newblock Surfactant-induced fingering phenomena beyond the critical micelle
  concentration.
\newblock \emph{Journal of Fluid Mechanics}, 564:\penalty0 105--138, 2006.

\bibitem[Kalogirou and Blyth(2019)]{kalogirou2019role}
Anna Kalogirou and MG~Blyth.
\newblock The role of soluble surfactants in the linear stability of two-layer
  flow in a channel.
\newblock \emph{Journal of Fluid Mechanics}, 873:\penalty0 18--48, 2019.

\bibitem[Oron et~al.(1997)Oron, Davis, and Bankoff]{oron1997long}
Alexander Oron, Stephen~H Davis, and S~George Bankoff.
\newblock Long-scale evolution of thin liquid films.
\newblock \emph{Reviews of modern physics}, 69\penalty0 (3):\penalty0 931,
  1997.

\bibitem[Jensen and Grotberg(1993)]{jensen1993spreading}
OE~Jensen and JB~Grotberg.
\newblock The spreading of heat or soluble surfactant along a thin liquid film.
\newblock \emph{Physics of Fluids A: Fluid Dynamics}, 5\penalty0 (1):\penalty0
  58--68, 1993.

\bibitem[Edmonstone and Matar(2004)]{edmonstone2004simultaneous}
BD~Edmonstone and OK~Matar.
\newblock Simultaneous thermal and surfactant-induced marangoni effects in thin
  liquid films.
\newblock \emph{Journal of colloid and interface science}, 274\penalty0
  (1):\penalty0 183--199, 2004.

\bibitem[Mavromoustaki et~al.(2012)Mavromoustaki, Matar, and
  Craster]{mavromoustaki2012dynamics}
A~Mavromoustaki, OK~Matar, and RV~Craster.
\newblock Dynamics of a climbing surfactant-laden film--i: base-state flow.
\newblock \emph{Journal of colloid and interface science}, 371\penalty0
  (1):\penalty0 107--120, 2012.

\bibitem[Hossain et~al.(2024{\natexlab{c}})Hossain, Ghosh, and
  Behera]{hossain2024odd}
Md~Mouzakkir Hossain, Sukhendu Ghosh, and Harekrushna Behera.
\newblock Odd-viscosity induced surfactant-laden shear-imposed viscous film
  over a slippery incline: a stability analysis.
\newblock \emph{Meccanica}, pages 1--21, 2024{\natexlab{c}}.

\bibitem[Yih(1991)]{yih1991stability}
Chia-Shun Yih.
\newblock Stability of liquid flow down an inclined plane.
\newblock In \emph{Selected Papers By Chia-Shun Yih: (In 2 Volumes)}, pages
  357--370. World Scientific, 1991.

\bibitem[Tilton and Cortelezzi(2008)]{tilton2008linear}
Nils Tilton and Luca Cortelezzi.
\newblock Linear stability analysis of pressure-driven flows in channels with
  porous walls.
\newblock \emph{Journal of Fluid Mechanics}, 604:\penalty0 411--445, 2008.

\bibitem[D{\'a}valos-Orozco(2023)]{davalos2023marangoni}
LA~D{\'a}valos-Orozco.
\newblock Marangoni stability of a thin liquid film falling down above or below
  an inclined thick wall with slip.
\newblock \emph{Meccanica}, 58\penalty0 (10):\penalty0 1909--1928, 2023.

\bibitem[Selim and Zakaria(2024)]{selim2024gravity}
RS~Selim and Kadry Zakaria.
\newblock Gravity-driven film flow of a power-law fluid over a wavy substrate
  with slip condition.
\newblock \emph{Journal of Nonlinear Mathematical Physics}, 31\penalty0
  (1):\penalty0 65, 2024.

\bibitem[Jeevahan et~al.(2018)Jeevahan, Chandrasekaran, Britto~Joseph,
  Durairaj, and Mageshwaran]{jeevahan2018superhydrophobic}
Jeya Jeevahan, M~Chandrasekaran, G~Britto~Joseph, RB~Durairaj, and GJJOCT
  Mageshwaran.
\newblock Superhydrophobic surfaces: a review on fundamentals, applications,
  and challenges.
\newblock \emph{Journal of Coatings Technology and Research}, 15:\penalty0
  231--250, 2018.

\bibitem[Chin et~al.(1986)Chin, Abernath, and Bertschy]{chin1986gravity}
RW~Chin, FH~Abernath, and JR~Bertschy.
\newblock Gravity and shear wave stability of free surface flows. part 1.
  numerical calculations.
\newblock \emph{Journal of Fluid Mechanics}, 168:\penalty0 501--513, 1986.

\bibitem[Kelly et~al.(1989)Kelly, Goussis, Lin, and Hsu]{kelly1989mechanism}
RE~Kelly, DA~Goussis, SP~Lin, and FK~Hsu.
\newblock The mechanism for surface wave instability in film flow down an
  inclined plane.
\newblock \emph{Physics of Fluids A: Fluid Dynamics}, 1\penalty0 (5):\penalty0
  819--828, 1989.

\bibitem[Hossain et~al.(2023)Hossain, Tsai, and Behera]{hossain2023instability}
Md~Mouzakkir Hossain, Chia-Cheng Tsai, and Harekrushna Behera.
\newblock Instability mechanism of shear-layered fluid in the presence of a
  floating elastic plate.
\newblock \emph{Physics of Fluids}, 35\penalty0 (2), 2023.

\bibitem[Samanta(2023)]{samanta2023elliptic}
Arghya Samanta.
\newblock An elliptic velocity profile-based two-equation model in viscous
  film.
\newblock \emph{Physics of Fluids}, 35\penalty0 (2), 2023.

\bibitem[Kalogirou and Blyth(2021)]{kalogirou2021instabilities}
Anna Kalogirou and MG~Blyth.
\newblock Instabilities at a sheared interface over a liquid laden with soluble
  surfactant.
\newblock \emph{Journal of Engineering Mathematics}, 129\penalty0 (1):\penalty0
  3, 2021.

\end{thebibliography}
\end{document}